\documentclass[aps,rmp,reprint,amsmath,amssymb,graphicx,longbibliography]{revtex4-1}

\usepackage{amssymb} 
\usepackage{amsmath}  
\usepackage{amsthm}
\usepackage{lgreek}
\usepackage{bm}
\usepackage{color}
\usepackage[usenames,dvipsnames]{xcolor}
\usepackage{graphicx}

\usepackage[squaren]{SIunits}

\begin{document}

\title{Active Particles in Complex and Crowded Environments}

\author{Clemens Bechinger}
\affiliation{(1) 2. Physikalisches Institut, Universit\"at Stuttgart, Pfaffenwaldring 57, 70569 Stuttgart, Germany}
\affiliation{(2) Max-Planck-Institut f\"ur Intelligente Systeme, Heisenbergstra{\ss}e 3, 70569 Stuttgart, Germany}

\author{Roberto Di Leonardo}
\affiliation{(3) Dipartimento di Fisica, Universit\`a ``Sapienza", I-00185, Roma, Italy}
\affiliation{(4) NANOTEC-CNR Institute of Nanotechnology, Soft and Living Matter Laboratory, I-00185 Roma, Italy}

\author{Hartmut L\"owen}
\affiliation{(5) Institut f\"ur Theoretische Physik II: Weiche Materie, Heinrich-Heine-Universit\"at D\"usseldorf, D-40225 D\"usseldorf, Germany}

\author{Charles Reichhardt}
\affiliation{(6) Theoretical Division, Los Alamos National Laboratory, Los Alamos, New Mexico 87545 USA}

\author{Giorgio Volpe}
\affiliation{(7) Department of Chemistry, University College London, 20 Gordon Street, London WC1H 0AJ, United Kingdom}

\author{Giovanni Volpe}
\email{giovanni.volpe@gu.se}
\affiliation{(8) Department of Physics, University of Gothenburg, SE-41296 Gothenburg, Sweden}
\affiliation{(9) Soft Matter Lab, Department of Physics, and UNAM --- National Nanotechnology Research Center, Bilkent University, Ankara 06800, Turkey}

\date{\today{}}

\begin{abstract}
Differently from passive Brownian particles, active particles, also known as self-propelled Brownian particles or microswimmers and nanoswimmers, are capable of taking up energy from their environment and converting it into directed motion. Because of this constant flow of energy, their behavior can only be explained and understood within the framework of nonequilibrium physics. In the biological realm, many cells perform directed motion, for example, as a way to browse for nutrients or to avoid toxins. Inspired by these motile microorganisms, researchers  have been developing artificial particles that feature similar swimming behaviors based on different mechanisms; these manmade micro- and nanomachines hold a great potential as autonomous agents for healthcare, sustainability, and security applications. With a focus on the basic physical features of the interactions of self-propelled Brownian particles with a crowded and complex environment, this comprehensive review will put the reader at the very forefront of the field, providing a guided tour through its basic principles, the development of artificial self-propelling micro- and nanoparticles, and their application to the study of nonequilibrium phenomena, as well as the open challenges that the field is currently facing.
\end{abstract}


\maketitle

\tableofcontents{}

\section{Introduction}\label{sec:intro}

Active matter systems  are able to take energy from their environment and drive themselves far from equilibrium \cite{ramaswamy2010mechnaics}. Thanks to this property, they feature a series of novel behaviors that are not attainable by matter at thermal equilibrium, including, for example, swarming and the emergence of other collective properties \cite{schweitzer2007brownian}. Their study provides great hope to uncover new physics and, simultaneously, to lead to the development of novel strategies for designing smart devices and materials. In recent years, a significant and growing effort has been devoted to advancing this field and to explore its applications in a diverse set of disciplines such as statistical physics \cite{ramaswamy2010mechnaics}, biology \cite{viswanathan2011physics}, robotics \cite{brambilla2013swarm}, social transport \cite{helbing2001traffic}, soft matter \cite{marchetti2013hydrodynamics}, and biomedicine \cite{wang2012nano}.

\begin{figure*}
\includegraphics[width=.75\textwidth]{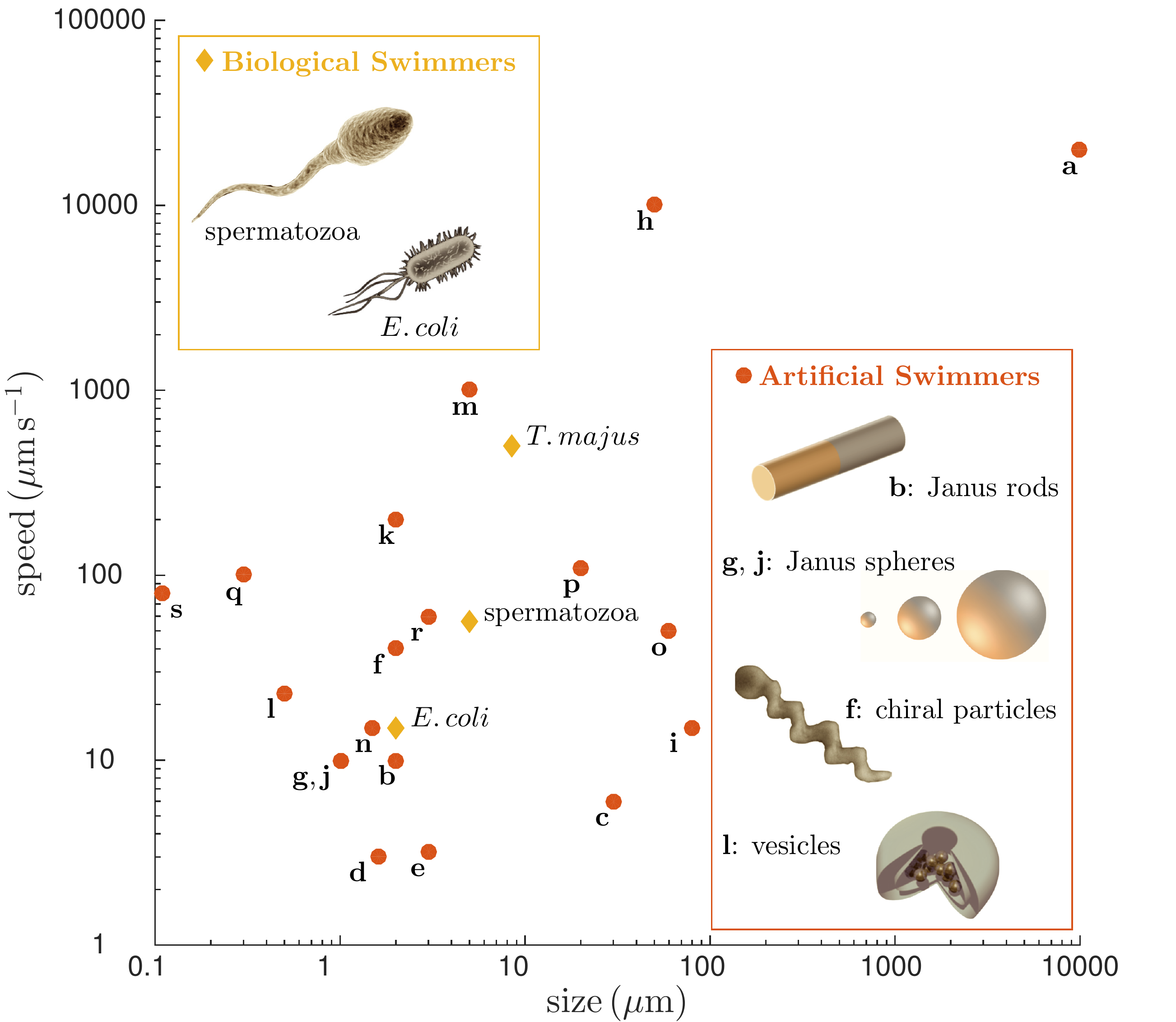}
\caption{(Color online) Self-propelled Brownian particles are biological or manmade objects capable of taking up energy from their environment and converting it into directed motion; they are micro- and nanoscopic in size and have propulsion speeds (typically) up to a fraction of a millimeter per second. The letters correspond to the artificial microswimmers in Table~\ref{tab:realizations}. The insets show examples of biological and artificial swimmers.  For the artificial swimmers four main recurrent geometries can be identified so far: Janus rods, Janus spheres, chiral particles, and vesicles.
\label{F1}
}
\end{figure*}

An important example of active matter is constituted by natural and artificial objects capable of self-propulsion. Self-propelled particles were originally studied to model swarm behavior of animals at the macroscale. \citet{reynolds1987flocks} introduced a `Boids model' to simulate the aggregate motion of flocks of birds, herds of land animals, and schools of fish within computer graphics applications; \citet{vicsek1995novel} then introduced his namesake model as a special case. In the Vicsek model, a swarm is modeled by a collection of self-propelling particles that move with a constant speed but tend to align with the average direction of motion of the particles in their local neighborhood \cite{czirok2000collective,chate2008modeling}. Swarming systems give rise to emergent behaviors, which occur at many different scales; furthermore, some of these behaviors are turning out to be robust and universal, e.g. they are independent of the type of animals constituting the swarm \cite{buhl2006disorder}. It has in fact become a challenge for theoretical physics to find minimal statistical models that capture these features \cite{toner2005hydrodynamics,bertin2009hydrodynamic,li2008minimal}.

Self-propelled Brownian particles, in particular, have come under the spotlight of the physical and biophysical research communities. These active particles are biological or manmade microscopic and nanoscopic objects that can propel themselves by taking up energy from their environment and converting it into directed motion \cite{ebbens2010pursuit}. On the one hand, self-propulsion is a common feature in microorganisms \cite{lauga2009hydrodynamics,cates2012diffusive,poon2013clarkia} and allows for a more efficient exploration of the environment when looking for nutrients or running away from toxic substances \cite{viswanathan2011physics}; a paradigmatic example is the swimming behavior of bacteria such as \emph{Escherichia coli} \cite{berg2004ecoli}. On the other hand, tremendous progress has recently been made towards the fabrication of artificial micro- and nanoswimmers that can self-propel based on different propulsion mechanisms; some characteristic examples of artificial self-propelled Brownian particles are provided in Fig.~\ref{F1} and Table~\ref{tab:realizations}.

\begin{table*}
\caption{Examples of experimentally realized artificial microswimmers and relative propulsion mechanisms. The letters in the first column correspond to the examples plotted in Fig.~\ref{F1}.\label{tab:realizations}}
\begin{ruledtabular}
\begin{tabular}{cp{.29\textwidth}p{.29\textwidth}p{.15\textwidth}p{.11\textwidth}p{.11\textwidth}p{.01\textwidth}}
\centering
&
Microswimmer 
& 
\centering
Propulsion mechanism 
& 
\centering
Medium & 
\centering
Dimensions 
& 
\centering
Max. Speed 
&\\
\hline
{\bf a}
&
\raggedright
PDMS platelets coated with Pt \cite{ismagilov2002autonomous}
&
\raggedright
Bubbles generated in a ${\rm H_2O_2}$ aqueous solution by asymmetric patterns of Pt
&
\centering
${\rm H_2O_2}$ aqueous meniscus
&
\centering
$1 \, {\rm cm}$
&
\centering
$2 \, {\rm cm\, s^{-1}}$
& \\
{\bf b}
&
\raggedright
Rod-shaped particles consisting of Au and Pt segments \cite{paxton2004catalytic}
&
\raggedright
Catalysis of oxygen at the Pt end of the rod
&
\centering
Near a boundary in ${\rm H_2O_2}$ acqueous solution
&
\centering
$2\,{\rm \mu m}$ (length),
$370\,{\rm nm}$ (width)
&
\centering
$10\,{\rm \mu m \, s^{-1}}$
& \\
{\bf c}
&
\raggedright
Linear chains of DNA-linked magnetic colloidal particles attached to red blood cells \cite{dreyfus2005microscopic}
&
\raggedright
External actuation of the flexible artificial flagella by oscillating magnetic fields
&
\centering
Aqueous solution
&
\centering
$30\,{\rm \mu m}$
&
\centering
$6\,{\rm \mu m \, s^{-1}}$
& \\
{\bf d}
&
\raggedright
Janus spherical particles with a catalytic Pt patch \cite{howse2007self}
&
\raggedright
Self-diffusiophoresis catalyzed by a chemical reaction on the Pt surface
&
\centering
${\rm H_2O_2}$ aqueous solution
&
\centering
$1.6\,{\rm \mu m}$
&
\centering
$3\,{\rm \mu m \, s^{-1}}$
&\\
{\bf e}
&
\raggedright
DNA-linked anisotropic doublets composed of paramagnetic colloidal particles \cite{tierno2008controlled}
&
\raggedright
Rotation induced by a rotating magnetic field
&
\centering
Near a boundary in aqueous solution
&
\centering
$3\,{\rm \mu m}$
&
\centering
$3.2\,{\rm \mu m \, s^{-1}}$
&\\
{\bf f}
&
\raggedright
Chiral colloidal propellers \cite{ghosh2009controlled}
&
\raggedright
External actuation by a magnetic field
&
\centering
Aqueous solution
&
\centering
$2\,{\rm \mu m}$ (length),
$250\,{\rm nm}$ (width)
&
\centering
$40\,{\rm \mu m \, s^{-1}}$
&\\
{\bf g}
&
\raggedright
Janus particles half-coated with Au \cite{jiang2010active}
&
\raggedright
Self-thermophoresis due to local heating at the Au cap
&
\centering
Aqueous solution
&
\centering
$1\,{\rm \mu m}$
&
\centering
$10\,{\rm \mu m \, s^{-1}}$
&\\
{\bf h}
&
\raggedright
Catalytic microjets \cite{sanchez2011superfast}
&
\raggedright
${\rm H_2O_2}$ catalysis on the internal surface of the microjet
&
\centering
${\rm H_2O_2}$ aqueous solution
&
\centering
$50\,{\rm \mu m}$ (length),
$1\,{\rm \mu m}$ (width)
&
\centering
$10\,{\rm mm \, s^{-1}}$
&\\
{\bf i}
&
\raggedright
Water droplets containing bromine \cite{thutupalli2011swarming}
&
\raggedright
Marangoni flow induced by a self-sustained bromination gradient along the drop surface
&
\centering
Oil phase containing a surfactant
&
\centering
$80\,{\rm \mu m}$
&
\centering
$15\,{\rm \mu m\,s^{-1}}$
& \\
{\bf j}
&
\raggedright
Janus particles with light-absorbing patches \cite{volpe2011microswimmers,buttinoni2012active,kuemmel2013circular}
&
\raggedright
Local demixing of a critical mixture due to heating associated to localized absorption of light
&
\centering
Critical mixture
(e.g. water-2,6-lutidine)
&
\centering
$0.1$ to $10\,{\rm \mu m}$
&
\centering
$10\,{\rm \mu m\,s^{-1}}$
& \\
{\bf k}
&
\raggedright
Rod-shaped particles consisting of Au and Pt (or Au and Ru) segments \cite{wang2012autonomous}
&
\raggedright
Self-acoustophoresis in a ultrasonic standing wave
&
\centering
Aqueous solution
&
\centering
$1$ to $3\,{\rm \mu m}$ (length),
$300\,{\rm nm}$ (width)
&
\centering
$200\,{\rm \mu m \, s^{-1}}$
& \\
{\bf l}
&
\raggedright
Pt-loaded stomatocytes  \cite{wilson2012autonomous}
&
\raggedright
Bubbles generated in a ${\rm H_2O_2}$ aqueous solution by entrapped Pt nanoparticles
&
\centering
${\rm H_2O_2}$ acqueous solution
&
\centering
$0.5\,{\rm \mu m}$
&
\centering
$23\,{\rm \mu m \, s^{-1}}$
& \\
{\bf m}
&
\raggedright
Colloidal rollers made of PMMA beads  \cite{bricard2013emergence}
&
\raggedright
Spontaneous charge symmetry breaking resulting in a net electrostatic torque
&
\centering
conducting fluid (hexadecane solution)
&
\centering
$5\,{\rm \mu m}$
&
\centering
$1\,{\rm mm \, s^{-1}}$
& \\
{\bf n}
&
\raggedright
Polymeric spheres encapsulating most of an antiferromagnetic hematite cube \cite{palacci2013living}
&
\raggedright
Self-phoretic motion near a boundary due to the decomposition of ${\rm H_2O_2}$ by the hematite cube when illuminated by ultraviolet light
&
\centering
Near a boundary in ${\rm H_2O_2}$ acqueous solution
&
\centering
$1.5\,{\rm \mu m}$
&
\centering
$15\,{\rm \mu m \, s^{-1}}$
& \\
{\bf o}
&
\raggedright
Water droplets \cite{izri2014self-propulsion}
&
\raggedright
Water solubilization by the reverse micellar solution
&
\centering
Oil phase with surfactants above the critical micellar solution
&
\centering
$60\,{\rm \mu m}$
&
\centering
$50\,{\rm \mu m \, s^{-1}}$
& \\
{\bf p}
&
\raggedright
Janus microspheres with Mg core, Au nanoparticles, and ${\rm TiO_2}$ shell layer \cite{li2014water}
&
\raggedright
Bubble thrust generated from the Mg-water reaction
&
\centering
Aqueous solution
&
\centering
$20\,{\rm \mu m}$
&
\centering
$110\,{\rm \mu m \, s^{-1}}$
& \\
{\bf q}
&
\raggedright
Hollow mesoporous silica Janus particles \cite{ma2015enzyme,ma2015catalytic}
&
\raggedright
Catalysis powered by Pt or by three different enzymes (catalase, urease, and glucose oxidase) 
&
\centering
Aqueous solution
&
\centering
$50$ to $500\,{\rm nm}$
&
\centering
$100\,{\rm \mu m \, s^{-1}}$
& \\
{\bf r}
&
\raggedright
Janus particles half-coated with Cr  \cite{nishiguchi2015mesoscopic}
&
\raggedright
AC electric field
&
\centering
Aqueous solution
&
\centering
$3\,{\rm \mu m}$
&
\centering
$60\,{\rm \mu m \, s^{-1}}$
& \\
{\bf s}
&
\raggedright
Enzyme-loaded polymeric vesicles \cite{joseph2016active}
&
\raggedright
Glucose catalysis powered by catalase and glucose oxidase
&
\centering
Aqueous solution
&
\centering
$0.1\,{\rm \mu m}$
&
\centering
$80\,{\rm \mu m \, s^{-1}}$
& \\
\end{tabular}
\end{ruledtabular}
\end{table*}

While the motion of passive Brownian particles is driven by equilibrium thermal fluctuations due to random collisions with the surrounding fluid molecules \cite{babic2005colloids}, self-propelled Brownian particles exhibit an interplay between random fluctuations and active swimming that drives them into a far-from-equilibrium state \cite{erdmann2000brownian,schweitzer2007brownian,hanggi2009artificial,hauser2015statistical}. Thus, their behavior can only be explained and understood within the framework of nonequilibrium physics \cite{cates2012diffusive}, for which they provide ideal model systems.

From a more applied perspective, active particles provide great hope to address some challenges that our society is currently facing --- in particular, personalized healthcare, environmental sustainability, and security \cite{nelson2010microrobots,wang2012nano,patra2013intelligent,abdelmohsen2014micro,gao2014environmental,ebbens2015active}. These potential applications can be built around the core functionalities of self-propelled Brownian particles, i.e. transport, sensing, and manipulation. In fact, these micro- and nanomachines hold the promise of performing key tasks in an autonomous, targeted, and selective way. The possibility of designing, using, and controlling micro- and nanoswimmers in realistic settings of operation is tantalizing as a way to localize, pick-up, and deliver nanoscopic cargoes in several applications --- from the targeted delivery of drugs, biomarkers, or contrast agents in healthcare applications \cite{nelson2010microrobots,wang2012nano,patra2013intelligent,abdelmohsen2014micro} to the autonomous depollution of water and soils contaminated because of bad waste management, climate changes, or chemical terroristic attacks in sustainability and security applications \cite{gao2014environmental}. 

The field of active matter is now confronted with various open challenges that will keep researchers busy for decades to come. First, there is a need to understand how living and inanimate active matter systems develop social and (possibly) tunable collective behaviors that are not attainable by their counterparts at thermal equilibrium. Then, there is a need to understand the dynamics of active particles in real-life environments (e.g. in living tissues and porous soils), where randomness, patchiness, and crowding can either limit or enhance how biological and artificial microswimmers perform a given task, such as finding nutrients or delivering a nanoscopic cargo. Finally, there is still a strong need to effectively scale down to the nanoscale our current understanding of active matter systems.

With this review, we provide a guided tour through the basic principles of self-propulsion at the micro- and nanoscale, the development of artificial self-propelling micro- and nanoparticles, and their application to the study of far-from-equilibrium phenomena, as well as through the open challenges that the field is now facing.

\section{Non-interacting Active Particles in Homogenous Environments}\label{sec:noninteracting}

Before proceeding to analyze the behavior of active particles in crowded and complex environments, we will set the stage by considering the simpler (and more fundamental) case of individual active particles in homogeneous environments, i.e. without obstacles or other particles. We will first introduce a simple model of an active Brownian particle,\footnote{We remark that the term ``active Brownian particle" has mainly been used in the literature to denote the specific, simplified model of active matter described in this section, which consists of repulsive spherical particles that are driven by a constant force whose direction rotates by thermal diffusion. In the following, we will use the term ``active Brownian particle" when we refer to this specific model and its straightforward generalizations (see Section~\ref{sec:chiral}), while we will use the terms ``active particle" or ``self-propelled particle" when we refer to more general systems.} which will permit us to understand the main differences between passive and active Brownian motion (Section~\ref{sec:versus}) and serve as a starting point to discuss the basic mathematical models for active motion (Section~\ref{sec:phenom}). We will then introduce the concepts of effective diffusion coefficient and effective temperature for self-propelled Brownian particles, as well as their limitations, i.e. differences between systems at equilibrium at a higher temperature and systems out of equilibrium (Section~\ref{sec:temp}). We will then briefly review biological microswimmers (Section~\ref{sec:bio}). Finally, we will conclude with an overview of experimental achievements connected to the realization of artificial micro- and nanoswimmers including a discussion of the principal experimental approaches that have been proposed so far to build active particles (Section~\ref{sec:art}).

\subsection{Brownian motion vs. active Brownian motion}\label{sec:versus}

\begin{figure*}[t!]
\includegraphics[width=\textwidth]{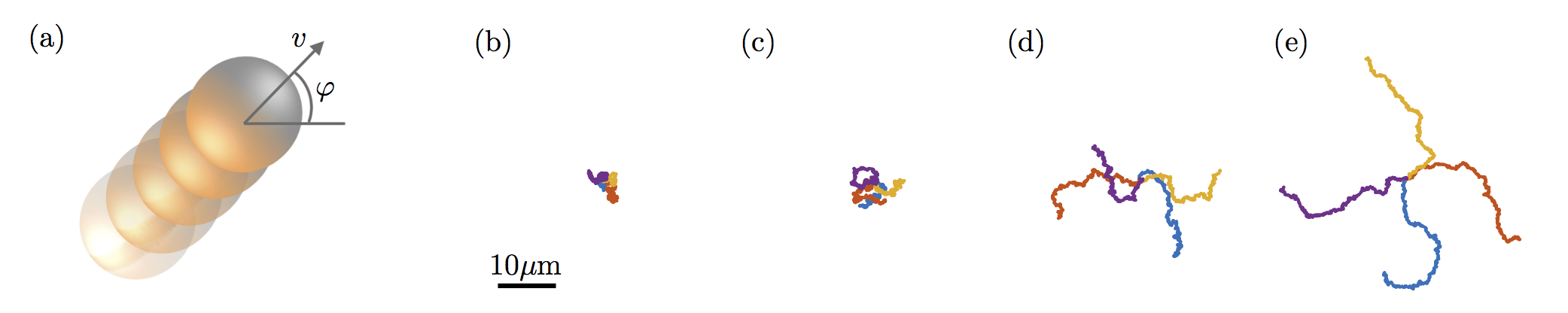}
\caption{(Color online) Active Brownian particles in two dimensions. (a) An active Brownian particle in water ($R = 1\,{\rm \mu m}$, $\eta = 0.001\,{\rm Pa \, s}$) placed at position $[x,y]$ is characterized by an orientation $\varphi$ along which it propels itself with speed $v$ while undergoing Brownian motion in both position and orientation. The resulting trajectories are shown for different velocities (b) $v=0\,{\rm \mu m\,s^{-1}}$ (Brownian particle), (c) $v=1\,{\rm \mu m\,s^{-1}}$, (d) $v=2\,{\rm \mu m\,s^{-1}}$, and (e) $v=3\,{\rm \mu m\,s^{-1}}$; with increasing values of $v$, the active particles move over longer distances before their direction of motion is randomized; four different 10-s trajectories are shown for each value of velocity. 
\label{F2}
}
\end{figure*}

In order to start acquiring some basic understanding of the differences between passive and active Brownian motion, a good (and pedagogic) approach is to compare two-dimensional trajectories of single spherical passive and active particles of equal (hydrodynamic) radius $R$ in a homogenous environment, i.e. where no physical barriers or other particles are present and where there is a homogeneous and constant distribution of the energy source for the active particle.

The motion of a passive Brownian particle is purely diffusive with translational diffusion coefficient
\begin{equation}\label{eq:Dt}
D_{\rm T} = \frac{k_{\rm B} T}{6\pi\eta R} \; ,
\end{equation}
where $k_{\rm B}$ is the Boltzmann constant, $T$ the absolute temperature, and $\eta$ the fluid viscosity. The particle also undergoes rotational diffusion with a characteristic time scale $\tau_{\rm R}$ given by the inverse of the particle's rotational diffusion coefficient
\begin{equation}\label{eq:Dr}
D_{\rm R} = \tau_{\rm R}^{-1} = \frac{k_{\rm B} T}{8\pi\eta R^3} \; .
\end{equation}
From the above formulas, it is clear that, while the translational diffusion of a particle scales with its linear dimension, its rotational diffusion scales with its volume. For example, for a particle with $R \approx 1\,{\rm \mu m}$ in water, $D_{\rm T} \approx 0.2\,{\rm \mu m^2 \, s^{-1}}$  and $D_{\rm R} \approx 0.17\,{\rm rad^2 \, s^{-1}}$ ($\tau_{\rm R}\approx6\,{\rm s}$), while, for a particle 10 times smaller ($R \approx 100\,{\rm nm}$), $D_{\rm T}\approx 2\,{\rm \mu m^2 \, s^{-1}}$ is one order of magnitude larger but $D_{\rm R} \approx 170\,{\rm rad^2 \, s^{-1}}$ is three orders of magnitude larger ($\tau_{\rm R} \approx 6\,{\rm ms}$).

In a homogeneous environment, the translational and rotational motions are independent from each other. Therefore, the stochastic equations of motion for a \emph{passive Brownian particle} in a two-dimensional space are
\begin{equation}\label{eq:pbm}
\left\{\begin{array}{ccl}
\displaystyle \dot{x} & = & \displaystyle \sqrt{2D_{\rm T}} \, \xi_x \\[6pt]
\displaystyle \dot{y} & = & \displaystyle \sqrt{2D_{\rm T}} \, \xi_y \\[6pt]
\displaystyle \dot{\varphi} & = & \displaystyle \sqrt{2D_{\rm R}} \, \xi_\varphi
\end{array}\right.
\end{equation}
where $[x,y]$ is the particle position, $\varphi$ is its orientation (Fig.~\ref{F2}a), and $\xi_x$, $\xi_y$, and $\xi_\varphi$ represent independent white noise stochastic processes with zero mean and correlation $\delta(t)$. Interestingly, the equations for each degree of freedom (i.e. $x$, $y$, and $\varphi$) are decoupled. Inertial effects are neglected because microscopic particles are typically in a low-Reynolds-number regime \cite{purcell1977life}. Some examples of the corresponding trajectories are illustrated in Fig.~\ref{F2}b.

For a \emph{self-propelled particle} with velocity $v$ instead, the direction of motion is itself subject to rotational diffusion, which leads to a coupling between rotation and translation. The corresponding stochastic differential equations are
\begin{equation}\label{eq:abm}
\left\{\begin{array}{ccl}
\displaystyle \dot{x} & = & v \displaystyle \cos\varphi + \sqrt{2D_{\rm T}} \, \xi_x \\[6pt]
\displaystyle \dot{y} & = & v \displaystyle \sin\varphi + \sqrt{2D_{\rm T}} \, \xi_y \\[6pt]
\displaystyle \dot{\varphi} & = & \displaystyle \sqrt{2D_{\rm R}} \, \xi_\varphi
\end{array}\right.
\end{equation}
Some examples of trajectories for various $v$ are shown in Figs.~\ref{F2}c, \ref{F2}d, and \ref{F2}e: as $v$ increases, we obtain active trajectories that are characterized by directed motion at short time scales; however, over long time scales the orientation and direction of motion of the particle are randomized by its rotational diffusion \cite{howse2007self}.

To emphasize the difference between Brownian motion and active Brownian motion, it is instructive to consider the average particle trajectory given the initial position and orientation fixed at time $t=0$, i.e. $x(0) = y(0) = 0$ and $\varphi(0) = 0$. In the case of passive Brownian motion, this average vanishes by symmetry, i.e. $\langle x(t) \rangle = \langle y(t) \rangle \equiv 0$, where $\langle ... \rangle$ represents the ensemble average. For an active particle instead, the average is a straight line along the $x$-direction (determined by the prescribed initial orientation),
\begin{equation}\label{eq:straightav}
\langle x(t) \rangle = \frac{v}{D_{\rm R}} \left[ 1-\exp(-D_{\rm R} t) \right] = v \tau_{\rm R} \left[ 1 - \exp\left(-\frac{t}{\tau_{\rm R}}\right) \right] \; ,
\end{equation}
while $ \langle y(t) \rangle \equiv 0$ because of symmetry. This implies that, on average, an active Brownian particle will move along the direction of its initial orientation for a finite \emph{persistence length}
\begin{equation}\label{eq:L}
L = \frac{v}{D_{\rm R}} = v \, \tau_{\rm R}\; ,
\end{equation}
before its direction is randomized.

The relative importance of directed motion versus diffusion for an active Brownian particle can be characterized by its \emph{P\'eclet number}
\begin{equation}\label{eq:pe}
{\rm Pe} \propto \frac{v}{\sqrt{D_{\rm T}D_{\rm R}}} \; ,
\end{equation}
where the proportionality sign is used because the literature is inconsistent about the value of the numerical prefactor. If ${\rm Pe}$ is small, diffusion is important, while, if ${\rm Pe}$ is large, directed motion prevails. 

The model for active Brownian motion described by Eqs.~(\ref{eq:abm}) can be straightforwardly generalized to the case of an active particle moving in three dimensions. In this case, the particle position is described by three Cartesian coordinates, i.e. $\left[x,y,z\right]$, and its orientation by the polar and azimuthal angles, i.e. $\left[\vartheta,\varphi\right]$, which perform a Brownian motion on the unit sphere \cite{carlsson2010algorithm}.

\subsection{Phenomenological models}\label{sec:phenom}

In this section, we will extend the simple model introduced in Section~\ref{sec:versus} to describe the motion of more complex (and realistic) active Brownian particles. First, we will introduce models that account for chiral active Brownian motion (Section~\ref{sec:chiral}). We will then consider more general models of active Brownian motion where reorientation occurs due to mechanisms other than rotational diffusion (Section~\ref{sec:orientation}) and where the active particles are non-spherical (Section~\ref{sec:non-spherical}). Finally, we will discuss the use of external forces and torques when modeling active Brownian motion (Section~\ref{sec:externalforces}) and we will provide some considerations about numerics (Section~\ref{sec:numerical}).

Before proceeding further, we remark that in this section the microscopic swimming mechanism is completely ignored; in particular hydrodynamic interactions are disregarded and only the observable effects of net motion are considered. While the models introduced here are phenomenological, they are very effective to describe the motion of microswimmers in homogenous environments. We can cast this point in terms of the difference between ``microswimmers" and ``active particles". Microswimmers are force-free and torque-free objects capable of self-propulsion in a (typically) viscous environment and, importantly, exhibit an explicit hydrodynamic coupling with the embedding solvent via flow fields generated by the swimming strokes they perform. Instead, active particles represent a much simpler concept consisting of self-propelled particles in an inert solvent, which only provides hydrodynamic friction and a stochastic momentum transfer. While the observable behavior of the two is the same in a homogenous environment and in the absence of interactions between particles, hydrodynamic interactions may play a major role in the presence of obstacles or other microswimmers. The simpler model of active particles delivers, however, good results in terms of the particle's behavior and is more intuitive. In fact, the self-propulsion of an active Brownian particle is implicitly modeled by using an effective force fixed in the particle's body frame. For this reason, in this review we typically consider active particles, while we will discuss hydrodynamic interactions in Section~\ref{sec:hydro} (see also \citet{golestanian2010hydrodynamics}, \citet{marchetti2013hydrodynamics} and \citet{elgeti2015physics} for extensive reviews on the role of hydrodynamics in active matter systems). We will provide a more detailed theoretical justification of why active particles are a good model for microswimmers in Section~\ref{sec:externalforces}.

\subsubsection{Chiral active Brownian motion}\label{sec:chiral}

\begin{figure}[b!]
\includegraphics[width=\columnwidth]{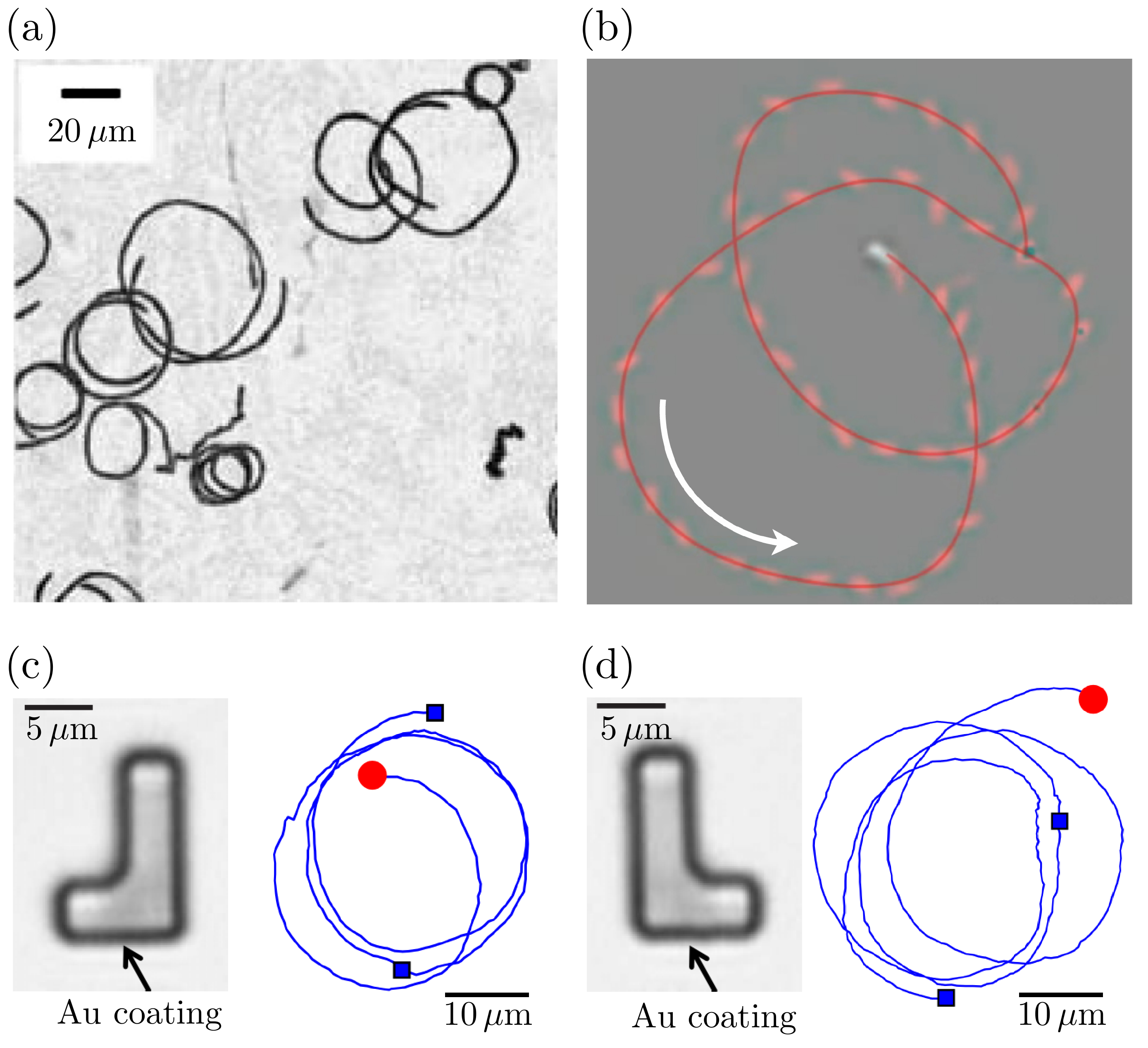}
\caption{(Color online) Biological and artificial chiral active Brownian motion. (a) Phase-contrast video microscopy images showing \emph{E. coli} cells swimming in circular trajectories near a glass surface. Superposition of 8-s video images.
From \citet{lauga2006swimming}.
(b) Circular trajectories are also observed for \emph{E. coli} bacteria swimming over liquid--air interfaces but the direction is reversed.
From \citet{dileonardo2011swimming}.
(c-d) Trajectories of a (c) dextrogyre and (d) levogyre artificial microswimmer driven by self-diffusiophoresis: in each plot, the red bullet correspond to the initial particle position and the two blue squares to its position after 1 and 2 minutes. The insets show microscope images of two different swimmers with the Au coating (not visible in the bright-field image) indicated by an arrow.
From \citet{kuemmel2013circular}.
\label{F3}
}
\end{figure}

\begin{figure*}[t!]
\includegraphics[width=.75\textwidth]{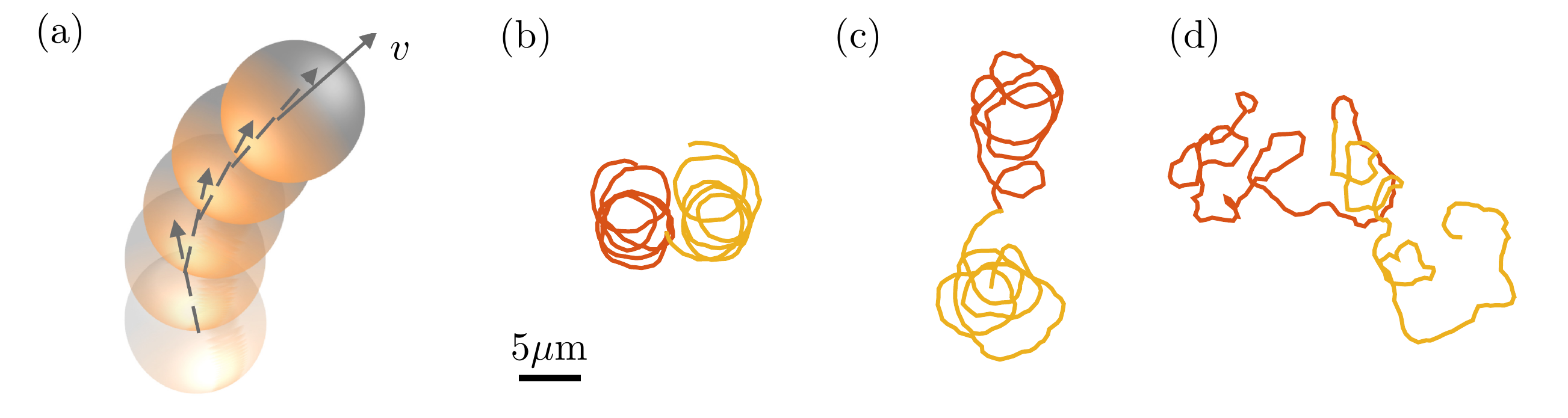}
\caption{(Color online) Chiral active Brownian motion in two dimensions. (a) A two-dimensional chiral active Brownian particle has a deterministic angular velocity $\omega$ that, if the particle's speed $v > 0$, entails a rotation around an effective external axis. (b-d) Sample trajectories of dextrogyre (red/dark gray) and levogyre (yellow/light gray) active chiral particles with $v = 30\,{\rm \mu m\,s^{-1}}$, $\omega = 10\,{\rm rad\,s^{-1}}$, and different radii ($R = 1000$, $500$ and $250\,{\rm nm}$ for (b), (c) and (d), respectively); as the particle size decreases, the trajectories become less deterministic because the rotational diffusion, responsible for the reorientation of the particle direction, scales according to $R^{-3}$ (Eq.~(\ref{eq:Dr})).
\label{F4}
}
\end{figure*}

Swimming along a straight line --- corresponding to the linearly directed Brownian motion considered until now --- is the exception rather than the rule. In fact, ideal straight swimming only occurs if the left--right symmetry relative to the internal propulsion direction is not broken; even small deviations from this symmetry destabilize any straight motion and make it chiral.  One can assign a chirality (or helicity) to the path, the sign of which determines whether the motion is clockwise (dextrogyre) or anti-clockwise (levogyre). The result is a motion along circular trajectories in two dimensions (circle swimming) and along helical trajectories in three dimensions (helical swimming).

The occurrence of microorganisms swimming in circles was pointed out more than a century ago by \citet{jennings1901significance} and, since then, has been observed in many different situations, in particular close to a substrate for bacteria \cite{berg1990chemotaxis,diluzio2005escherichia,lauga2006swimming,hill2007hydrodynamic,shenoy2007kinematic,schmidt200non-gaussian} and spermatozoa \cite{woolley2003motility,riedel2005self,friedrich2008stochastic}. Likewise, helical swimming in three dimensions has been observed for various bacteria and sperm cells \cite{jennings1901significance,brokaw1958chemotaxis,brokaw1959random,crenshaw1996look,fenchel1999motile,thar2001true,mchenry2003kinematics,corkidi2008tracking,jekely2008mechanism}. Figures~\ref{F3}a and \ref{F3}b show examples of \emph{E. coli} cells swimming in circular trajectories near a glass surface and at a liquid--air interface, respectively. Examples of non-living but active particles moving in circles are spherical camphors at an air--water interface \cite{nakata1997self-rotation} and chiral (L-shaped) colloidal swimmers on a substrate \cite{kuemmel2013circular}. Finally, trajectories of deformable active particles \cite{ohta2009deformable} and even of completely blinded and ear-plugged pedestrians \cite{obata2005fluctuations} can possess significant circular characteristics.

The origin of chiral motion can be manifold. In particular, it can be due to an anisotropy in the particle shape, which leads to a translation--rotation coupling in the hydrodynamic sense \cite{kraft2013brownian}, or an anisotropy in the propulsion mechanism; \citet{kuemmel2013circular} studied experimentally an example where both mechanisms are simultaneously present (Figs.~\ref{F3}c and \ref{F3}d). Furthermore, even a cluster of non-chiral swimmers, which stick together by direct forces \cite{redner2013reentrant}, by hydrodynamics, or just by the activity itself \cite{palacci2013living,buttinoni2013dynamical}, will in general lead to situations of total non-vanishing torque on the cluster center \cite{kaiser2015active}, thus leading to circling clusters \cite{schwarz2012phase}. Finally, the particle rotation can be induced by external fields; a standard example is a magnetic field perpendicular to the plane of motion exerting a torque on the particles \cite{cebers2011diffusion}. We remark that, even though the emergence of circular motion can be often attributed to hydrodynamic effects, in this section we will focus on a phenomenological description and leave the proper hydrodynamic description to Section~\ref{sec:boundaries}.

For a two-dimensional chiral active Brownian particle (Fig.~\ref{F4}a), in addition to the random diffusion and the internal self-propulsion modeled by Eqs.~(\ref{eq:abm}), the particle orientation $\varphi$ also rotates with angular velocity $\omega$, where the sign of $\omega$ determines the chirality of the motion. The resulting set of equations describing this motion in two dimensions is \cite{vanteeffelen2008dynamics,mijalkov2013sorting,volpe2014simulation}
\begin{equation}\label{eq:chiral}
\left\{\begin{array}{ccl}
\displaystyle \dot{x} & = & \displaystyle v \cos\varphi + \sqrt{2D_{\rm T}} \, \xi_x \\[6pt]
\displaystyle \dot{y} & = & \displaystyle v \sin\varphi + \sqrt{2D_{\rm T}} \, \xi_y \\[6pt]
\displaystyle \dot{\varphi} & = & \displaystyle \omega + \sqrt{2D_{\rm R}} \, \xi_\varphi
\end{array}\right.
\end{equation}
Some examples of trajectories are shown in Figs.~\ref{F4}b, \ref{F4}c, and \ref{F4}d for particles of decreasing radius. As the particle size decreases, the trajectories become less deterministic because the rotational diffusion, responsible for the reorientation of the particle direction, scales according to $R^{-3}$ (Eq.~(\ref{eq:Dr})). The model given by Eqs.~(\ref{eq:chiral}) can be straightforwardly extended to the helicoidal motion of a three-dimensional chiral active particle following an approach along the lines of the discussion at the end of Section~\ref{sec:versus}.

It is interesting to consider how the noise-averaged trajectory given in Eq.~(\ref{eq:straightav}) changes in the presence of chiral motion. In this case, the noise-averaged trajectory has the shape of a logarithmic spiral, i.e. a {\it spira mirabilis} \cite{vanteeffelen2008dynamics}, which in polar coordinates is written as
\begin{equation}
\rho \propto \exp \left[- D_{\rm R}( \varphi - \varphi(0))/\omega \right] \; ,
\end{equation}
where $\rho$ is the radial coordinate and $\varphi$ is the azimuthal coordinate. In three dimensions, the noise-averaged trajectory is a concho-spiral \cite{wittkowski2012self-propelled}, which is the generalization of the logarithmic spiral. Stochatic helical swimming was  recently investigated in \emph{Colonial Choanoflagellates} \cite{kirkegaard2016motility}.

\begin{figure*}[t!]
\includegraphics[width=.75\textwidth]{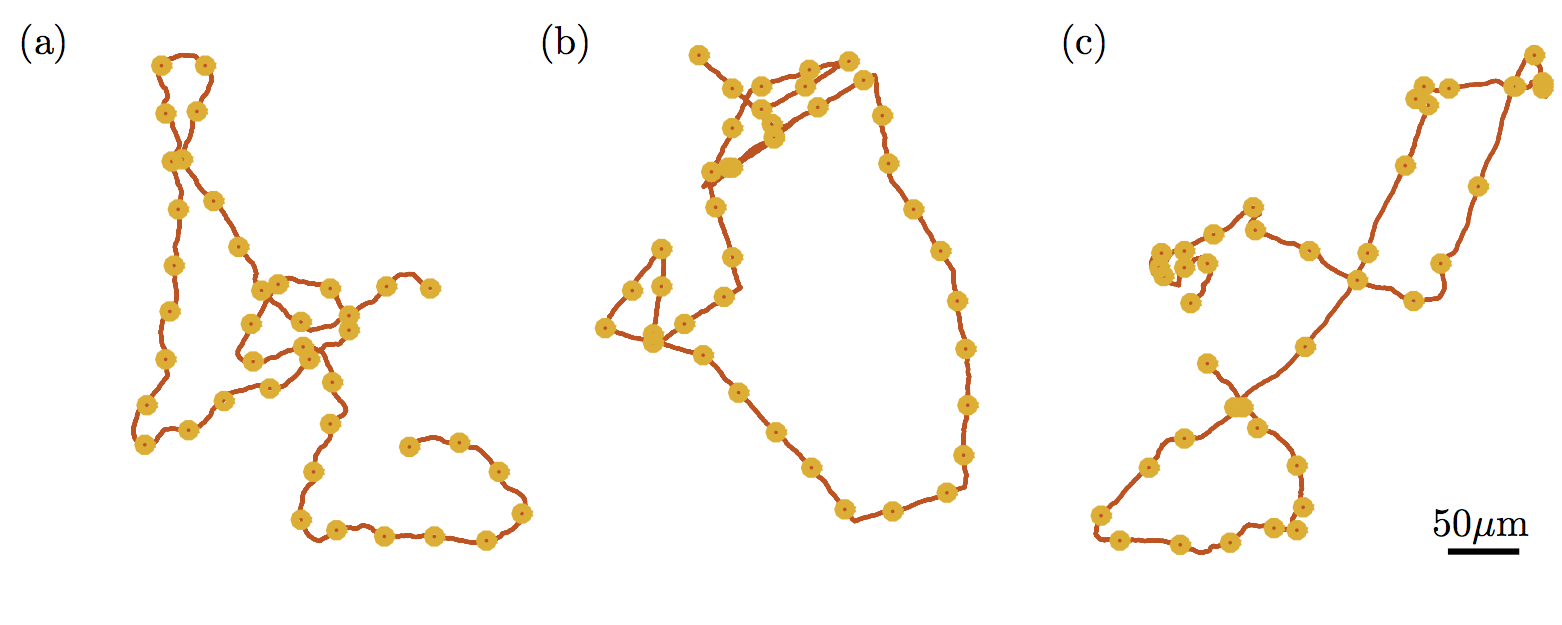}
\caption{(Color online) Sample trajectories of active Brownian particles corresponding to different mechanisms generating active motion: (a) rotational diffusion dynamics; (b) run-and-tumble dynamics; and (c) Gaussian noise dynamics. The dots correspond to the particle position sampled every $5\,{\rm s}$.
\label{F5}
}
\end{figure*}

\subsubsection{Models for active particle reorientation}\label{sec:orientation}

The simple models presented so far, and in particular the one discussed in Section~\ref{sec:versus}, consider an active  particle whose velocity is constant in modulus and whose orientation undergoes free diffusion. This type of dynamics --- to which we shall refer as \emph{rotational diffusion dynamics} (Fig.~\ref{F5}a) --- is often encountered in the case of self-propelling Janus colloids \cite{howse2007self,buttinoni2012active,palacci2013living}. There are, however, other processes that generate active Brownian motion; here we will consider, in particular, the \emph{run-and-tumble dynamics} and the \emph{Gaussian noise dynamics} \cite{koumakis2014directed}. More general models include also velocity- and space-dependent friction \cite{babel2014swimming,taktikos2011modeling,romanczuk2012active}. It has also been recently speculated that finite-time correlations in the orientational dynamics can affect the swimmer's diffusivity \cite{ghosh2015memory}.

The run-and-tumble dynamics (Fig.~\ref{F5}b) were introduced to describe the motion of \emph{E. coli} bacteria \cite{berg1979movement,schnitzer1990strategies,berg2004ecoli}. They consist of a random walk that alternates linear straight runs at constant speed with Poisson-distributed reorientation events called tumbles. Even though their microscopic (short-time) dynamics are different, their long-time diffusion properties are equivalent to those of the rotational diffusion dynamics described in Section~\ref{sec:versus} \cite{tailleur2008statistical,cates2013active,solon2015active}. 

In the Gaussian noise dynamics (Fig.~\ref{F5}c), the active particle velocity (along each direction) fluctuates as an Ornstein--Uhlenbeck process \cite{uhlenbeck1930theory}. This is, for example, a good model for the motion of colloidal particles in a bacterial bath, where multiple interactions with the motile bacteria tend to gradually change the direction and amplitude of the particle's velocity, at least as long as the concentration is not so high to give rise to collective phenomena \cite{wu2000particle}.

Finally, we can also consider the interesting limit of the rotational diffusion dynamics when the rotational diffusion is zero, or similarly in the run-and-tumble dynamics when the run time is infinite. In this case, the equations of motion of the active particle contain no stochastic terms and the particle keeps on moving ballistically along straight lines until it interacts with some obstacles or other particles. Such a limit is reached, e.g., for sufficiently large active colloids or for active colloids moving through an extremely viscous fluid.

\subsubsection{Non-spherical active particles}\label{sec:non-spherical}

The models presented until now --- in particular, Eqs.~(\ref{eq:abm}) and (\ref{eq:chiral}) ---  are valid for spherical active particles. However, while most active particles considered in experiments and simulations are spherically or axially symmetric, many bacteria and motile microorganisms considerably deviate from such ideal shapes, and this strongly alters their swimming properties.

In order to understand how we can derive the equations of motion for non-spherical active Brownian particles, it is useful to rewrite in a vectorial form the model presented in Section~\ref{sec:versus} for the simpler case of a spherical active particle:
\begin{equation}\label{eq:effectiveforce}
\gamma \dot{\bf r} = F \hat{\bf e} + \boldsymbol{\xi} \; ,
\end{equation}
where $\gamma =6\pi \eta R$ denotes the particle's Stokes friction coefficient (for a sphere of radius $R$ with sticky boundary conditions at the particle surface), ${\bf r}$ is its position vector, $F$ is an effective force acting on the particle, $\hat{\bf e}$ is its orientation unit vector, and $\boldsymbol{\xi}$ is a random vector with zero mean and correlation $2 k_{\rm B} T \gamma \, \mathbb{I} \, \delta(t)$, where $\mathbb{I}$ is the identity matrix in the appropriate number of dimensions. If the particle's orientation does not change, i.e. $\hat{\bf e}(t)\equiv\hat{\bf e}(0)$ (e.g. being fixed by an external aligning magnetic field), the particle swims with a self-propulsion speed $v = F/\gamma$ along its orientation $\hat{\bf e}$ and its trajectory is trivially given by ${\bf r}(t) = {\bf r}(0) + v t \, \hat{\bf e}(0)$. If the particle orientation can instead change, e.g. if $\hat{\bf e}$ is subject to rotational diffusion, the particle will perform active Brownian motion.

We can now generalize these simple considerations for a spherical particle to more complex shapes, as systematically discussed in \citet{tenhagen2015self-propulsion}. When the particle has a rigid anisotropic shape, the resulting equations of motion can be written in compact form as 
\begin{equation}\label{eq:v0gen2}
\mathbb{H} \cdot {\bf V} = {\bf K} + \boldsymbol{\chi} \; ,
\end{equation}
where $\mathbb{H}$ is the \emph{grand resistance matrix} or \emph{hydrodynamic friction tensor} (see also Section~\ref{sec:microhydrodynamics}) \cite{happel1991low,fernandes2002brownian}, ${\bf V}=[{\bf v}, \boldsymbol{\omega}]$ is a  generalized  velocity with $\boldsymbol{v}$ and $\boldsymbol{\omega}$ the particle's translational and angular velocities, ${\bf K}=[{\bf F},{\bf T}]$ is a generalized force with ${\bf F}$ and ${\bf T}$ the effective force and torque acting on the particle, and $\boldsymbol{\chi}$ is a random vector with correlation $2 k_{\rm B} T \, \mathbb{H} \,\delta(t)$.
Equation~(\ref{eq:v0gen2}) is best understood in the body frame of the moving particle where $\mathbb{H}$, ${\bf K}$, and ${\bf V}$ are constant, but it can also be transformed to the laboratory frame \cite{wittkowski2012self-propelled}. In the deterministic limit (i.e. $\boldsymbol{\chi}={\bf 0}$), the particle trajectories are straight lines if $\boldsymbol{\omega}=\boldsymbol{0}$, and circles in two dimensions (or helices in three dimensions) if $\boldsymbol{\omega}\neq\boldsymbol{0}$ \cite{friedrich2009steering,wittkowski2012self-propelled}. In the opposite limit when ${\bf K}=0$, we recover the case of a free Brownian particle, which however features nontrivial dynamical correlations \cite{makino2004brownian,fernandes2002brownian,kraft2013brownian,cichoki2015brownian}.

\subsubsection{Modeling active motion with external forces and torques}\label{sec:externalforces}

Equation~(\ref{eq:effectiveforce}) describes the motion of a spherical active particle using an effective ``internal" force $F = \gamma v$ fixed in the particle's body frame. $F$ is identical to the force acting on a hypothetical spring whose ends are bound to the microswimmers and to the laboratory \cite{takatori2014swim}; hence, it can be  directly measured, at least in principle. While this force can be viewed as a special force field ${\bf F}({\bf r}, \hat{\bf e}) = F \hat{\bf e}$ experienced by the particle, it is clearly non-conservative, i.e. it cannot be expressed as a spatial gradient of a scalar potential. The advantage in modeling self-propulsion by an effective driving force is that this force can be straightforwardly added to all other existing forces, e.g. body forces from real external fields (like gravity or confinement), forces stemming from the interaction with other particles, and random forces mimicking the random collisions with the solvent. This keeps the model simple, flexible and transparent. This approach has been followed by  many recent works; see, e.g., \citet{chen2006rotating}, \citet{peruani2006nonequilibrium}, \citet{li2008minimal}, \citet{vanteeffelen2008dynamics}, \citet{mehandia2008collective}, \citet{wensink2008aggregation}, \citet{tenhagen2011dynamics}, \citet{angelani2011active}, \citet{kaiser2012capture}, \citet{kaiser2013capturing}, \citet{yang2012using}, \citet{wittkowski2012self-propelled}, \citet{wensink2012emergent}, \citet{mccandlish2012spontaneous}, \citet{bialke2012crystallization}, \citet{redner2013structure}, \citet{reichhardt2013active}, \citet{elgeti2013wall}, \citet{mijalkov2013sorting}, \citet{fily2014freezing}, \citet{costanzo2014motility} and \citet{wang2014diffusion}.

These simple considerations for a spherical active particle can be generalized to more complex situations, such as to Eqs.~(\ref{eq:v0gen2}) for non-spherical active particles. In general, the following considerations hold to decide whether a model based on effective forces and torques can be applied safely.
On the one hand, the effective forces and torques can be used if we consider a single particle in an unbounded fluid whose propulsion speed is a generic explicit function of time \cite{babel2014swimming} or of the particle's position \cite{magiera2015trapping}.
On the other hand, body forces and torques  arising, e.g., from an external field or from (non-hydrodynamic) particle interactions can simply be added to the effective forces and torques, under the sole assumption that the presence of the body forces and torques should not affect the self-propulsion mechanism itself; a classical counter-example to this assumption are bimetallic nanorods driven by electrophoresis in an external electric field \cite{paxton2004catalytic,paxton2006catalytically}, as the external electric field perturbs the transport of ions through the rod and the screening around it, and thus significantly affects its propagation \cite{brown2014ionic}.

In order to avoid a potential confusion, we remark that the use of effective forces does not imply that the solvent flow field is modeled correctly; contrarily, the flow field is not considered at all. When the propagation is generated by a non-reciprocal mechanical motion of different parts of the swimmer, any internal motion should fulfill Newton's third law such that the total force acting on the swimmer is zero at any time. As we will see more in detail in Section~\ref{sec:hydro}, this implies that the solvent velocity field ${\bf u}({\bf r})$ around a swimmer does not decay as a force monopole (i.e. ${\bf u}({\bf r}) \propto 1/r$, as if the particle were dragged by a constant external force field), but (much faster) as a force dipole (i.e. ${\bf u}({\bf r}) \propto 1/r^2$). The notion of an effective internal force, therefore, seems to contradict this general statement that the motion of a swimmer is force-free. The solution of this apparent contradiction is that the modeling via an effective internal force does {\it not} resolve the solvent velocity field but is just a coarse-grained effective description for swimming with a constant speed along the particle trajectory. Therefore, the concept of effective forces and torques is of limited utility when the solvent flow field, which is generated by the self-propelled particles, has to be taken into account explicitly. This applies, for example, to the far field of the solvent flow that governs the dynamics of a particle pair (and discriminates between pullers and pushers \cite{downton2009simulation}), to the hydrodynamic interaction between a particle and an obstacle \cite{kreuter2013transport,takagi2014hydrodynamic,chilukuri2014impact,sipos2015hydrodynamic}, and to the complicated many-body hydrodynamic interactions in a dense suspension of swimmers \cite{kapral2008multiparticle,gompper2009multi-scale,alexander2009hydrodynamics,reigh2012synchronization}. Nonetheless, there are various situations where hydrodynamic interactions do not play a major role; this is the case for dry active matter \cite{marchetti2013hydrodynamics}, for effects close to a substrate where lubrication is dominating, and for highly crowded environments where the hydrodynamic interactions can cancel if no global flow is built up \cite{wioland2014confinement}.

\subsubsection{Numerical considerations}\label{sec:numerical}

Numerically, the continuous-time solution to the set of stochastic differential equations given by Eqs.~(\ref{eq:abm}), as well as for the other equations presented in this section, can be obtained by approximating it with a set of finite difference equations \cite{ermak1978brownian,volpe2013simulation,volpe2014simulation}. Even though in most practical applications the simple first-order scheme works best, care has to be taken to choose the time step small enough; higher-order algorithms can also be employed to obtain faster convergence of the solution \cite{honerkamp1993stochastic,kloeden1999numerical} and to deal correctly with interactions with obstacles or other particles \cite{behringer2011hard,behringer2012brownian}.

\subsection{Effective diffusion coefficient and effective temperature}\label{sec:temp}

\begin{figure}[b]
\includegraphics[width=\columnwidth]{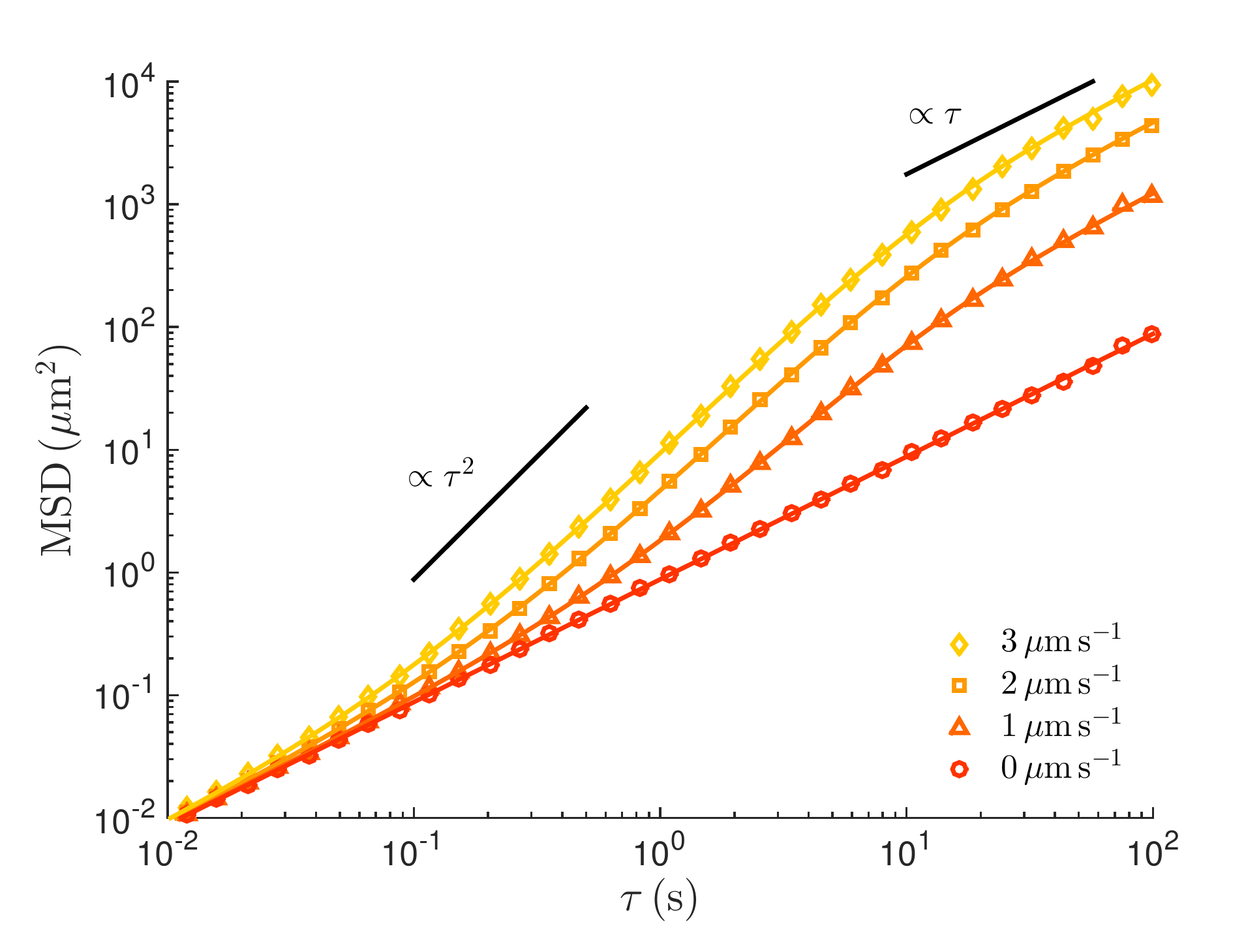}
\caption{(Color online) Mean square displacement (MSD) of active Brownian particles and effective diffusion coefficients. Numerically calculated (symbols) and theoretical (lines) MSD for an active Brownian particle with velocity $v=0\,{\rm \mu m\,s^{-1}}$ (circles), $v=1\,{\rm \mu m\,s^{-1}}$ (triangles), $v=2\,{\rm \mu m\,s^{-1}}$ (squares), and $v=3\,{\rm \mu m\,s^{-1}}$ (diamonds). For a passive Brownian particle ($v=0\,{\rm \mu m\,s^{-1}}$, circles) the motion is always diffusive (${\rm MSD}(\tau) \propto \tau$), while for an active Brownian particle the motion is diffusive with diffusion coefficient $D_{\rm T}$ at very short time scales (${\rm MSD}(\tau) \propto \tau$ for $\tau \ll \tau_{\rm R}$), ballistic at intermediate time scales (${\rm MSD}(\tau) \propto \tau^2$ for $\tau \approx \tau_{\rm R}$), and again diffusive but with an enhanced diffusion coefficient at long time scales (${\rm MSD}(\tau) \propto \tau$ for $\tau \gg \tau_{\rm R}$).
\label{F6}
}
\end{figure}

As we have seen in Section~\ref{sec:versus}, when the speed $v$ of an active particle increases in a homogeneous environment (no crowding and no physical barriers), its trajectories (Fig.~\ref{F2}) are typically dominated by directed motion on short time scales and by an enhanced random diffusion at long times, the latter due to random changes in the swimming direction \cite{howse2007self}. These qualitative considerations can be made more precise by calculating the mean square displacement ${\rm MSD}(\tau)$ of the motion for both passive and active particles at different values of $v$, as shown in Fig.~\ref{F6}. The ${\rm MSD}(\tau)$ quantifies how a particle moves away from its initial position and can be calculated directly from a trajectory. For a passive Brownian particle, the ${\rm MSD}(\tau)$ in two dimensions is 
\begin{equation}\label{eq:msdBM}
{\rm MSD}(\tau) =  4 D_{\rm T}  \tau \; ,
\end{equation}
which is valid for times significantly longer than the \emph{momentum relaxation} time $\tau_{\rm m} = m/\gamma$ of the particle, where $m$ is the mass of the particle.\footnote{To be more precise, the theoretical ${\rm MSD}(\tau)$ for a passive Brownian particle is given by the Ornstein--Uhlenbeck formula \cite{uhlenbeck1930theory}, which in two dimensions reads
\begin{equation*}
{\rm MSD}(\tau) =  
4 D_{\rm T}  \tau
+
\frac{4 k_{\rm B} T }{m} \tau_{\rm m}^2
\left[ 
e^{-\tau/\tau_{\rm m}} -1 
\right] \; .
\end{equation*}
At long time scales, the ${\rm MSD}(\tau)$ of a passive Brownian particle is therefore linear in time with a slope controlled by the particle's diffusion coefficient $D_{\rm T}$. This occurs for $\tau \gg \tau_{\rm m}$, where $\tau_{\rm m}$ for small colloidal particles is of the order of microseconds. In a liquid environment, furthermore, also the hydrodynamic memory of the fluid, i.e. the mass of the fluid displaced together with the particle, must be taken into account and can, in fact, significantly increase the effective momentum relaxation time \cite{lukic2005direct,franosch2011resonances,pesce2014long}.}

For an active particle instead, the theoretical ${\rm MSD}(\tau)$ is given by \cite{franke1990galvanotaxis,howse2007self,martens2012probability}\footnote{Note that Eq.~(\ref{eq:msdAM}) is formally equal to the Ornstein--Uhlenbeck formula for the MSD of a Brownian particle with inertia, which describes the transition from the ballistic regime to the diffusive regime, although at a much shorter time scale than for active particles \cite{uhlenbeck1930theory}.}
\begin{equation}\label{eq:msdAM}
{\rm MSD}(\tau) = \left[ 
4 D_{\rm T} + 2 v^2 \tau_{\rm R}
\right] \tau
+
2 v^2 \tau_{\rm R}^2
\left[ 
e^{-\tau/\tau_{\rm R}} -1 
\right] \; .
\end{equation}
This expression for the MSD holds whenever the active speed component is characterized by an  exponential decay; this is the case, in particular, for all models for active particle reorientation we considered in Section~\ref{sec:orientation}, i.e. rotational diffusion, run-and-tumble dynamics, and Gaussian noise dynamics. At very short time scales, i.e. $\tau \ll \tau_{\rm R}$, this expression becomes ${\rm MSD}(\tau) = 4 D_{\rm T} \tau$ and, thus, the motion is diffusive with the typical Brownian short-time diffusion coefficient $D_{\rm T}$; this diffusive short-time regime indeed can be seen in experiments if the strength of self-propulsion is not very large \cite{zheng2013non-gaussian}. At slightly longer time scales, i.e. $\tau \approx \tau_{\rm R}$, ${\rm MSD}(\tau) = 4 D_{\rm T} \tau + 2 v^2 \tau^2$ so that the motion is superdiffusive. At much longer time scales, i.e. $\tau \gg \tau_{\rm R}$, ${\rm MSD}(\tau) = \left[ 4 D_{\rm T} + 2 v^2\tau_{\rm R} \right] \tau$ and, thus, the ${\rm MSD}(\tau)$ is proportional to $\tau$, since the rotational diffusion leads to a randomization of the direction of propulsion and the particle undergoes a random walk whose step length is the product of the propelling velocity $v$ and the rotational diffusion time $\tau_{\rm R}$ (equal to the persistence length given by Eq.~(\ref{eq:L})). This leads to a substantial enhancement of the effective diffusion coefficient over the value $D_{\rm T}$, which corresponds to $D_{\rm eff} = D_{\rm T} + {1 \over 2} v^2 \tau_{\rm R}$. 

One might be tempted to think that the stationary states of active Brownian systems could resemble equilibrium states at a higher effective temperature
\begin{equation}
T_{\rm eff} = \frac{\gamma D_{\rm eff}}{k_{\rm B}} = T + \frac{\gamma v^2 \tau_{\rm R}}{2 k_{\rm B}} \; .
\end{equation}
This simple picture of active particles as hot colloids may be correct  in some simple situations, such as dilute non-interacting active particles that sediment in a uniform external force field \cite{tailleur2009sedimentation,palacci2010sedimentation, maggi2013motility}. However, as soon as interactions become important or external fields are inhomogeneous, one observes phenomena like clustering in repulsive systems or rectification effects \cite{koumakis2014directed,volpe2014simulation} that are not compatible with the picture of a quasi-equilibrium state at one effective temperature \cite{argun2016experimental}.

\subsection{Biological microswimmers}\label{sec:bio}

Various kinds of biological microswimmers exist in nature, e.g. bacteria \cite{berg1972chemotaxis,berg1990chemotaxis,berg2004ecoli}, unicellular protozoa \cite{machemer1972ciliary,blake197mechanics}, and spermatozoa \cite{riedel2005self,woolley2003motility}. Typically, the planktonic swimming motion of these microorganisms is generated by flagella or cilia powered by molecular motors \cite{lauga2012dance,poon2013clarkia,elgeti2015physics,alizadehrad2015simulating}. Alternative methods, such as crawling or swarming, do not involve swimming in a fluid but they rather require cells to move on a substrate, or through a gel or porous material.

While many properties of the motion of biological microswimmers can be understood in terms of effective Langevin equations, which we have seen in Section~\ref{sec:phenom}, several models have been proposed to understand in more detail their microscopic mechanisms. \citet{lighthill1952squirming} introduced a model for squirmers in a viscous fluid. The Lighthill model assumes that the movement of a spherical particle, covered by a deformable spherical envelope, is caused by an effective slip velocity between the particle and the solvent; this model was corrected to describe the metachronal wave-like beat of cilia densely placed on the surface of a microorganism by a progressive waving envelope \cite{blake1971spherical}. 
Finally, several approaches have addressed the question of the swimming velocity and rate of dissipation of motile microorganisms with different shapes. Analytically tractable examples of these biological microswimmers consist of point particles connected by active links exerting periodic forcing \cite{felderhof2006swimming,najafi2004simple}, spherical particles that self-propel due to shape modulation of their surface \cite{felderhof2014optimal}, assemblies of rigid spheres that interact through elastic forces \cite{felderhof2014swimming}, or one big sphere propelled by a chain of three little spheres through hydrodynamic and elastic interactions \cite{felderhof2014collinear}; the latter configuration in particular has been realized and studied using optically trapped particles \cite{leoni2009basic}.

\subsection{Artificial microswimmers}\label{sec:art}

Various methods have been developed to realize artificial microswimmers that can reproduce the swimming behavior of motile biological cells making use of diverse propulsion mechanisms. As seen in Section~\ref{sec:intro}, these manmade self-propelling particles in fact hold the great promise to change the way in which we perform several tasks in, e.g., healthcare and environmental applications \cite{nelson2010microrobots,wang2012nano,patra2013intelligent,abdelmohsen2014micro,gao2014environmental}. We refer the reader to Table~\ref{tab:realizations} and Fig.~\ref{F1} for examples of experimentally realized active particles. 

The basic idea behind the self-propulsion of micro- and nanoparticles is that breaking their symmetry leads to propulsion through various phoretic mechanisms. Independent of the specific propulsion mechanism, the absence of inertial effects requires non-reciprocal driving patterns in Newtonian liquids \cite{purcell1977life}. Indeed, the temporal motion of flagella or beating cilia found in biological systems follows non-reciprocal and periodic patterns (see also Section~\ref{sec:microhydrodynamics}). Accordingly, this also needs to be taken into account in the design of synthetic microswimmers.

The first demonstration of the concept of swimming powered by asymmetrical chemical reactions can be traced back to the seminal work by Whitesides and co-workers who made millimeter-scale chemically-powered surface swimmers \cite{ismagilov2002autonomous}, while the pioneering demonstration of microswimming in bulk was done by \citet{paxton2004catalytic}, who reported the propulsion of conducting nanorod devices. Subsequently, in an attempt to realize a synthetic flagellum, \citet{dreyfus2005microscopic} fabricated a linear flexible chain of colloidal magnetic particles linked by short DNA segments; such chains align and oscillate with an external rotating magnetic field, and closely resemble the beating pattern of flagella; moreover, the strength of their swimming speed can be controlled by the rotation frequency of the field.

In general, microswimmers can be powered by two main categories of propulsion mechanisms \cite{ebbens2010pursuit}: they can be powered by local conversion of energy (e.g. catalytic processes); or they can be driven by external (e.g. electric, magnetic, acoustic) fields. In this context, it is now important to remark that a distinction exists between \emph{internally} driven active matter and particles that are brought out of equilibrium by \emph{external} fields: while microswimmers powered by these two mechanisms feature a motion that can be described with similar effective models (see, e.g., Sections~\ref{sec:versus} and \ref{sec:phenom}), they present quite different micrscopic details in their interaction with their environment (see, e.g., their hydrodynamic properties discussed in Section~\ref{sec:hydro}). In some cases, a combination of both is possible, e.g. an external field may be required to induce local energy conversion.

In this section, we will first discuss the main physical principles of the propulsion mechanisms based on local energy conversion (Section~\ref{sec:local}) and external fields (Section~\ref{sec:external}). We will then introduce the main experimental methods that are used to build a very successful class of artificial microswimmers, i.e. Janus particles (Section~\ref{sec:methods}).

\subsubsection{Propulsion by local energy conversion}\label{sec:local}

A versatile method to impose propulsion forces onto colloidal particles is the use of phoretic transport due to the generation of chemical, electrostatic, or thermal field gradients. When such gradients are generated externally, passive colloidal particles move of phoretic motion: for example, when colloids are exposed to an electrolyte concentration gradient, they migrate towards the higher salt regions \cite{ebel1988diffusiophoresis}. Therefore, if a particle generates its own local gradient, a self-phoretic motion can take place \cite{golestanian2007designing}. 

The self-generation of gradients by a particle requires some type of asymmetry in its properties, e.g. its shape, material, or chemical functionalization. Based on such considerations, first \citet{paxton2004catalytic}  and then \citet{fournier-bidoz2005synthetic} observed that gold--platinum (Au--Pt) and gold--nickel (Au--Ni) microrods displayed considerably enhanced directed motion in hydrogen peroxide (${\rm H_2O_2}$) solutions. An electro-kinetic model seems to be consistent with most experimental observations: the bimetallic microrod is considered as an electrochemical cell that supports an internal electrical current in order to maintain a redox reaction at its two extremities, where protons are created (Pt/Ni end) and consumed (Au end); due to the flux of protons along the rod, a fluid flow is generated that moves the rod. We remark that other mechanisms have also been suggested to explain the motion of these microrods, including the formation of oxygen bubbles \cite{ismagilov2002autonomous}; this, however, would suggest the motion of Au--Pt microrods to be in the direction of the Au end (i.e. opposite to the site where the oxygen bubbles are created), which is in disagreement with experimental observations.

Bubble formation in ${\rm H_2O_2}$ aqueous solutions as the dominant driving mechanism has been observed in tubular structures of catalytic materials. The internal catalytic wall of these \emph{microjets} (consisting of, e.g., Pt) decomposes ${\rm H_2O_2}$ into ${\rm H_2 O}$ and ${\rm O_2}$; the produced ${\rm O_2}$ accumulates inside the tube and forms gas bubbles, which are ejected from one tube extremity, thus causing the propulsion of the microjet in the opposite direction \cite{solovev2009catalytic,solovev2010magnetic}.

Biologically active swimmers have also been created by functionalizing a conductive fiber with glucose oxidase and bilirubin oxidase: in the presence of glucose, a redox reaction takes place leading to a proton flux and thus to a bioelectrochemical self-propulsion \cite{mano2005bioelectrochemical}.

In contrast to electrically conductive systems, which are essential for the above driving mechanisms, propulsion can also be achieved with dielectric particles (e.g. made of silica, polystyrene, or melamine). The majority of such systems is based on so-called \emph{Janus particles} (named after the two-faced Roman god), where dielectric colloids are partially coated with thin layers of catalytic materials like Pt or palladium (Pd) \cite{golestanian2005propulsion}. When such particles are immersed in an aqueous solution enriched with ${\rm H_2O_2}$, they locally decompose it into ${\rm H_2 O}$ and ${\rm O_2}$, and thus create a local concentration gradient that eventually leads to self-diffusiophoresis. This concept, which was originally pioneered by \citet{howse2007self}, has been very successful, and has been used and modified by many other groups worldwide. Instead of Pt or Pd, hematite has also been used as a catalyst; this has the advantage of permitting one to control the ${\rm H_2O_2}$ decomposition using light: in fact the hematite catalyzes the ${\rm H_2O_2}$ decomposition only when illuminated with blue light \cite{palacci2013living}. The details of the catalytic processes involved in the ${\rm H_2O_2}$ decomposition are quite complex and subject to current investigation; for example, the propulsion strength and direction show a strong dependence on added salt and ionic surfactants \cite{brown2014ionic}.

When metal-coated Janus particles are illuminated with strong laser light, temperature gradients along the particles can also form due to the selective heating of the metallic cap; this leads to a strong self-thermophoretic motion that can be controlled by the incident laser power, as it has been shown in the case of Au-capped colloidal particles \cite{jiang2009manipulation,jiang2010active,bregulla2013individually}. However, the need for high intensity gradients can also lead to optical forces \cite{jones2015optical}, which can interfere with the self-diffusionphoresis mechanisms.

In contrast to thermophoretic effects, which require sufficiently strong light intensity, much smaller intensity is required when Janus particles are immersed in binary liquid mixtures with a lower critical point. When the bath temperature is kept sufficiently close to the critical temperature, even at small illumination intensities, light absorption at the cap leads to local heating, which results in a local phase separation and in a diffusiophoretic motion due to a concentration gradient across the particle \cite{volpe2011microswimmers,buttinoni2012active}. Due to the much smaller light intensities (compared to the thermophoretic mechanisms described above), optical forces are typically negligible. This propulsion mechanism has recently been theoretically investigated by \citet{wurger2015self} and \citet{samin2015self}.

In addition to phoretic forces, Marangoni stresses can also generate particle propulsion. Experimentally, this has been demonstrated, e.g., in a system of water droplets (containing bromine) suspended in an oil phase and stabilized by a surfactant \cite{thutupalli2011swarming,schmitt2013swimming}. Due to the spontaneous reaction of bromide with the surfactant (bromination), a self-sustained bromination gradient along the drop surface is generated that eventually leads to a Marangoni flow and thus propulsion. A similar mechanism has been also found in pure water droplets that are stabilized in an oil phase with surfactants above the critical micellar concentration \cite{izri2014self-propulsion}. Capillary forces can be also exploited for self-propulsion through local heating produced by light-absorption. Using this mechanism, asymmetric micro-gears, suspended at a liquid-air interface, can spin at hundreds rpm under wide field illumination with incoherent light \cite{maggi2015micromotors}.

Finally, we remark that there are other driving mechanisms that do not rely on asymmetric particles, but where a spontaneous symmetry breaking occurs. For example, the reactive droplets studied in \citet{thutupalli2011swarming} produce a spontaneous symmetry breaking of the chemical reaction, which then determines the direction of the propulsion velocity. \citet{izri2014self-propulsion} reported the spontaneous motion in a system consisting of pure water droplets in an oily surfactant medium. \citet{bricard2013emergence,bricard2015emergent} demonstrated that self-propulsion can emerge as the result of a spontanous symmetry breaking of the electric charge distribution around a colloidal particle immersed in a conducting fluid and in presence of an electrical field.

\subsubsection{Propulsion by external fields}\label{sec:external}

In addition to mechanisms based on local energy conversion, a swimming motion can be achieved by the periodic non-reciprocal geometrical deformation or reorientation of the swimmer's body, which can be achieved by applying some (time-dependent) external fields that induce forces and mechanical torques on the object. For example, the use of a rotating magnetic field leads to the deformation of semi-flexible rods \cite{dreyfus2005microscopic} and results in a movement resembling that of biological flagella (e.g. in bacteria and spermatozoa); a numerical study of this system was presented by \citet{gauger2006numerical}. In the case of rigid but chiral magnetic objects (propellers), a rotating magnetic field leads to the rotation of the propeller and thus to a swimming motion that can be fully controlled in three dimensions by using triaxial Helmholtz coils \cite{ghosh2009controlled}. Alternatively, the application of vertical alternating magnetic fields to a dispersion of magnetic microparticles at a liquid--air interface leads to the formation of magnetic ``snakes" due to the coupling between the liquid's surface deformation and the collective response of the particles \cite{snezhko2009self-assembled}. An elliptically polarized rotating magnetic field can generate the dynamic assembly of microscopic colloidal rotors, which, close to a confining plate, start moving because of the cooperative flow generated by the spinning particles acting as a hydrodynamic ``conveyor belt" \cite{martinez2015colloidal,martinez2015magnetic}.

Forces can also be applied to microscopic objects by excitation of ultrasound waves. Using the interaction of suspended objects with acoustic waves, the levitation, propulsion, rotation, and alignment of metallic microrods have been achieved \cite{wang2012autonomous}: their directional motion is due to a self-acoustophoretic mechanism, whose driving strength depends sensitively on the shape asymmetry of the microrods (e.g. the curvature at their end).

Finally, the application of alternating electric fields to Janus particles has been predicted to lead to an unbalanced liquid flow due to induced-charge electro-osmosis \cite{ramos1998ac,ajdari2000pumping}. Directed motion has indeed been observed when Au-coated colloidal particles in NaCl solutions are subject to uniform fields with frequencies in the range  from $100\,{\rm Hz}$ to $10\,{\rm kHz}$ \cite{gangwal2008induced-charge}.

\subsubsection{Synthesis of Janus particles}\label{sec:methods}

Janus particles are special types of micro- and nanoparticles whose surfaces have two or more distinct physical and/or chemical properties \cite{golestanian2007designing}. The simplest realization is achieved by dividing the particle into two distinct parts, each either made of a different material or bearing different functional groups. For example, a Janus particle may have one half of its surface covered by hydrophilic groups and the other half by hydrophobic groups. This gives the particle unique properties related to its asymmetric structure and/or functionalization, which can in particular be exploited to obtain self-propulsion. A recent comprehensive review by \citet{walther2013janus} covers all aspects from synthesis and self-assembly to novel physical properties and applications of Janus particles.

The synthesis of Janus nanoparticles requires the ability to selectively create each side of a particle with different chemical properties in a reliable, high-yield, and cost-effective way. Currently, three major methods are used \cite{lattuada2011synthesis}:
\begin{description}
\item[Masking] The two main masking techniques that are commonly employed to produce Janus particles are evaporative deposition and suspension at the interface between two fluid phases \cite{lattuada2011synthesis}. In \emph{evaporative deposition} \cite{love2002fabrication}, homogeneous particles are placed on a surface (e.g. a coverslip) in such a way that only one hemisphere is exposed, typically forming a monolayer colloidal crystal; they are subsequently placed in an evaporation chamber, where they are covered on one side with some materials, typically a metal (e.g. Al, Au, Pt) or carbon; and finally they are released in aqueous solution (e.g. by sonication). In the \emph{phase separation} technique \cite{gu2005heterodimers}, the particles are placed at the interface between two phases in a liquid environment; subsequently, one side is exposed to a chemical that changes the particles' properties; and finally the Janus particles are released in a homogeneous solution. This latter technique scales particularly well to the nanoscale.
\item[Self-assembly] This can be achieved by the use of block copolymers or selective absorption \cite{lattuada2011synthesis}. \emph{Block copolymers} are made up of blocks of different polymerized monomers \cite{erhardt2001janus}. \emph{Competitive absorption} involves two substrates that phase-separate due to one or more opposite physical or chemical properties; when these substrates are mixed with colloidal particles (e.g. Au nanoparticles), they maintain their separation and form two phases with the particles in the middle \cite{vilain2007study}.
\item[Phase separation] This method involves the mixing of two or more incompatible substances that then separate into their own domains while still part of a single nanoparticle. This method can produce Janus nanoparticles made of two inorganic or organic substances \cite{carbone2010colloidal,lattuada2011synthesis}.
\end{description}

Non-spherical Janus particles with simple shapes, e.g. rods and cylinders, can be produced with methods similar to the ones discussed above. For example, in the block-copolymer method, the production of Janus spheres, cylinders, sheets, and ribbons is possible by adjusting the molecular weights of the blocks in the initial polymer and also the degree of cross-linking. More complex shapes can be obtained using \emph{microphotolithography}, which has been used, e.g., to produce the L-shaped particles shown in Figs.~\ref{F3}c and \ref{F3}d \cite{kuemmel2013circular}, or \emph{glancing angle deposition}, which has been used, e.g., to produce the chiral colloidal propellers shown in Fig.~\ref{F1}f \cite{ghosh2009controlled}.

\section{Hydrodynamics}\label{sec:hydro}

As we have seen in the previous section, active particles use a wide variety of mechanisms to achieve self-propulsion in liquid environments. In fact, often they are microswimmers that are capable of perturbing the surrounding fluid in a way that generates a net displacement of their own body. The structure of the dynamical laws governing  fluid motion is therefore crucial to understand propulsion of individual swimmers. Furthermore, as an active particle moves in a fluid, a complex fluid flow pattern travels along with it and can influence the motion of nearby active or passive floating bodies. Moreover, whenever this flow pattern is distorted from its bulk structure by the presence of obstacles like confining walls, swimming speed and direction can change leading to phenomena like swimming in circles or trapping by obstacles. The present section is structured in four main parts discussing hydrodynamic effects involved in generating propulsion in the bulk of a fluid (Section~\ref{sec:microhydrodynamics}), particle--particle hydrodynamic interactions (Section~\ref{sec:ppinter}), hydrodynamic couplings to confining walls (Section~\ref{sec:boundaries}), and swimming in a non-Newtonian medium (Section~\ref{sec:viscoelastic}).

\subsection{Microhydrodynamics of self-propulsion}\label{sec:microhydrodynamics}

\begin{figure}[b]
\includegraphics[width=\columnwidth]{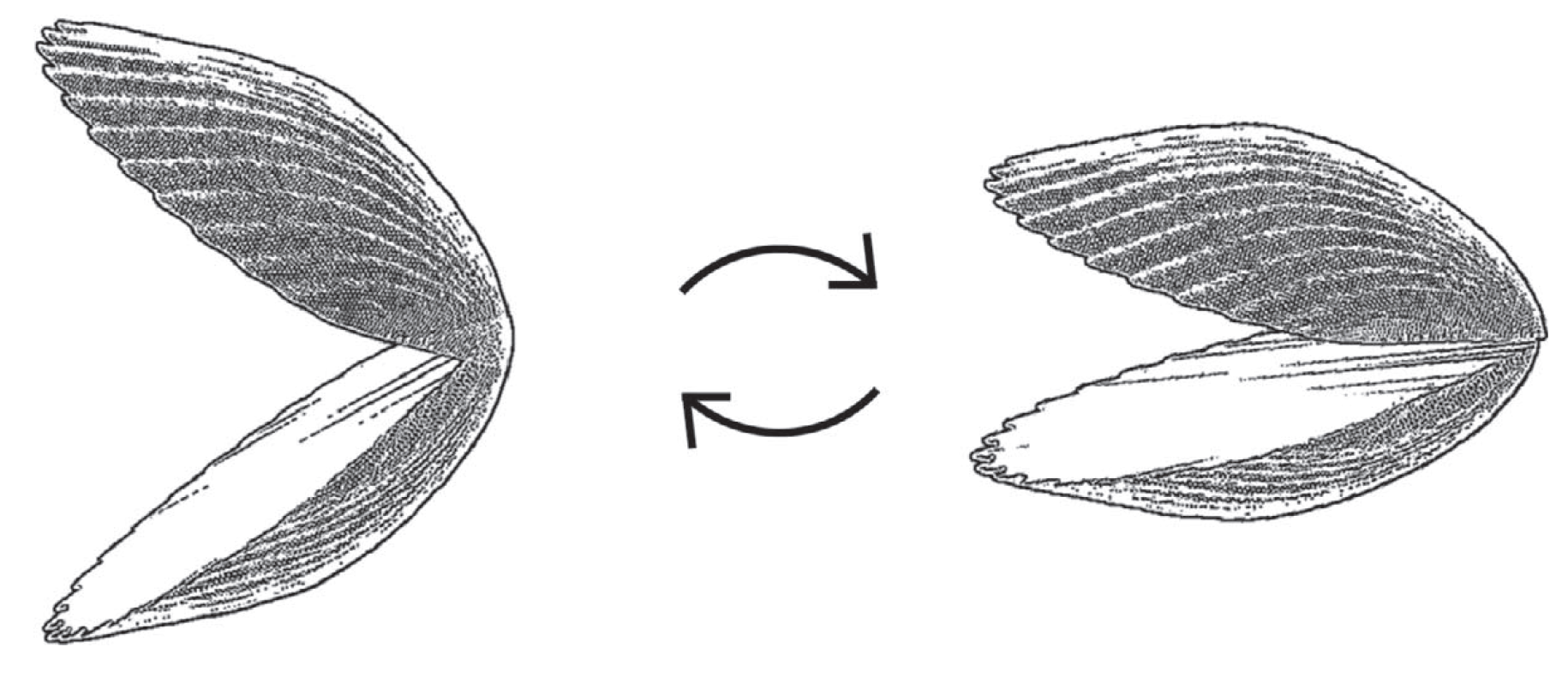}
\caption{The scallop theorem. Schematic drawing of a scallop performing a reciprocal motion: by \emph{quickly} closing its two shells, the scallop ejects two fluid jets behind its hinge so that its body recoils in the opposite direction; then, by \emph{slowly} opening the shells, the scallop comes back to its initial shape with little net displacement. The scallop theorem states that such a reciprocal motion cannot generate propulsion in a (viscous) fluid at a low-Reynolds-number regime \cite{purcell1977life}. From \citet{qiu2014swimming}.
\label{F7}
}
\end{figure}

Swimming at the macroscopic scale heavily relies on the inertia of the surrounding fluid. For example, one of the most simple swimming mechanism is that employed by scallops, which swim by opening and closing their two valves, as shown in Fig.~\ref{F7}. However, swimming strategies of this kind, i.e. involving a reciprocal shape deformation, are doomed to failure at microscopic scales \cite{purcell1977life} (at least in purely viscous liquids, as we will see in Section~\ref{sec:viscoelastic}), where fluid flows obey kinematic reversibility.

Slow  flows of incompressible viscous fluids are governed by Stokes equations:
\begin{equation}\label{eq:stokes}
\left\{\begin{array}{rcc}
\eta\nabla^2\mathbf u-\nabla p=0\\
\nabla\cdot\mathbf u=0
\end{array}\right.
\end{equation}
where $\eta$ is the fluid viscosity, ${\bf u}$ is the velocity of the flow, and $\nabla p$ is the gradient of the pressure. A flow is slow enough for Eqs.~(\ref{eq:stokes}) to be valid when the typical speed $v \ll v_{\rm c} = \eta/ (\rho L)$, where $\rho$ is the fluid density and $L$ is a characteristic length scale. The value of the characteristic speed $v_{\rm c}$ is on the order of $1\,{\rm m\, s^{-1}}$ for water when $L \sim1\,{\rm \mu m}$, and grows as the system size is reduced. The adimensional ratio between the two speeds is the \emph{Reynolds number}
\begin{equation}
{\rm Re} = \frac{v}{v_{\rm c}} = \frac{\rho L v}{\eta} \, ,
\end{equation}
which is usually interpreted as the relative importance of inertial forces in comparison with viscous forces. Generally, ${\rm Re} \ll 1$ in all problems involving self-propulsion at the microscale, where typical flow speeds rarely exceed $100\,{\rm \mu m\, s^{-1}}$ (see Fig.~\ref{F1} and Table~\ref{tab:realizations}). Considering, for example, an \emph{E. coli} bacterium swimming in water, $L \approx 1\,{\rm \mu m}$, $v \approx 30\,{\rm \mu m\,s^{-1}}$, $\eta = 0.001\,{\rm Pa\, s}$, and $\rho  = 1000\,{\rm kg\,m^{-3}}$, so that ${\rm Re} \approx 3 \times 10^{-5} \ll 1$.

A solid object that is dragged through the fluid by external forces imposes a boundary condition to the flow field given by the rigidity condition
\begin{equation}\label{eq:bc}
\mathbf u={\bf v}+\boldsymbol\omega \times \boldsymbol{\rho}
\end{equation}
where ${\bf v}$ and $\boldsymbol{\omega}$ are the instantaneous translational and rotational speed of the object, and $\boldsymbol\rho$ is a generic point on the body surface. The resulting flow and pressure fields can be obtained by solving Stokes equations (Eqs.~(\ref{eq:stokes})) with the above boundary conditions (Eq.~(\ref{eq:bc})). By integrating the stress tensor $\boldsymbol\sigma=\eta\left(\nabla{\bf u}+\nabla{\bf u}^{\rm T}\right)$ over the object's boundary, one obtains the total force and torque acting on the body:
\begin{equation}\label{eq:Stress1}
\mathbf F=\oint_S \boldsymbol\sigma \cdot \hat{\bf n}\,dS
\end{equation}
and
\begin{equation}\label{eq:Stress2}
\mathbf T=\oint_S \boldsymbol{\rho}\times\boldsymbol\sigma \cdot \hat{\bf n}\,dS \; ,
\end{equation}
where $S$ is a surface enclosing the body, and $\hat{\bf n}$ is the unit vector perpendicular to the surface. We remark here that, as we will see later, microswimmers are force-free and torque-free objects. From the linearity of Stokes equation the above integrations result in a linear relationship:
\begin{equation}\label{resistance}
\left[\begin{array}{c}
{\bf F}\\
{\bf T}
\end{array}\right]
=
\left[\begin{array}{cc}
{\bf A} & {\bf B}\\
{\bf B}^{\rm T} & {\bf D}
\end{array}\right]
\left[\begin{array}{c}
{\bf v}\\
\boldsymbol{\omega}
\end{array}\right] \; ,
\end{equation}
where ${\bf A}$, ${\bf B}$, and ${\bf D}$ are tensors determined by the shape and orientation of the object. In particular, ${\bf A}$ represents the translational resistance tensor connecting the linear speed ${\bf v}$ to the force ${\bf F}$ that is needed to move a body in a purely translatory motion with speed ${\bf v}$;  similarly, ${\bf D}$ connects angular speed $\boldsymbol{\omega}$ to the torque $\mathbf T$ in pure rotations; and ${\bf B}$ and ${\bf B}^{\rm T}$ represent the coupling between rotational and translational motion. For example, for a sphere of radius $R$, ${\bf A}=6\pi\eta R\,\mathbb{I}$, $\mathbb{D}=8\pi\eta R^3\,\mathbb{I}$, and ${\bf B} = {\bf B}^{\rm T} = {\bf 0}$. In the most general case, however, ${\bf A}$ and ${\bf D}$ are non-isotropic. Furthermore, for chiral bodies, i.e. bodies that do not possess three mutually orthogonal planes of symmetry, ${\bf B}$ and ${\bf B}^{\rm T}$ do not vanish. Therefore, a rotating chiral body can be used as a propeller generating a thrust force ${\bf B} \cdot \boldsymbol{\omega}$. The resistance tensor can be calculated analytically only in very few simple cases \cite{kim2005microhydrodynamics}. Approximation schemes with various degrees of complexity exist for slender bodies \cite{lighthill1976flagellar}. Different schemes have been compared in the particular case of a helical propeller by \citet{rodenborn2013propulsion}. Direct experimental measurements of the resistance coefficients of a real flagellar bundle is possible by using optical tweezers \cite{chattopadhyay2006swimming,bianchi2015polar}.

For example, swimming bacteria like {\it E. coli} use a helical bundle of flagellar filaments as a propeller for the cell body. The whole cell can be thought as composed of two rigid bodies that translate as a single rigid object but can rotate with different speeds around a common axis. Assuming that the only non-vanishing speed components are those lying on the cell axis, we can write two scalar resistance equations for the cell body and the flagellar bundle: 
\begin{equation}
\left\{\begin{array}{ccl}
F_{\rm b} & = & A_{\rm b} v\\
F_{\rm f} & = & A_{\rm f} v+B_{\rm f}\omega_{\rm f}
\end{array}\right.
\end{equation}
where subscripts ``b" and ``f" refer respectively to cell body or flagellar bundle. Since no external forces are applied on the swimmer, $F_{\rm b}+F_{\rm f}=0$ and the swimming speed is obtained as 
\begin{equation}
v = \frac{B_{\rm f}\,\omega_{\rm f}}{A_{\rm b}+A_{\rm f}} \; .
\end{equation}
Imposing the torque-free condition one can similarly obtain the body rotation frequency. 

The simplified picture of a microswimmer composed of two counter-rotating rigid units holds only for swimming procaryotes like bacteria whose flagella are passive filaments connected to the cell body via a rotary motor \cite{berg2003rotary}. More generally, the swimming problem can be formulated starting from the common feature that active particles perturb the flow field by imposing specific boundary conditions for ${\bf u}(\boldsymbol{\rho})$ where $\boldsymbol{\rho}$ is the position vector of a generic point on the particle surface  \cite{lauga2009hydrodynamics}. Self-propelling particles use two main mechanisms for imposing boundary conditions for the flow field over their surface: 
\begin{enumerate}
\item they can modify their shape and displace fluid at contact with sticky boundary conditions (\emph{swimmers});
\item they can produce a gradient in some thermodynamic quantity resulting in a phoretic slip velocity that is tangential to the particle surface (\emph{squirmers}).
\end{enumerate}
The first strategy is mainly adopted by biological active particles that swim by waving or rotating flagella. The second mechanism typically applies to artificial self-propelling particles, like Janus colloids, that use some form of energy stored in the environment to maintain a stationary local gradient of, e.g., concentration or  temperature. In both cases the hydrodynamic swimming problem can be formulated in general as follows:
\begin{enumerate}

\item solve Stokes equations with prescribed boundary conditions for the flow field on the particle surface (Eq. (\ref{eq:bc})), obtaining ${\bf u}({\bf r})$;

\item integrate the viscous stress over the particle surface to get the total force and torque acting on the particle (Eqs.~(\ref{eq:Stress1}) and (\ref{eq:Stress2})); 
\item impose force-free (${\bf F}={\bf 0}$) and torque-free (${\bf T}={\bf 0}$) conditions, and solve for the rigid speeds ${\bf v}$ and $\boldsymbol{\omega}$. 

\end{enumerate}
The self-propulsion speed ${\bf v}(t)$ at a given time $t$ is then a function of the instantaneous values of the boundary speed distribution ${\bf u}(\boldsymbol\rho)$ evaluated at the same time $t$. While most squirmers impose a stationary slip velocity on their boundary resulting in a constant ${\bf v}$, all swimmers necessarily deform in a cyclic (but not reciprocal) way resulting in a time-varying propulsion speed that cycles with the period $T_{\rm p}$ of the shape deformation. It can be shown that, since Stokes equations are linear and do not have any explicit dependence on time, the total net displacement after a cycle, i.e. $\int_0^{T_{\rm p}} {\bf v}(t)\,dt$, vanishes if the shape deformation is reciprocal, i.e. if the cycle is closed by retracing the same sequence of shapes in reverse order \cite{lauga2009hydrodynamics}. \citet{najafi2004simple} proposed a simple one-dimensional low-Reynolds-number swimmer consisting of three spheres linked by rigid rods whose lengths can change periodically but not reciprocally. Recently, \citet{bet2016efficient} derived a generalized scallop theorem that allows one to optimize the speed, power, and efficiency of a swimmer by altering its geometry. As a consequence of the scallop theorem a micron-sized scallop cannot swim at low Reynolds number but only move back and forth with no net displacement. Hydrodynamic reversibility, however, breaks down in non-Newtonian fluids (see Section~\ref{sec:viscoelastic}) allowing self-propulsion with reciprocal deformations \cite{qiu2014swimming}.

\subsection{Particle--particle hydrodynamic interactions}\label{sec:ppinter}

\begin{figure*}[ht]
\includegraphics[width=\textwidth]{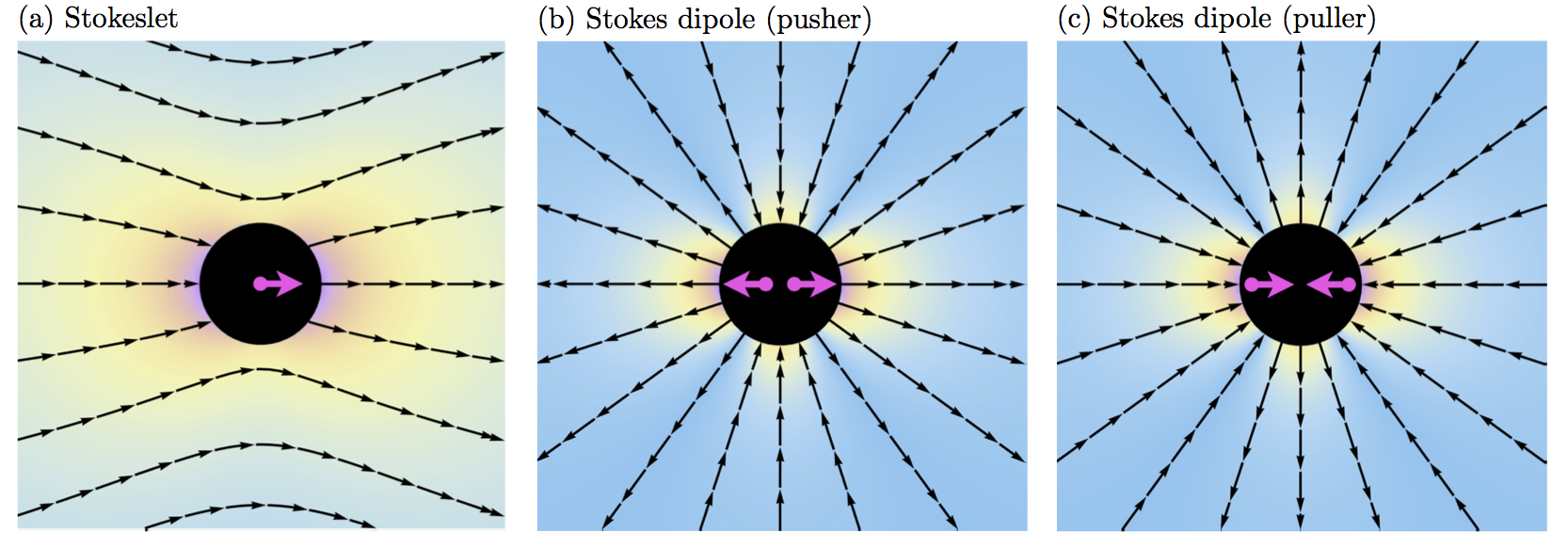}
\caption{(Color online) Flow singularities.
(a) Stokeslet corresponding to the far field of an active particle driven by an external force.
(b-c) Stokes dipoles corresponding to active particles driven by an internal force corresponding to (b) pushers and (c) pullers, both moving horizontally.
\label{F8}
}
\end{figure*}

Both forced colloids and active particles generate a flow that obeys Stokes equations (Eqs.~(\ref{eq:stokes})) subject to the appropriate boundary conditions for the flow field. 
Although analytical solutions can be obtained only in some very simple cases, the structure of the far-field flows can be discussed on a very general basis and, as we will see in the following, can be used to anticipate, at least qualitatively, the peculiar interactions that active particles manifest between themselves or with external obstacles and confining walls. 

Since Stokes equations are linear, flow fields can be conveniently expressed as the superposition of singular solutions \cite{pozrikidis1992boundary}. The dominant singularity for particles driven by external forces is the Stokes flow generated by a point force, also known as the \emph{stokeslet}.  For a point force of magnitude $F$ and direction $\hat{\bf e}$ the stokeslet flow at a position ${\bf r} = r \hat{\bf r}$ from the force origin is given by
\begin{equation}\label{eq:stokeslet}
{\bf u}_{\rm s}(\mathbf r)=\frac{F}{8\pi\eta\,r}\left[\,(\hat{\bf e}\cdot\hat{\bf r}) \hat{\bf r}+\hat{\bf e}\,\right] \; .
\end{equation}
The corresponding field is presented in Fig.~\ref{F8}a. The magnitude of a stokeslet decays as the inverse distance from the force origin and describes with very good approximation hydrodynamic couplings between externally driven colloidal particles \cite{quake1999direct,dileonardo2007eigenmodes} up to very short interparticle distances. Importantly, active particles propelled by internal forces are instead force-free so that the leading term cannot be a stokeslet. 

The next singularity is the \emph{Stokes dipole}, which can be obtained from the stokeslet by differentiation and represents the far-field flow generated by two nearby and opposite point forces. There is a close analogy with the multipolar expansion of electric fields, although in the hydrodynamic case the field sources are vector forces rather than scalar charges. Therefore, while electric dipoles are represented by vectors, force dipoles are instead tensors, which can be symmetric when the two point forces are parallel to their separation direction, or antisymmetric when they point in opposite orthogonal direction to their separation direction. However, since self-propelled particles need to be torque-free, the antisymmetric part, that would correspond to a point torque, vanishes and the far-field flow is usually dominated by a symmetric force dipole where forces and separation direction are parallel with the unit vector $\hat{\bf e}$:
\begin{equation}\label{eq:dipoleflow}
{\bf u}_{\rm d}({\bf r})=\frac{P}{8\pi\eta\,r^2}\left[\,3(\hat{\bf e}\cdot\hat{\bf r})^2-1\,\right]\hat{\mathbf r} \; ,
\end{equation}
where $P$ is the magnitude of the force dipole having the dimensions of force times length. In flagellated swimming bacteria like \emph{E. coli} the dipole arises from having most of the drag on the cell body side and pushing the fluid along the swimming direction while an opposite force is  generated on the fluid all along the rotating flagellar bundle. The resulting flow pattern, shown in Fig.~\ref{F8}b, is purely radial and pushes fluid in the forward and backward directions, while a lateral inward flow guarantees mass conservation. Swimmers of this kind are called \emph{pushers} ($P>0$) as opposed to \emph{pullers} ($P<0$) like the algae \emph{Chamydomonas reinhardtii}, where two waving flagella propel a cell body that lags behind in the swimming direction (Fig.~\ref{F8}c).\footnote{It has been shown experimentally that the flow field around Chamydomonas is not a simple dipole \cite{drescher2010direct,guasto2010oscillatory}.}

Flows around swimming bacteria have been observed by microparticle imaging velocimetry \cite{drescher2011fluid}. The measured flow fields were well approximated by a dipole of stokeslet with magnitude $F=0.4\,{\rm pN}$ separated by a distance $d=2\,{\rm \mu m}$. The magnitude of the force is consistent with the expected thrust force generated by the flagellar bundle, which is connected to the swimming speed $v$ by $F \approx \eta d v$. The generated flow field is approximately $v\,(d/r)^2$ representing a small perturbation over the free swimming speed $v$ of a nearby swimmer located at a distance $r$ larger than $d$. These perturbations could however manifest themselves as small couplings in the velocities of nearby swimming cells that are advected by the flows generated by neighbors. It was shown that, if one neglects correlations in the orientation of nearby swimmers, velocity couplings arise from quadrupolar terms, while the effect of dipolar flows cancels out on average \cite{liao2007pair}. \citet{baskaran2009statistical}, starting with a minimal physical model of a stroke-averaged swimmer in a fluid, derived a continuum description of a suspension of active organisms that incorporates fluid-mediated, long-range hydrodynamic interactions among the swimmers.

\begin{figure}[h]
\includegraphics[width=\columnwidth]{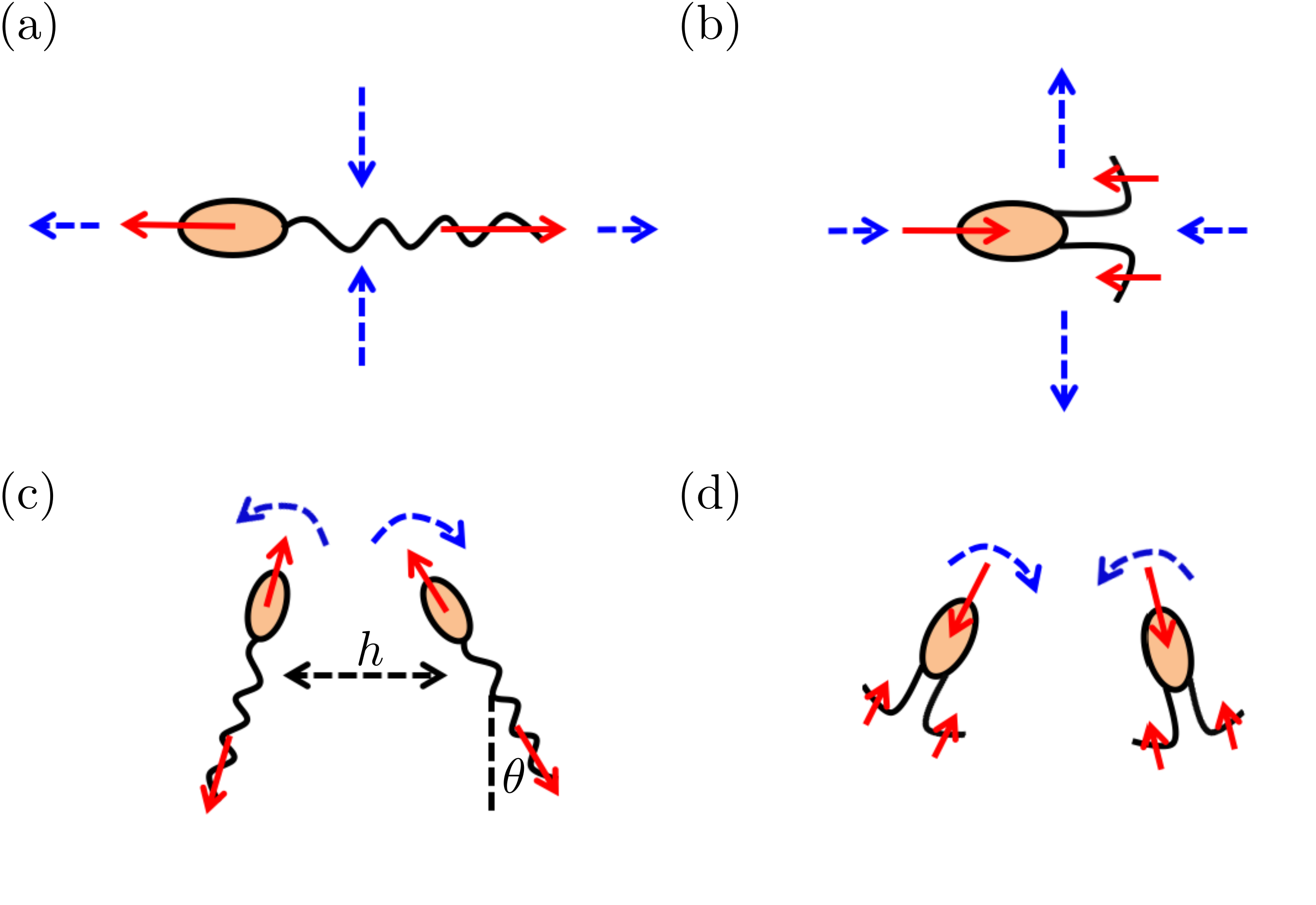}
\caption{(Color online) Flow fields created by swimmers at low Reynolds numbers. 
(a) Pushers have a positive force-dipole ($P > 0$) and induce an outgoing flow field directed along their swimming direction (repulsion) and an incoming flow field from their sides (attraction). In all panels, the red solid arrows represent the local forcing from the swimmer on the surrounding fluid and the blue dashed arrows represent the fluid flows around the swimmer. (b) Pullers have a negative force dipole ($P < 0$), inducing an incoming flow field along their swimming direction and an outgoing flow field along their sides. (c) Two pushers on a converging course reorient each other, tending toward a configuration where they are parallel and swimming side-by-side ($h$ is the separation distance between the swimmers, and $\theta$ is the angle between the swimmers direction and the direction normal to their separation). (d) Two pullers on a diverging course reorient each other, tending toward a configuration in which they are antiparallel and swimming away from each other.
From \citet{lauga2009hydrodynamics}.
\label{F9}
}
\end{figure}

Dipolar flow fields have also an associated vorticity  given by
\begin{equation}\label{eq:dipolevorticity}
\boldsymbol{\Omega}({\bf r})
=
\nabla\times{\bf u}_{\rm d}
=
\frac{3 P}{4\pi\eta\,r^3}(\hat{\bf e}\cdot\hat{\mathbf r})(\hat{\bf e}\times\hat{\mathbf r}) \; .
\end{equation}
Two nearby swimmers will therefore reorient each other through advection by the vorticity field produced by the neighbor. Since the vorticity field vanishes when either $\hat{\mathbf e}\cdot\hat{\mathbf r}=0$ or $\hat{\mathbf e}\times\hat{\mathbf r}=0$, two possible equilibrium relative orientation exist: the first one, where the swimming direction is  orthogonal to the separation distance, is only stable for pushers ($P>0$, Figs.~\ref{F9}a and \ref{F9}c); the second one, with $\hat{\mathbf e}$ parallel to $\mathbf r$, becomes stable for pullers ($P<0$, Figs.~\ref{F9}b and \ref{F9}d).  Interestingly, in both cases, the equilibrium relative orientation gives rise to reciprocally attractive flows (Eq.~(\ref{eq:dipoleflow})).
The reorienting action exerted by nearby swimmers contributes to the bending of swimming trajectories.   
There are other mechanisms contributing to the reorientation of active particles, such as Brownian rotational diffusion, tumbles in flagellar locomotion, and cell--cell collisions through steric forces.
The first two mechanisms do not depend on density and are always present even in diluted suspensions. In concentrated samples, however, cell--cell scattering, either through hydrodynamic interactions or direct contact, may become the dominant reorientation mechanism. Often, at those concentrations where hydrodynamic interactions become relevant, average interparticle distances are so small that higher-order singularities may become predominant \cite{drescher2011fluid,liao2007pair}.

\subsection{Hydrodynamic coupling to walls}\label{sec:boundaries}

\begin{figure*}
\includegraphics[width=\textwidth]{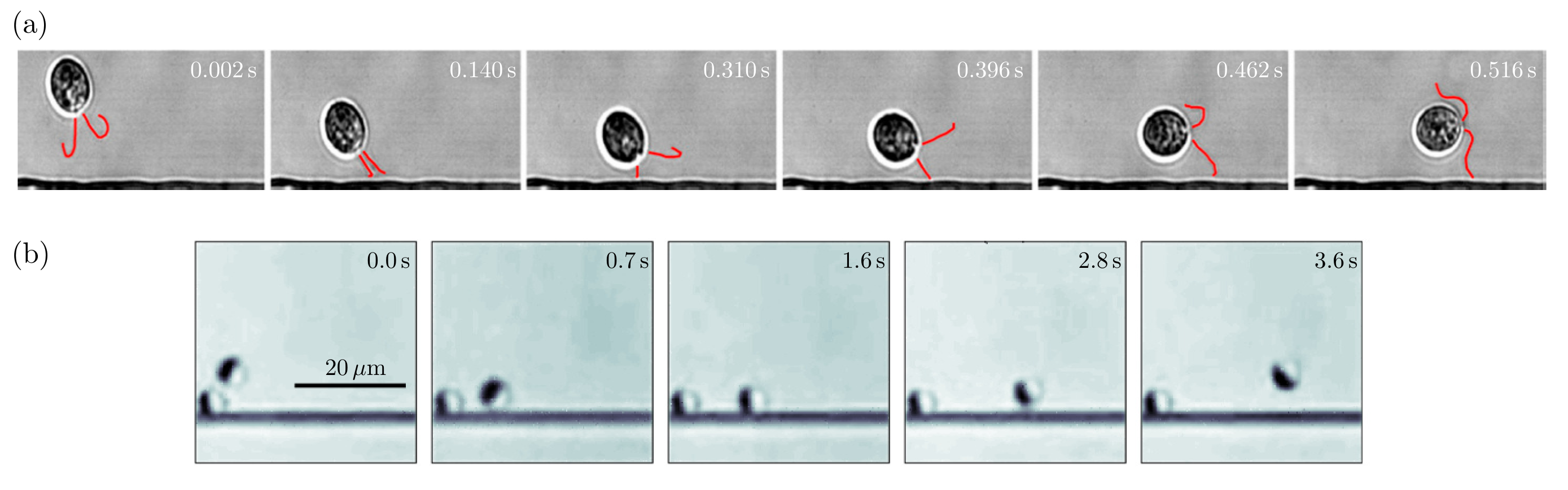}
\caption{(Color online)
Scattering of microswimmers by a wall:
(a) the surface scattering of \emph{Chlamydomonas reinhardtii} is governed by ciliary contact interactions (cilia manually marked in red), from \citet{kantsler2013ciliary};
and (b) time series of snapshots demonstrating the approach to, contact with, and detachment form a wall of a self-propelled Janus particle, from \citet{volpe2011microswimmers}.
\label{F10}
}
\end{figure*}

\begin{figure}[h]
\includegraphics[width=\columnwidth]{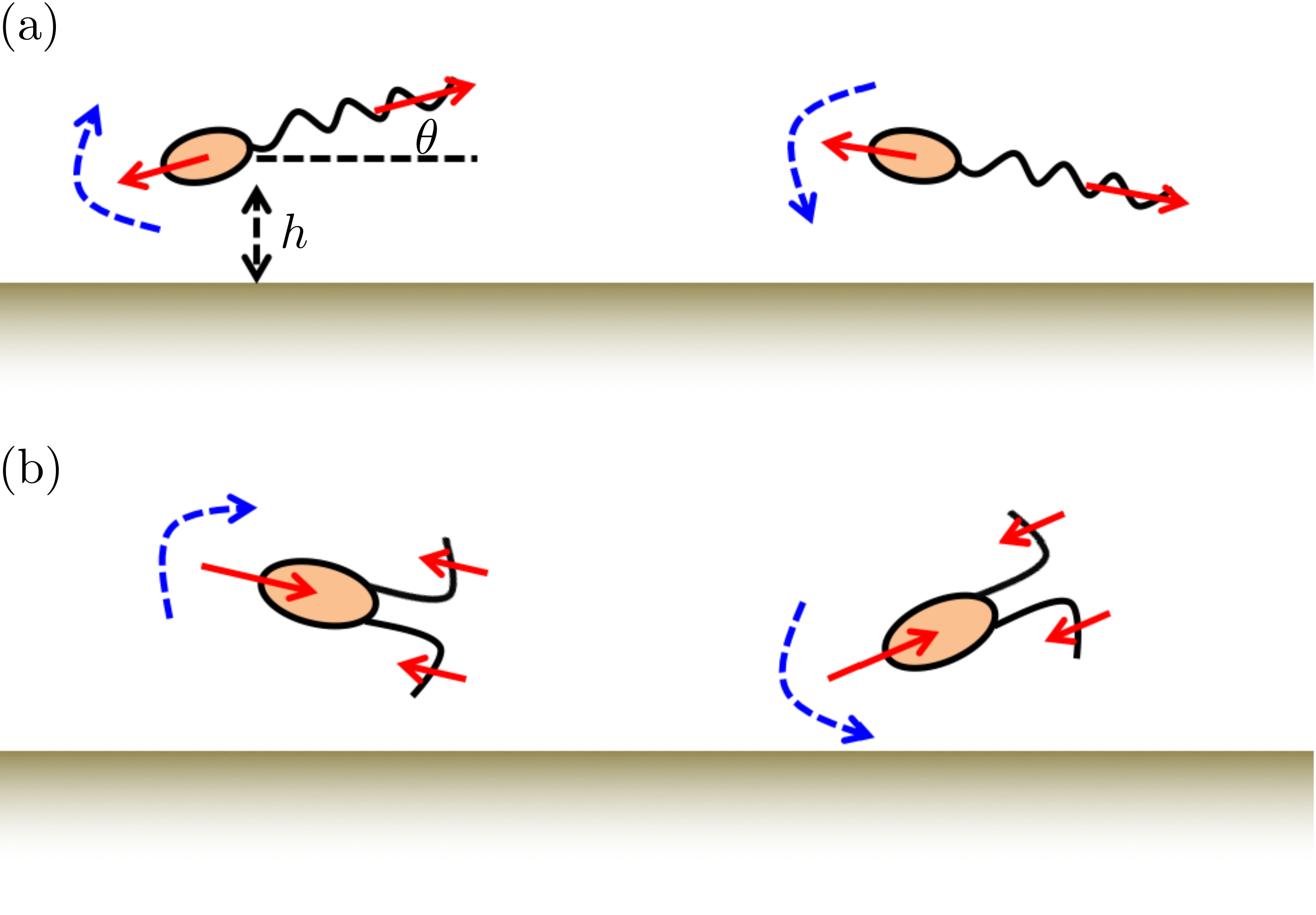}
\caption{(Color online) Wall-induced rotation of microswimmers. A microswimmer is located at a distance $h$ from a solid surface and oriented at an angle $\theta$ with respect to the direction parallel to the surface: (a) pushers are reoriented hydrodynamically in the direction parallel to the surface (equilibrium, $\theta = 0$, see also Fig.~\ref{F9}c); (b) pullers are reoriented in the direction perpendicular to the surface (equilibrium, $\theta =  \pm\pi/2$, see also Fig.~\ref{F9}d). The red solid arrows represent local forcing from the microswimmer on the surrounding fluid; the blue dashed arrows represent the torque acting on the microswimmer.
From \citet{lauga2009hydrodynamics}.
\label{F11}
}
\end{figure}

Active particles may behave very differently when approaching a solid boundary. Although far-field hydrodynamics is often capable of anticipating the correct behavior \cite{spagnolie2012hydrodynamics}, most reorientation dynamics take place when the particle is in close contact with the boundary. Near-field effects, steric interactions, and direct flagellar contact dynamics \cite{kantsler2013ciliary} become predominant in this regime giving rise to  a broad range of behaviors, from stable wall-trapping (Fig.~\ref{F31}b) to wall-scattering (Fig.~\ref{F10}). Furthermore, active particles may exhibit rheotaxis (i.e. movement in response to a flow) near walls \cite{uspal2015rheotaxis,uspal2015self}.

Far-field predictions can be obtained considering the flow generated by the image singularities that need to be collocated inside a bounding wall in order to satisfy the no-slip boundary condition at the wall surface \cite{spagnolie2012hydrodynamics}.
In fact, the leading order effect of image singularities is qualitatively the same as that obtained by replacing the wall with a specular image swimmer located on the opposite side of the wall surface (see Figs.~\ref{F9}c and \ref{F9}d). Therefore, in a very similar way as discussed for swimmer--swimmer interactions, far-field reflected flows will align pushers in a direction that is parallel to the wall surface and attract them to the wall (Fig.~\ref{F11}a). The stable swimming direction for pullers is parallel to the surface normal, leading again to an attractive image flow (Fig.~\ref{F11}b). 

Although far-field predictions may correctly describe how active particles approach a wall, the actual collision dynamics and the fate of the resulting trajectory will depend on detailed near-field hydrodynamics and contact interactions. Swimming bacteria like \textit{E. coli} display a remarkable tendency to swim in close contact with walls once they reach them \cite{frymier1995three}.
Hydrodynamic effects have been proposed to be at the origin of this wall entrapment via two distinct mechanisms: 
the first one is via far-field reflected flows  \cite{berke2008hydrodynamic,drescher2011fluid} as discussed above;
the second mechanism involves hydrodynamic torques that arise in anisotropic bodies swimming in close contact with a wall and leading to tilted swimming, i.e. with the cell swimming direction pointing into the confining wall \cite{vigeant2002reversible, spagnolie2012hydrodynamics}. It has been suggested 
that steric repulsion and rotational Brownian motion might be already enough to reproduce the observed accumulation of bacteria in the proximity of solid flat walls  \cite{li2009accumulation,volpe2014simulation}. However, recent experiments have demonstrated that stable trapping is also observed around cylindrical pillars ruling out steric effects as the sole origin for wall entrapment in \textit{E. coli} bacteria \cite{sipos2015hydrodynamic}. 
While pusher swimmers like \textit{E. coli} mostly interact with the wall through the cell body that moves ahead of the flagella, puller swimmers like \textit{C. reinhardtii}, moving with beating flagella ahead of the body, display more complex interaction dynamics with confining walls \cite{kantsler2013ciliary,contino2015microalgae}: near-field effects and direct ciliary contact lead to wall-scattering, rather than wall-trapping, with the cells escaping from the wall at a characteristic angle that does not depend on the initial direction of approach to the wall. Direct flagellar contact is also important in wall interactions for sperm cells due to the large amplitude of flagellar waveforms \cite{kantsler2013ciliary}.
In swimming bacteria like \textit{E. coli}, the chiral nature of the flagella, when combined with the presence of a nearby wall, also results in a tendency to swim along clockwise circular trajectories above solid surfaces \cite{berg1990chemotaxis, frymier1995three, lauga2006swimming}. Such a mechanism can be exploited to direct bacterial motions in microchannels \cite{diluzio2005escherichia}, or to design microfluidic devices that can sort bacteria according to their motility or size \cite{hulme2008using}. When the boundary condition on the wall changes from no-slip on a solid wall to the almost perfect slip on a liquid--air interface, the direction of circular swimming is reversed \cite{dileonardo2011swimming}.

\subsection{Non-Newtonian media}\label{sec:viscoelastic}

\begin{figure}[t]
\includegraphics[width=\columnwidth]{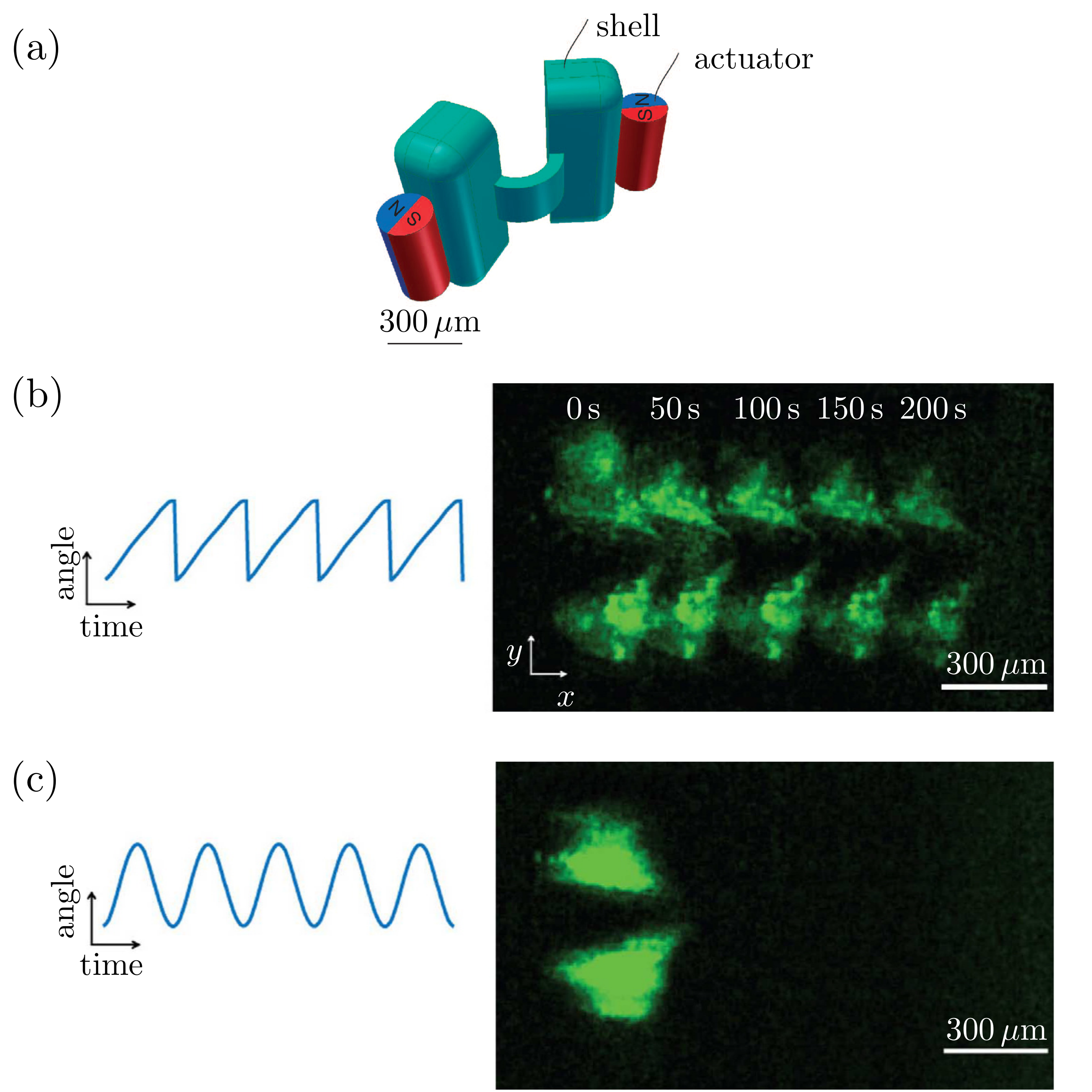}
\caption{(Color online) Reciprocal swimming in a low-Reynold-number non-Newtonian medium. (a) 3D model of a submillimeter-sized ``micro-scallop" capable of swimming in a low-Reynold-number non-Newtonian fluid. (b-c) Displacement of the micro-scallop in a shear-thickening and a shear-thinning fluid. (b) Forward net displacement of the micro-scallop in a shear-thickening fluid and asymmetric actuation (blue curve). The image is a time-lapse composite picture of five frames at intervals of $50\,{\rm s}$, with the net displacement occurring along the $x$ direction. (c) Corresponding image of the micro-scallop in shear-thickening fluid with symmetric actuation (blue curve) and no discernible net displacement. From \citet{qiu2014swimming}.
\label{F12}
}
\end{figure}

Until now we have considered living microorganisms and artificial particles swimming in Newtonian liquids at low Reynolds numbers (${\rm Re} \ll 1$). However, most biological fluids, where also artificial microswimmers will be required to operate in future envisaged biomedical applications, are non-Newtonian ~\cite{lauga2009hydrodynamics,nelson2014micro}, e.g. blood, synovial and cerebrospinal fluids, vitreous humour, mucus, and saliva \cite{fung1981biomechanics,nelson2014micro}. Despite this, propulsion of microswimmers in non-Newtonian fluids remains relatively unexplored.

Generally speaking, the physical properties of a fluid can be described by its viscosity $\eta$, which represents the relationship between the shear stress $\tau_{\rm s}$ and the local shear velocity $\frac{\partial u}{\partial y}$ in a fluid along one direction $x$:
\begin{equation}\label{eq:viscosity}
\tau_{\rm s} = \eta \frac{\partial u_x}{\partial y} \; ,
\end{equation}
where $u_x$ is the fluid velocity along the $x$-direction and the shear velocity is the gradient of the velocity along the perpendicular direction $y$. In a Newtonian fluid, $\eta$ is constant so that the viscous stresses arising from its flow, at every point, are linearly proportional to the local shear rates. Non-Newtonian media, instead, are any fluids whose flow properties differ in any way from this case (i.e. where $\eta$ changes): the viscosity $\eta$ can change in different ways, and different types of non-Newtonian fluids can be defined, such as \emph{shear-thinning} and \emph{shear-thickening} fluids where $\eta$ decreases or increases with the shear rate, as well as \emph{viscoelastic} and \emph{viscoplastic} fluids. For example, most of the fluids in the human body are viscoelastic (e.g. sputum, mucus, and vitreous humour) \cite{fung1981biomechanics} with many of them featuring a shear-thinning behavior (e.g. saliva, blood, and synovial fluid) \cite{qiu2014swimming,nelson2014micro}. In the study of viscoelastic fluids, it is often useful to introduce the \emph{Weissenberg number}, which is a dimensionless number that compares the viscous forces to the elastic forces; it is often given as
\begin{equation}
{\rm Wi} = \dot{\gamma} \lambda \, ,
\end{equation}
where $\dot{\gamma}$ is the shear rate and $\lambda$ is the stress relaxation time of the fluid.

The behavior of non-Newtonian media is very different from that of simple Newtonian media, such as water, because of their very different microscopic organization, which typically includes the presence of molecules, microparticles, or other complex macromolecular structures. The motion of swimmers will, therefore, also be affected by the physical properties of the fluid in which they are placed: early experiments, for example, showed that \emph{E. coli} and other types of bacteria can swim more efficiently in high-viscosity gel-forming fluids rather than in water \cite{schneider1974effect,berg1979movement}. The explanation of these experiments is still a matter of debate: the current standard model postulates the presence of bacteria-sized pores that allow them relative easy passage \cite{berg1979movement,magariyama2002mathematical}; more recent studies suggest that the fast-rotating bacterial flagellum gives rise to a lower local viscosity in its vicinity \cite{martinez2014flagellated}. In the low-Reynolds-number regime, Newtonian fluids are characterized by instantaneous and time-reversible flows that are described by the time-independent Stokes equations; as seen above with the scallop theorem, a consequence of this fact is that a swimmer will not be able to move if it is just executing geometrically reciprocal motion (i.e. a sequence of changes in its shapes that are perfectly identical when time-reversed) \cite{purcell1977life}: locomotion at low Reynolds numbers therefore generally requires nonreciprocal actuation of the swimmer that is achieved in nature, e.g., by breaking time-reversal symmetry with rotating helices \cite{turner2000real} and with cilia that show flexible oar-like beats \cite{brokaw1965non-sinusoidal}. However, theoretical and experimental work showed that breaking time-reversal symmetry is no longer a requirement in non-Newtonian fluids \cite{fu2009swimming,montenegro-johnson2013physics,keim2012fluid,qiu2014swimming}, where motion by periodic body-shape changes is possible when backward and forward strokes occur at different rates \cite{montenegro-johnson2013physics,qiu2014swimming}. Since the scallop theorem no longer holds in complex non-Newtonian fluids, it is possible to design and build novel swimmers that specifically operate in these complex fluids: for example, fluid elasticity can be used to either enhance or retard propulsion in non-Newtonian fluids \cite{lauga2007propulsion,leshansky2009enhanced,teran2010viscoelastic,liu2011force-free,espinosa-garcia2013fluid,schamel2014nanopropellers}. \citet{qiu2014swimming}, in particular, reported a symmetric ``micro-scallop" (Fig.~\ref{F12}a), a single-hinge microswimmer that can propel itself in a shear-thickening fluid by reciprocal motion even at low Reynolds numbers when the activation is asymemtric (Fig.~\ref{F12}b), but not when it is symmetric (Fig.~\ref{F12}c).

\section{Interacting Particles}\label{sec:interacting}

We will now move towards the study of interacting active particles. This is both fundamentally interesting and important in terms of potential applications because various classes of active particles, such as bacteria and artificial microswimmers, are more realistically  found in crowded environments, where they interact with both passive particles and other active particles. We will first give an overview of the kinds of interactions that take place between colloidal particles, also introducing swarming models (Section~\ref{sec:interactions}). We will then proceed to review two main aspects: we will first discuss dense suspensions of active Brownian particles, where interesting transitions, such as clustering and self-jamming, can appear as a function of particle density or activity (Section~\ref{sub:interacting1}); and we will then discuss the interactions between active and passive particles, where tantalizing phenomena, such as phase separation and active-depletion forces, can emerge (Section,~\ref{sub:interacting2}).

\subsection{Classification of particle interactions}\label{sec:interactions}

As we have seen in Sections~\ref{sec:noninteracting} and \ref{sec:hydro}, the dynamical properties of isolated active particles are rather well understood and can be modeled in various ways, ranging from particle-based descriptions, where active particles can be assumed to be point particles, disks, ellipses, or otherwise-shaped objects that obey an overdamped equation of motion, to full continuum-based models, where hydrodynamic effects are explicitly taken into account. In all these models, the motion of a single active particle in a homogenous environment is ballistic at short times and diffusive at long times. However, the presence of other active particles in the surroundings leads to mutual interactions, which not only change the single-particle dynamics but also lead to the emergence of cooperative phenomena such as dynamic clustering or phase separation.

Active particles are often subject to the same interaction forces as particles at thermal equilibrium, but the final effects on the particles' dynamics can be strikingly different. We will start by summarizing the most important effective interactions between passive colloids.
First, colloidal stabilization can be achieved sterically, which is modelled by an excluded-volume interaction such that two colloids are treated as impenetrable objects that cannot overlap in space.  Steric interactions are often numerically implemented in the following way: when a displacement makes two particles overlap, the particles are separated by moving each one half the overlap distance along their center-to-center axis. In the numerical implementation of Brownian dynamics simulations, the hard-core interaction is typically softened to avoid discontinuities at overlap. Therefore soft repulsive interactions, such as a steep Yukawa model or a truncated-and-shifted Lennard--Jones potential, are frequently employed to describe steric interactions. With the necessary adjustments, these considerations can be extended to particles with more complex shapes \cite{kirchhoff1996dynamical}.
Second, charge-stabilized suspensions are described by the traditional Derjaguin--Landau--Verwey--Overbeek (DLVO) theory \cite{derjaguin1941theory,verwey1947theory}, which involves a repulsive electrostatic part (often described by an effective pairwise screened Coulomb interaction) and an attractive part stemming from mutual van-der-Waals interactions. Other kinds of interactions can also emerge such as hydrodynamic interactions (discussed in Section~\ref{sec:hydro}).

More complex models for the interactions can be considered for non-spherical particles by adding torques, bending effects, or rotations, which affect the direction of the propulsion or motor force \cite{vicsek2012collective}. Furthermore, multiple particles can be connected together by springs or other potentials to create active polymers or active membranes.

\subsubsection{Aligning interactions, Vicsek model, and swarming}\label{sec:vicsek}

We will now turn our attention to systems where there are interactions capable of aligning the motion of active particles. In practice, several types of interactions can lead to alignment. For example, alignement can be the result of hydrodynamic interactions between swimmers, as we have seen in Section~\ref{sec:ppinter} and Fig.~\ref{F9}, steric interactions between elongated particles and even between self-propelled hard disks \cite{lam2015self}, and hydrodynamics and electrostatic between rolling colloids \cite{bricard2013emergence}.

Aligning interactions are particularly important in active matter systems, as they can lead to collective motion and swarming. One of the best-known and used collective motion models is the \emph{Vicsek model} \cite{vicsek1995novel,czirok1997spontaneously}. In its original version, the particles move with a constant velocity and interact only through an alignment term, whereby the direction of motion of particle $n$ is adjusted based on the average direction of motion of all neighboring particles within a flocking radius. A generalized version of the Vicsek model can be derived from the discrete-time version of Eq.~(\ref{eq:abm}), which describes an active Brownian particle in two dimensions, by modifying the equation for the orientation of the particle, obtaining
\begin{equation}
\left\{\begin{array}{ccl}
\displaystyle x_n(t+\Delta t) & = & x_n(t) + v \Delta t\displaystyle \cos\varphi_n + \sqrt{2D_{\rm T} \Delta t} \, \xi_{x,n} \\[6pt]
\displaystyle y_n(t+\Delta t) & = & y_n(t) + v \Delta t \displaystyle \sin\varphi_n + \sqrt{2D_{\rm T} \Delta t} \, \xi_{y,n} \\[6pt]
\displaystyle \varphi_n(t+\Delta t) & = & \displaystyle \left< \varphi_m (t) \right>
\end{array}\right.
\end{equation}
where the averaging is over all the particles within a flocking radius. The Vicsek model exhibits a phase transition as a function of increasing particle density from undirected motion to a state where all the particles move in the same direction 
and bands of particles appear\cite{gregoire2004onset,chate2008collective,ramaswamy2010mechnaics,ginelli2010relevance,solon2015phase}. 
Furthermore, the Vicsek model can be modified to include steric interactions between particles, which cause crystallization at high densities, as well as interactions with barriers or with a substrate \cite{drocco2012bidirectional}.

\subsection{Collective behaviors of active particles}\label{sub:interacting1}

In this section we will consider dense suspensions of active particles. First, we will see how the presence of interacting active particles can lead to the formation of clusters, also know as ``living crystals" (Section~\ref{sec:lc}). Then, we will see how the presence of active matter can change the rheological properties of the medium, inducing, in particular, active self-jamming (Section~\ref{sec:sj}) and active turbulence (Section~\ref{sec:turbulence}).

\subsubsection{Clustering and living crystals}\label{sec:lc}

\begin{figure}[b]
\includegraphics[width=\columnwidth]{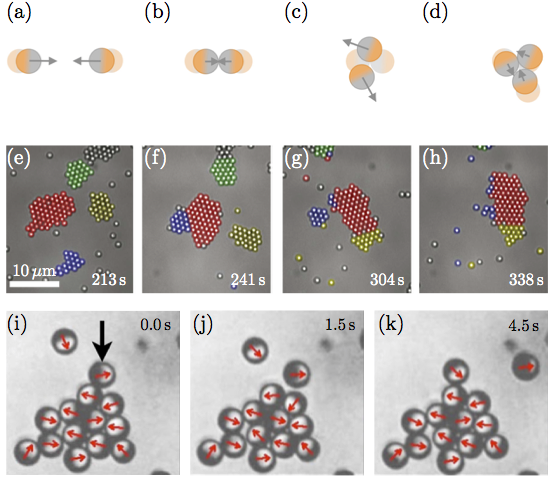}
\caption{(Color online) Clustering and living crystals.
(a-d) Qualitative explanation of the clustering process: (a) when two active particles collide head-on, (b) they block each other and form a two-particle cluster; (c) the cluster breaks apart on the time scale of the rotational diffusion; (d) depending on the particle speed and density, another particle might collide before the two-particle cluster has broken apart and, thus, the cluster grows into a three-particle cluster.
(e-h) Clusters assembled from a homogeneous distribution of active particles (solid area fraction $\phi=0.14$). The false colors show the time evolution of particles belonging to different clusters. The clusters are not static but rearrange, exchange particles, and merge. 
From \citet{palacci2013living}.
(i-k) Consecutive snapshots of a cluster of active Janus particles. The small red/dark gray arrows indicate the projected orientations of the caps. Particles along the rim mostly point inwards. The particle indicated by the large black arrow in (i), leaves the cluster (j) and is replaced by another particle (k). From \citet{buttinoni2013dynamical}.
\label{F13}
}
\end{figure}

Dilute suspension of passive colloidal particles do not spontaneously form clusters unless there are strong attractive interactions. This is nevertheless possible in suspensions of active particles even in the presence of purely repulsive interactions. This was first predicted theoretically \cite{tailleur2008statistical,fily2012athermal,redner2013reentrant} and has recently been observed experimentally \cite{palacci2013living,buttinoni2013dynamical}.

A simple qualitative explanation of this phenomenon is shown in Figs.~\ref{F13}a-d: when two active particles collide, they block each other due to the persistence of their motion (sequence of Figs.~\ref{F13}a to \ref{F13}b); such a two-particle cluster breaks when one of the two particles points away, which happens on a time scale comparable to the particle reorientation time (e.g. the rotational diffusion time $\tau_{\rm R}$ given by Eq.~(\ref{eq:Dr})) (Fig.~\ref{F13}c); since the mean time between collisions is controlled by the particle speed and density, depending on these parameters another particle might collide before the two-particle cluster has broken, forming a three-particle cluster (Fig.~\ref{F13}d); this leads to metastable clusters of a few particles or, if the mean collision time falls below a certain value, to the growth of ever larger clusters via the setting in of a dynamical instability.

These qualitative considerations have in fact been confirmed by experiments where one observes the formation of clusters at intermediate particle densities \cite{theurkauff2012dynamic,ginot2015nonequilibrium} (see also the theoretical work by \citet{pohl2014dynamic,pohl2015self}). Due to the steady particle collisions, these clusters are subjected to strong changes in size and shape with their average size increasing with the particle activity. Similar dynamic clusters of finite size were reported for hematite swimmers, as shown in Figs.~\ref{F13}e-h \cite{palacci2013living}: in these experiments, the presence of attractive diffusiophoretic interactions could be directly measured. Comparison of the experiments with simulations suggested the presence of a diffusiophoretic aggregation mechanism which was caused by the interaction of the concentration profiles around each particle.

Experiments with active suspensions where diffusiophoretic interactions are neglibible also show clustering and phase separation into dense clusters and a dilute gas phase at higher particle concentrations, as shown in Figs.~\ref{F13}g-k \cite{buttinoni2013dynamical}.  These experimental observations were corroborated by numerical simulations of a minimal model, where only pure (short-ranged) repulsive interactions between the particles were considered and the behavior rationalized in terms of the self-trapping of active particles \cite{bialke2013microscopic}.

The formation of active crystals was also observed as the result of collective dynamics of fast swimming bacteria \cite{chen2015dynamic,petroff2015fast-moving}. 

Theoretical work has highlighted the connection between clustering and phase separation. \citet{tailleur2008statistical} introduced the concept of motility-induced phase separation (for a recent review see \citet{cates2015motility}). Assuming a density dependent motility $v(\phi)$, there is an instability from a perturbed homogeneous suspension of swimmers if the gradient of $dv(\phi)/d\phi$ is sufficiently negative such that $dv(\phi)/d\phi / v(\phi) < -1/\phi$. Microscopic approaches combined with an instability analysis start from the Smoluchowski equation \cite{bialke2013microscopic}. For large length scales, clustering of active Brownian spheres can be mapped onto phase separation of a passive system with attractive interactions \cite{speck2014effective,cates2015motility}.

We finally remark that hydrodynamic near fields acting between squirmers can play a crucial role in determining their aggregation and collective motion \cite{zottl2014hydrodynamics}; interestingly, in the presence of a harmonic trapping potential, they can form a self-assembled fluid pump at large enough P{\'e}clet numbers \cite{hennes2014self}.

\subsubsection{Self-jamming and active microrheology}\label{sec:sj}

Jamming occurs when the density of a loose ensemble of particles such as grains or bubbles becomes high enough that the system can support a shear as if it were a solid.  This phenomenon has been intensely studied in the contexts of granular matter, colloids, and emulsions \cite{van2009jamming,berthier2011microscopic,coulais2014ideal}; and in certain cases jamming has been shown to have properties consistent with a phase transition \cite{liu2010jamming,reichhardt2014aspects}. It is thus interesting to ask whether active matter systems can exhibit jamming-like features in the dense limit. 

It might seem that increasing the activity would tend to suppress jamming. However, as we have seen in the previous section, active matter systems can form a self-clustering state in which the particles inside a cluster are locally jammed.  \citet{henkes2011active} considered the active jamming of self-propelled soft disks.  They observed an aligned phase at low densities and a transition to a jammed phase for higher densities.  Even within the jammed state, the activity induces large correlated motions of the particles. \citet{berthier2013non} and \citet{berthier2014nonequilibrium} analyzed the collective dynamics of self-propelled particles in the high-density regime, where passive particles undergo a kinetic arrest to an amorphous glassy state, and found that the critical density for dynamic arrest continuously shifts to higher densities with increasing activity.

\begin{figure}[t]
\includegraphics[width=\columnwidth]{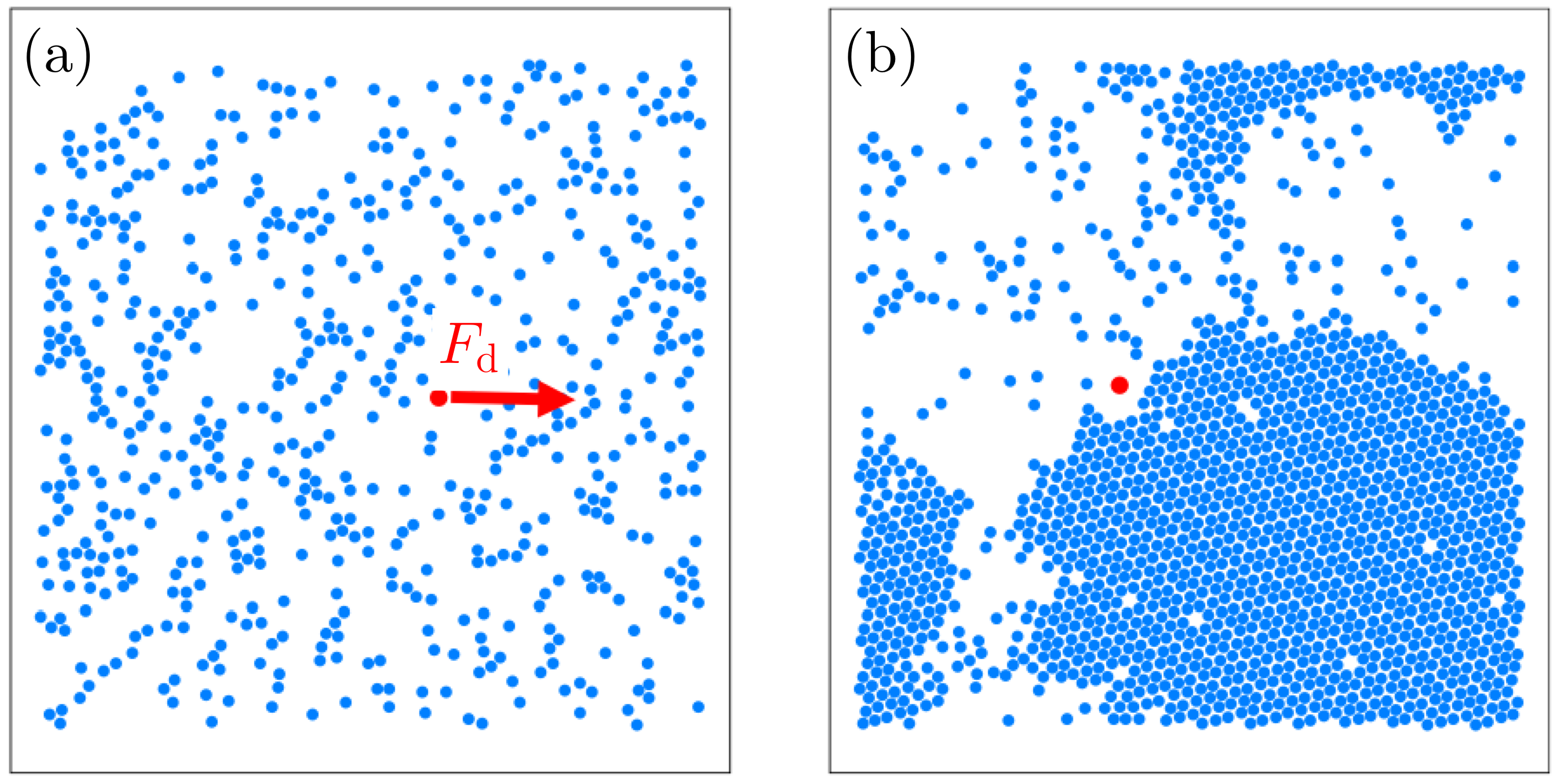}
\caption{(Color online) Self-jamming. 
(a) Particle locations for an active matter system (blue/dark gray particles) in a uniform liquid state at density $\phi\approx0.2$ with an externally driven probe particle (red/light gray particle). The arrow indicates the direction of the probe driving force $F_{\rm d}$. 
(b) Phase-separated cluster state at $\phi \approx 0.5$, consisting of a high-density phase with local crystal ordering coexisting with a low-density liquid phase.
From \citet{reichhardt2015active}.
\label{F14}
}
\end{figure}

One method for exploring the onset of jamming in a (passive or active) system is through microrheology. A probe particle is driven through a medium of other particles, and changes in the effective viscosity (or fluctuations of the probe motion) can be used to quantify changes in the medium \cite{squires2005simple,candelier2009creep,olson-reichhardt2010fluctuations}: if the medium is gas-like (Fig.~\ref{F14}a), the effective viscosity felt by the probe will be well-defined with relatively small fluctuations, while, if the particles in the medium are jammed (Fig.~\ref{F14}b), the effective viscosity will fluctuate more strongly and eventually acquire a dichotomic distribution (corresponding to motion either in the gas or in the solid phase of the active particle solution). 

\citet{foffano2012colloids} numerically studied a colloid driven through an active nematic bath, finding that the drag on the colloid was non-Stokesian and not proportional to the radius of the colloidal particle; they even observed instances of negative drag in which the particle moves in a direction opposite to an externally applied driving force.

\citet{reichhardt2015active} numerically studied a probe particle driven with a fixed external driving force $F_{\rm d}$ through a bath of run-and-tumble disks whose run length is fixed. At low densities ($\phi\approx0.2$), the active particles are in a liquid state (Fig.~\ref{F14}a), while, as the density increases,  there is a transition to a phase-separated or clustered state (Fig.~\ref{F14}b for $\phi=0.5$). Previous studies of an active probe particle near a jamming transition also showed that the probe particle exhibits avalanche behaviors, which were argued to provide evidence that the jamming transition is a continuous phase transition with critical properties \cite{redner2013reentrant,mognetti2013living}.  In passive systems, jamming is associated with a specifical critical density $\phi_{\rm c}$, while in active systems the clustered states self-organize into locally jammed regions with local density $\phi_{\rm c}$, indicating that active systems can exhibit critical fluctuations well below the bulk jamming density of passive systems.

Even though the results available until now are only numerical, it should be possible to translate these to experiments. For example, it is in principle experimentally feasible to study the drag on a driven probe particle placed in an active bath of swimming bacteria or active colloids as the bath conditions are varied.

\subsubsection{Active turbulence}\label{sec:turbulence}

A particularly interesting manifestation of collective behavior in microscopic active matter systems is the emergence of turbulent motion with the continuous formation and decay of whirls, jets, and vortices in dense solutions of active particles \cite{mendelson1999organized,dombrowski2004self,riedel2005self,sokolov2007concentration,saintillan2007orientational,cisneros2007fluid,breier2014spontaneous}. The observation of active turbulent patterns in a microscopic active system can be traced back to the experimental work of \citet{mendelson1999organized}, who observed the organization of swimming cells of \emph{Bacillus subtilis} into short-lived dynamic patterns. The formation of these patterns seems to be ubiquitous among active matter systems since similar active turbulence has been reported in other active systems on widely varying length and time scales, from suspensions of microtubules, filaments, and molecular motors \cite{schaller2010polar,sanchez2012spontaneous,sumino2012large}, to topological defects in nematic vesicles \cite{keber2014topology}, vortices in active nematics \cite{giomi2015geometry}, agitated granular matter \cite{narayan2007long}, and schools of fish and flocks of birds \cite{vicsek2012collective}. 

Normally turbulence is a consequence of inertia, which is usually negligible at micrometric and smaller length scales; hence, the active turbulence seen in the microscopic active systems needs a different explanation, although, to date, a unified description of the formation and structure of these patterns remains lacking \cite{ishikawa2011energy,wensink2012meso,dunkel2013fluid}. The full characterization of the active turbulent spectra needs first of all a clear justification of power-law scaling over several decades and the corresponding derivation of new classes of non-universal exponents; an important step forward in this direction was recently done by \citet{bratanov2015new}.

\subsection{Mixtures of active and passive particles}\label{sub:interacting2}

\begin{figure*}
\includegraphics[width=\textwidth]{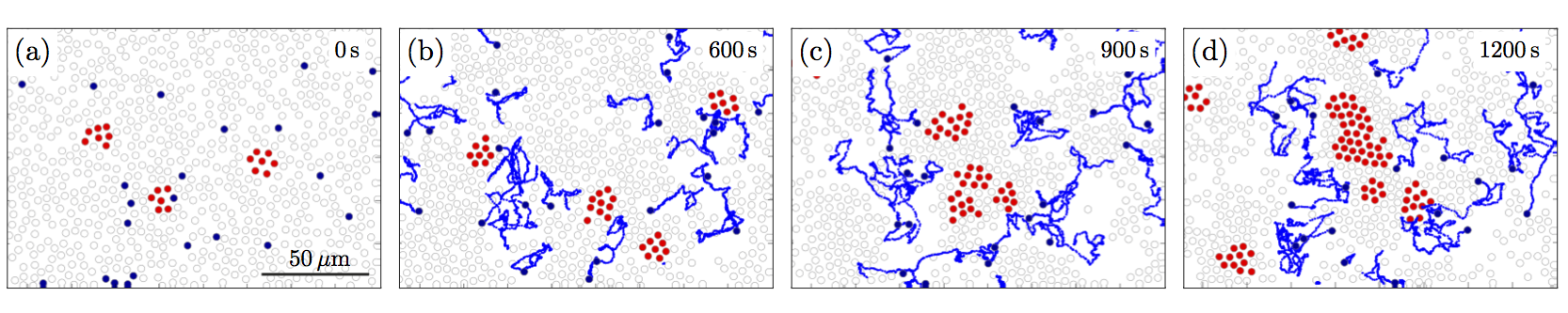}
\caption{(Color online) Doping of a passive particle solution with active particles.
Experimental snapshots of the temporal evolution of a mixture of passive ($\phi_{\rm p} = 0.40$) and active ($\phi_{\rm a} \approx 0.01$) particles at (a) $0$, (b) $600$, (c) $900$, and (d) $1200\,{\rm s}$ for P\'eclet number ${\rm Pe} \approx 20$. The passive particles belonging to clusters are represented as red/light gray circles, while those not belonging to clusters are represented as open circles. Active particles are shown as blue/dark gray circles and their trajectories over $300\,{\rm s}$ prior to each snapshot are represented as solid lines.
From \citet{kuemmel2015formation}.
\label{F15}
}
\end{figure*}

In this section we consider the interaction between passive and active particles. We will start by considering the effect that relatively few active particles can have on the properties of a solution of passive particles (Section~\ref{sec:doping}). We will then consider the case of a mixture with similar numbers of active and passive particles, where one can observe phase separation and turbulent behaviors (Section~\ref{sec:5050}). Afterwards, we will consider the limit where a few passive particles are immersed in a dense suspension of active particles: we will therefore introduce the concept of active bath (Section~\ref{sec:activebath}), active-particle-powered directed motion and gears (Section~\ref{sec:gears}), and active depletion and active-depletion forces (Section~\ref{sec:casimir}). Finally, we will consider the role of the shape and flexibility of the passive particles on the effects of an active bath (Section~\ref{sec:flexible}). Interesting reviews on some aspects of the topics covered in this section can be found in \citet{bialke2015active}, \citet{cates2015motility}, and \citet{elgeti2015physics}.

\subsubsection{Active doping}\label{sec:doping}

Doping of colloidal suspensions with a very small amount of active particles can strongly influence their properties and dynamics. 

\citet{ni2013pushing,ni2014crystallizing} showed with numerical simulations that the crystallization of hard-sphere glasses can be dramatically promoted by doping the system with small amounts of active particles.

\citet{kuemmel2015formation} demonstrated with experiments and numerical simulations that the structure and dynamics of a suspension of passive particles is strongly altered by adding a very small ($<1\%$) number of active particles. Figure~\ref{F15} illustrates the typical temporal changes in a colloidal suspension when doped with active particles; the active particles herd the passive particles favoring the formation of metastable clusters. Above a minimum passive particle concentration, it is possible to observe the formation of isolated dynamic clusters of passive colloids, which are surrounded by active particles. At higher passive particle concentrations, such activity-induced clusters start to merge and undergo further compression. When exceeding the threshold for spontaneous crystallization, active particles are found to accumulate at the interfacial regions between crystalline domains, where they lead to surface melting. 

Using numerical simulations, \citet{van2016fabriating} showed that active dopants can provide a route to removing grain boundaries in polycrystals: since, as we have seen above, active dopants both generate and are attracted to defects, such as vacancies and interstitials, they tend to cluster at grain boundaries; thus, the active particles both broaden and enhance the mobility of the grain boundaries, causing rapid coarsening of the crystal domains; finally, the remaining defects can be made to recrystallize by turning off the activity of the dopants, resulting in a large-scale single-domain crystal.

\subsubsection{Phase separation and turbulent behavior}\label{sec:5050}

A few numerical studies looked into the behavior of dense mixtures with approximately comparable numbers of both active and passive particles. These investigations have reported the emergence of interesting novel phenomena including active-passive segregation between rod-like particles \cite{mccandlish2012spontaneous}, emergence of flocking and turbulence \cite{hinz2014motility}, and phase separation \cite{yang2012using,das2014phase,yang2014aggregation,stenhammar2015activity,takatori2015theory,farage2015effective,tung2015micro}.

\citet{su2015suppressing} performed a numerical study of self-propelled rods interacting with passive particles where the passive particles also interact with each other via a Yukawa repulsion that can be tuned from weak to strong.  When the coupling between the passive particles is weak, they feature a turbulent behavior, while a strong coupling suppresses the turbulent behavior and allows the formation of a nearly triangular lattice.

\subsubsection{Active baths}\label{sec:activebath}

\begin{figure*}
\includegraphics[width=\textwidth]{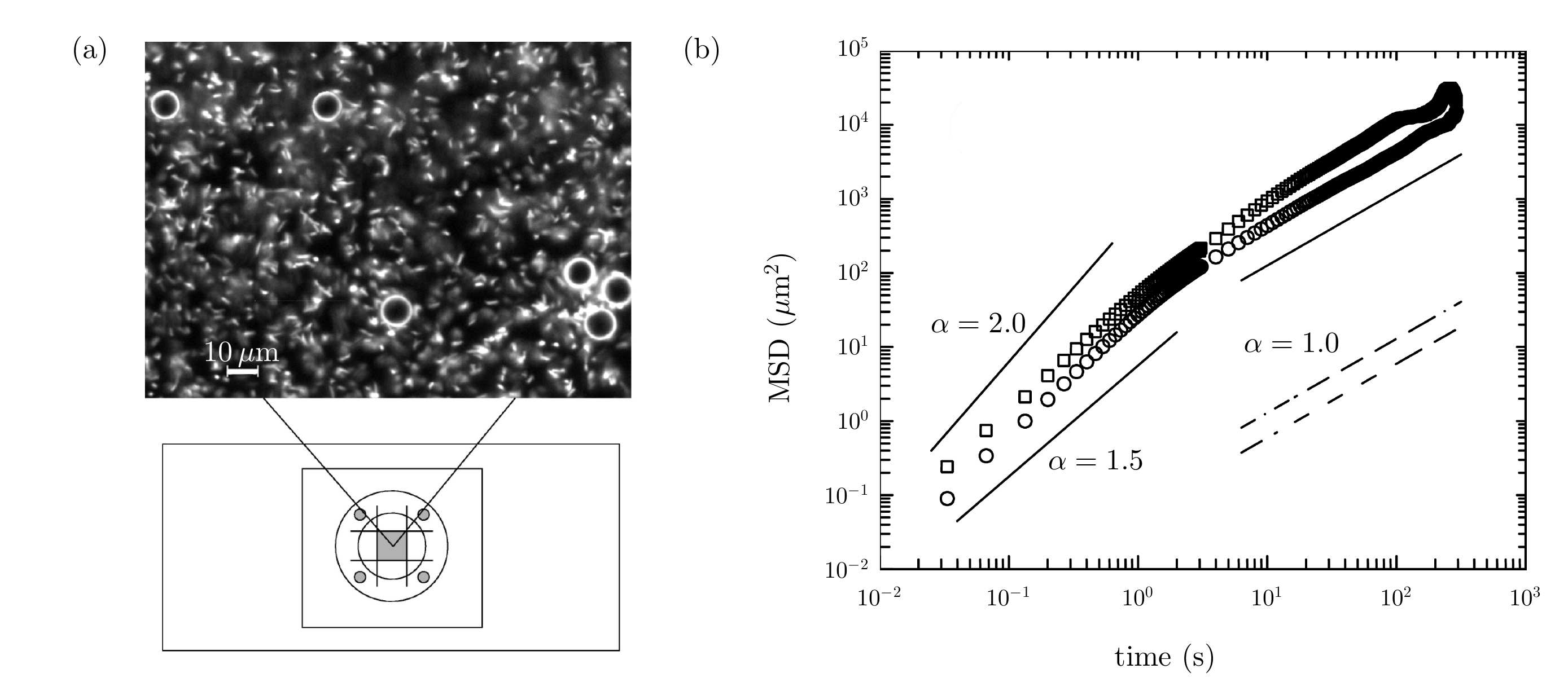}
\caption{Superdiffusion of passive particles in an active bath.
(a) Experimental setup and fluorescence image of a solution with \emph{E. coli} mixed with 10-${\rm \mu m}$ polystyrene spheres.
(b) MSD measurements of polystyrene spheres with diameters $4.5$ (squares) and $10\,{\rm \mu m}$ (circles) at a particle concentration of $\approx 10^{-3}\,{\mu^{-2}}$ in an active bath. The solid lines with slopes $\alpha=2.0$, $1.5$, and $1.0$ are guides to the eyes. The two dashed lines correspond to the thermal diffusion of 4.5-${\rm \mu m}$ and 10-${\rm \mu m}$ particles.
From \citet{wu2000particle}.
\label{F16}
}
\end{figure*}

\begin{figure}[h!]
\includegraphics[width=\columnwidth]{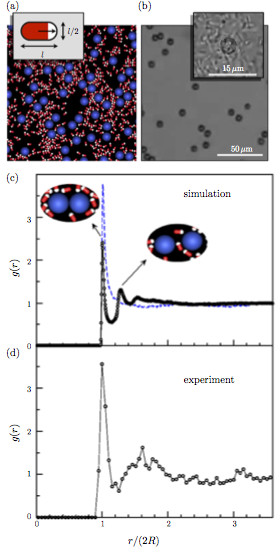}
\caption{(Color online) Interactions mediated by an active bath.
(a) Snapshot of a simulation of passive (spherical) particles immersed in an active bath. The active particles are motile bacteria (represented by spherocylinders with white heads pointing in the direction of self-propulsion). 
(b) Microscopy image from experiment. The bacteria are visible in the enlarged image in the inset.
(c) Simulated and (d) experimental radial distribution functions $g(r)$. Typical two-bead configurations corresponding to the two first peaks are also shown in the insets in (c). Distances are in units of particle diameter $2R$ (bacteria width is about $0.25$ in such a unit). The dashed line is the $g(r)$ obtained in simulations with a mixture of bacteria interacting with only one of the particles, i.e. without depletion.
From \citet{angelani2011effective}.
\label{F17}
}
\end{figure}

A passive particle is in an \emph{active bath} when it is in an environment where a wealth of active particles are present, e.g. motile bacteria, as shown in Fig.~\ref{F16}a \cite{wu2000particle}. These particles will exert non-thermal forces on the passive object so that it will experience non-thermal fluctuations and will behave widely different from a passive Brownian particle in a thermal bath.  For example, a passive colloidal particle in a dense bacterial bath behaves like an active particle due to multiple interactions with the self-propelled bacteria: the measured mean-square displacements indicate superdiffusion at short times and normal diffusion at long times \cite{wu2000particle}, as shown in Fig.~\ref{F16}b; however, the characteristic time at which this transition occurs is not related to the rotational diffusion, but rather to the density and activity of the bacteria acting as active particles. The long time diffusivity of passive tracers can be orders of magnitude larger than their thermal counterparts \cite{wu2000particle, mino2011enhanced} and also displays exponential tails in the distribution of displacements \cite{leptos2009dynamics, lin2011stirring}.  Active particles in the bath can contribute to the enhanced diffusivity of passive tracers through both hydrodynamic interactions \cite{gregoire2001active, thiffeault2010stirring, lin2011stirring, pushkin2013fluid, pushkin2014stirring} and direct contact forces \cite{gregoire2001active, valeriani2011colloids}.

The presence of an active bath can also significantly influence the microscopic thermodynamics of a particle. For example, \citet{argun2016experimental} has considered the applicability of Jarzynski equality in the presence of an active bath, finding that it fails because of the presence of non-Boltzmann statistics. This observation points towards a new direction in the study of nonequilibrium statistical physics and stochastic thermodynamics, where also the environment is far from equilibrium.

Moving a step further, it is also interesting to consider the interactions that emerge between passive particles in an active bath. \citet{angelani2011effective} conducted simulations and experiments of passive particles immersed in a bath of run-and-tumble swimming bacteria. Figure~\ref{F17}a shows a snapshot of the simulation where the bacteria are represented as spherocylinders with white-colored ``heads" indicating their direction of swimming and the passive particles are modeled as spheres.  Figure~\ref{F17}b shows a corresponding experimental snapshot. As the level of activity of the bacterial bath increases, a short range ``attraction" emerges between passive particles as evidenced by a pronounced peak in the radial distribution function in Fig.~\ref{F17}c. Figure~\ref{F17}d indicates the radial distribution function obtained in experiment, in good agreement with the simulation.
The origin of this short-range attraction is a nonequilibrium effect arising from the persistent character of the fluctuating forces exerted by bacteria.

\subsubsection{Directed motion and gears}\label{sec:gears}

\begin{figure}
\includegraphics[width=\columnwidth]{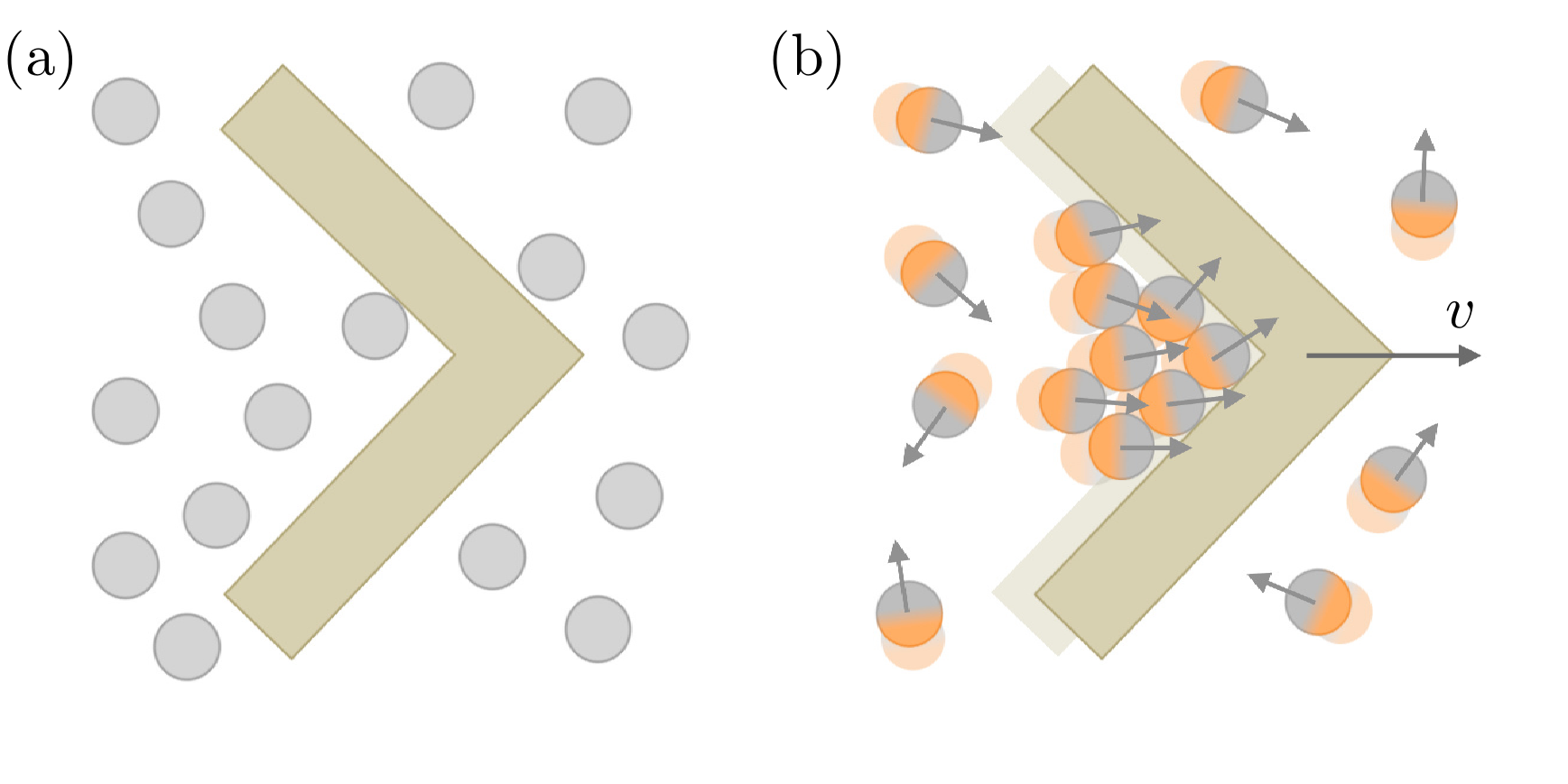}
\caption{(Color online) Motion of a passive wedge in an active bath.
(a) If the wedge is surrounded by passive particles, there is no directed motion, because the pressure due to the passive particles is equal on all sides of the object. 
(b) However, in an active bath, the active particles accumulate in the corner of the wedge, producing its net propulsion.
\label{F18}
}
\end{figure}

\begin{figure*}
\includegraphics[width=\textwidth]{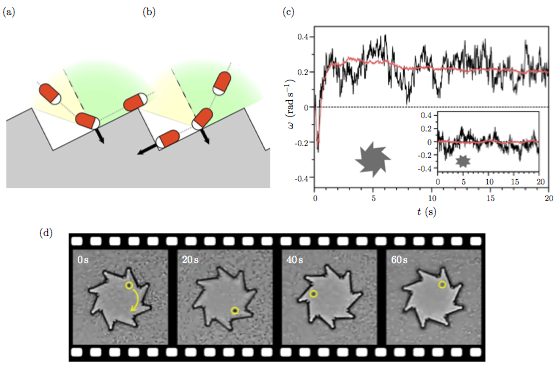}
\caption{(Color online) Bacteria-driven micromotors.
(a,b) Sketch of the collision of a single bacterium with the rotor boundary: (a) bacteria coming from the left area with respect to the normal leave the gear, while (b) bacteria from the right get stuck at the corner exerting a torque on the rotor. The arrows represent the forces exerted by the bacteria on the rotor.
(c) Angular velocity $\omega$ of the micromotor as a function of time: the black line refers to a single run; the red (lighter) line is the average over 100 independent runs. After a short transient regime (due to the initial configuration of bacteria), a fluctuating velocity around a mean value $\omega_0 \approx 0.21\,{\rm rad\,s^{-1}}$ is observed. Inset: same as main plot for a symmetrically shaped micromotor, which does not rotate (in average).
From \citet{angelani2009self}.
(d) A nanofabricated asymmetric gear ($48\,{\rm \mu m}$ external diameter, $10\,{\rm \mu m}$ thickness) rotates clockwise at $1\,{\rm rpm}$ when immersed in an active bath of motile \emph{E. coli} cells, visible in the background. The gear is sedimented at a liquid--air interface to reduce friction. The circle points to a black spot on the gear that is used for visual angle tracking.
From \citet{dileonardo2010bacterial}.
\label{F19}
}
\end{figure*}

Differently from spherical particles, asymmetric passive particles placed in an active bath can feature a directed motion. For example, we can consider the case of a V-shaped wedge, such as the one shown in Fig.~\ref{F18}. If the wedge is surrounded by passive particles (Fig.~\ref{F18}a), there is no directed motion, because the pressure due to the passive particles is equal on all sides of the object. However, in an active bath, the active particles accumulate in the cusp of the wedge, producing its net propulsion (Fig.~\ref{F18}b). This simple model system has been studied by various groups both numerically and experimentally \cite{angelani2010geometrically,wensink2014controlling,mallory2014brownian,smallenburg2015swim}. \citet{angelani2010geometrically} employed asymmetric wedges to create bacteria-powered shuttles. \citet{kaiser2014transport} performed experiments and simulations of a passive wedge placed in a bacterial bath and found that the wedge can undergo directed motion. Maximal efficiency occurs for an intermediate bacterial density,  so that a turbulent bath maximizes the speed of the carrier. This can be qualitatively understood by considering two limits: for high dilution, the pushing effect is small since only few swimmers are contributing; for high density, the bath is jammed, which leaves no mobility for the carrier; hence, there must be an optimal density that maximizes the transport efficiency in the intermediate turbulent state. As a function of the wedge apex angle the carrier speed attains a maximum at about $90^\circ$ \cite{kaiser2014transport}. \citet{mallory2014brownian} numerically studied a system of asymmetric tracers (partial circles with varied curvature) placed in a bath of self-propelled Janus particles and found that even low densities of active particles can induce pronounced directed motion of the tracers. U-shaped particles were found to be optimal to obtain enhanced transport  \cite{smallenburg2015swim}. Typically, the tracer speed increases with increasing persistence time of the individual self-propelled particles.

Going beyond a simple wedge, it is also possible to consider self-starting cogwheels that exhibits a spontaneous rotation when in an active bath \cite{angelani2009self,dileonardo2010bacterial,sokolov2010swimming}. These bacterial-driven micromotors work as active-matter ratchets, whose key operational feature is the trapping of the self-propelled particles in the funnel tips, against which the particles exert their motor force.  Purely thermal particles will not accumulate in the funnel tips but will instead diffuse uniformly throughout the system.  In the active matter case, the particles exert sufficient force on the cogwheel to move it, indicating that active matter can be used to perform work on appropriately shaped objects. \citet{angelani2009self} performed a numerical study where motile bacteria were modeled as rod-shaped particles undergoing run-and-tumble dynamics and large mobile objects in the shape of asymmetric gears were placed in the bacterial bath. Figures~\ref{F19}a and \ref{F19}b illustrate the microscopic particle-wall interactions responsible for the rotation of the gear: in Fig.~\ref{F19}a, a particle enters the frame from the left, runs along the wall of the gear (outlined in gray), and is shunted back into the bulk; in Fig.~\ref{F19}b, instead, a particle entering from the right is trapped at the corner of the gear and continues swimming into the corner, generating a torque that drives the rotation of the gear. Thus, the active particles collect in the inner corners of the gear teeth and produce a net torque on it, gradually rotating it in one direction as illustrated by the plot of the angular velocity $\omega$ versus time in Fig.~\ref{F19}c, and producing a realization of a bacteria-powered motor.  When a symmetric gear is used instead, as shown in the inset of Fig.~\ref{F19}c, it does not rotate since the forces exerted on it by the active bath are symmetric. \citet{dileonardo2010bacterial} confirmed these results experimentally as shown in Fig.~\ref{F19}d, showing that an asymmetric gear rotates. \citet{sokolov2010swimming} also performed experiments on asymmetric gears placed in a bath of swimming bacteria and emphasized that the collective effect of many swimming bacteria can considerably enhance the rotation rate.  At high bacteria concentrations, however, the motility of the bacteria decreases, causing a decrease in the rotation rate of the gear. Rectification of asymmetric gears has also been studied for the case where the active particles are small robots \cite{li2013asymmetric} or Janus particles \cite{maggi2016self}; in the latter case, it has been shown that a perfectly ordered micromotor can self-assemble when the edge lengths of the micro-gear are properly chosen with respect to Janus particle diameter.

\subsubsection{Active depletion}\label{sec:casimir}

\begin{figure*}
\includegraphics[width=\textwidth]{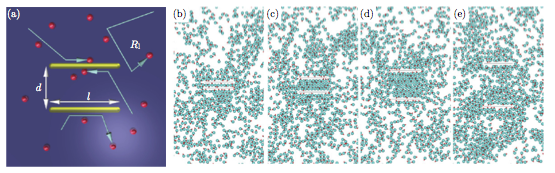}
\caption{(Color online)  Active-depletion interactions between two plates in an active bath.
(a) A schematic of the system containing run-and-tumble particles (spheres) with some particle trajectories indicated by lines and arrows. The run length is $R_{\rm l}$. The two parallel walls (bars) of length $l$ are separated by a distance $d$. When a particle moves along a wall, it imparts a force against the wall.
From \citet{ray2014casimir}.
(b-e) Typical snapshots of systems for density distributions with a wall-to-wall distance $d/R$ equal to (b) $2.1$, (c) $5$, (d) $10$, and (e) $20$, where $R$ is the radius of the active particles. If the particles are modeled as hard spheres, repulsive as well as attractive active-depletion forces can emerge.
From \citet{ni2015tunable}.
\label{F20}
}
\end{figure*}

\begin{figure}
\includegraphics[width=\columnwidth]{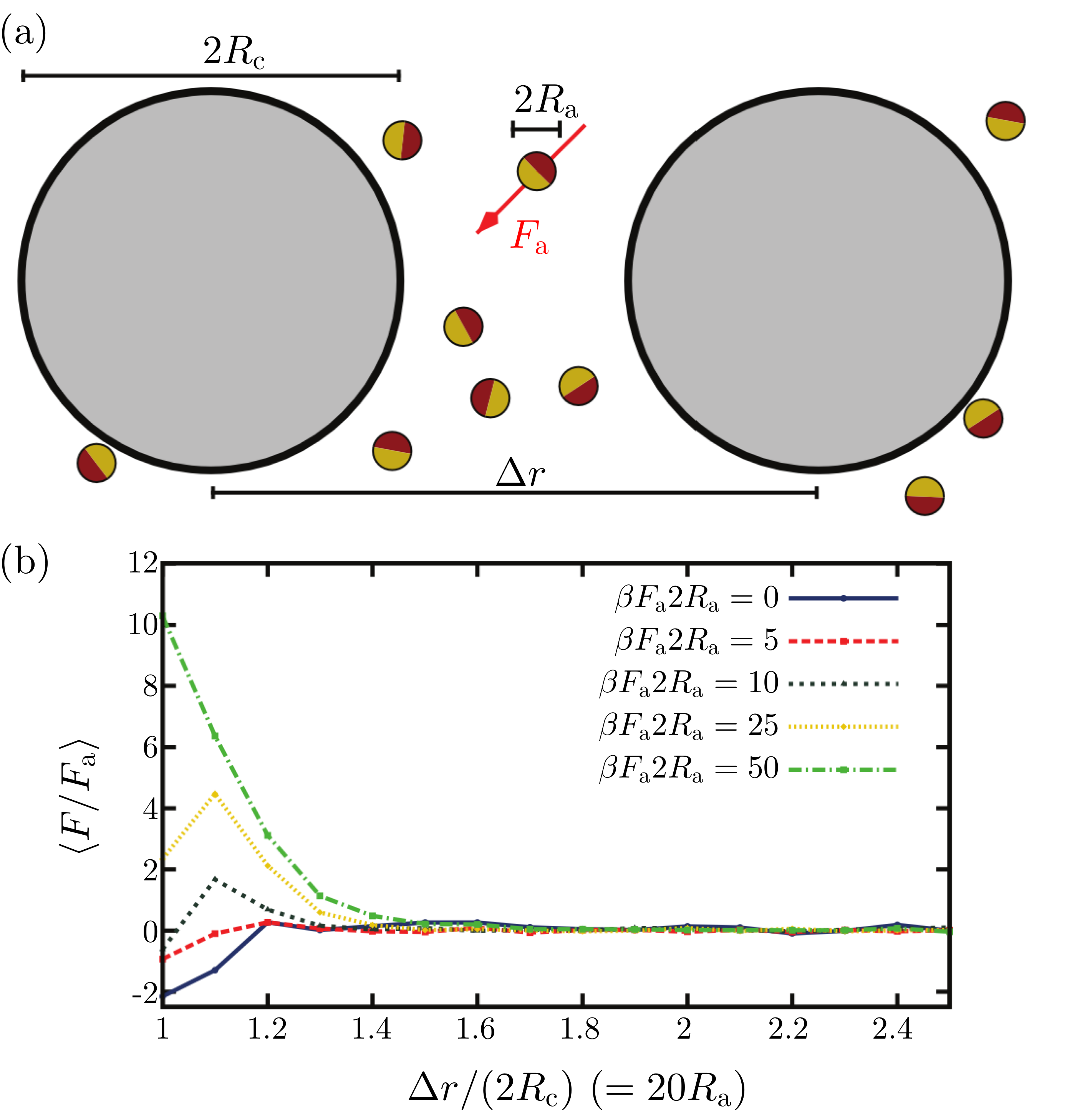}
\caption{(Color online) Emergence of repulsive active-depletion forces in active baths.
(a) Schematic representation of two colloidal disks in a bath of active particles. The persistent force $F_{\rm a}$ acts along a defined axis, which is shown by the arrow as well as the colors, where red/dark gray corresponds to the back of the particle and yellow/light gray to its front. 
(b) Effective rescaled forces $\langle F/F_{\rm a} \rangle$ experienced by two colloidal disks as a function of their separation for different values of depletant's activity for $R_{\rm c} = 10 R_{\rm a}$, where $R_{\rm c}$ is the radius of the colloids and $R_{\rm a}$ is the radius of the active particles ($\phi = 0.1$). Rescaling is only applied as long as $\beta F_{\rm a} 2 R_{\rm a} \neq 0$. Positive values correspond to a repulsion, which clearly dominates any depletion-driven interaction when the bath is active. The larger the active force and the larger the colloid-to-depletant size ratio, the stronger the repulsion.
From \citet{harder2014role}.
\label{F21}
}
\end{figure}

Going beyond the directed motion (or rotation) that an active bath can produce on a single passive object, it is also interesting to consider what kind of interactions emerge between multiple passive particles in the presence of an active bath. We can speak of \emph{active depletion}, as this case is the natural generalization of the depletion forces arising in passive baths \cite{asakura1954interaction}.

\citet{ray2014casimir} studied one of the simplest configurations, i.e. the case of two parallel plates surrounded by a bath of run-and-tumble active particles. As shown schematically in Fig.~\ref{F20}a, the plates are separated by a distance $d$ and the net force acting on the plates can be measured as $d$ is varied. The active particles are not allowed to interact with each other but only with the walls, corresponding to a very low active particle density regime.\footnote{\citet{ray2014casimir} also considered the effect of including particle--particle interactions in the dilute limit and found that the attraction between the plates persists but is reduced in magnitude.} Particles can only enter the region between the plates by approaching from the sides over a limited range of angles, which becomes more limited as $d$ decreases. In contrast, particles can reach the outer edges of the plates from any angle. This generates an effective active-depletion force that pushes the two plates together. It is also possible to exploit  this shadowing effect by constructing very long walls with a very small aperture in order to preferentially trap particles between the two plates, resulting in a negative (i.e. repulsive) effective force.  This opens the possibility that objects with carefully chosen geometries, when placed in an active bath, could experience a crossover from attractive to repulsive forces as their spacing $d$ diminishes, producing a tunable mimic of an inter-atomic potential.

\citet{ni2015tunable} studied a similar geometry consisting of two plates and active particles modeled as repulsively interacting colloidal particles whose direction of motion undergoes a gradual rotational diffusion rather than sudden run-and-tumble changes.  For a density of $\phi=0.4$, well below the passive crystallization density of $\phi=0.9$, they found that the force between the plates has an oscillatory nature, with a net repulsive force.  This effect arises due to the formation of a bridge between the two plates composed of a densely packed, partially crystalline cluster of active particles, while in the bulk no clusters are present, as illustrated in Figs.~\ref{F20}b-e.  As we have seen in Section~\ref{sec:sj}, in interacting active matter systems without walls or obstacles, there is a transition from a liquid state to a phase-separated cluster state when the particle density or activity level is high enough.  The walls act as nucleation sites for clusters of active particles even for particle densities well below the clean phase-separated regime.  When the nucleated clusters grow large enough, they form a bridge between the two walls and the repulsive interactions between the particles in the cluster produce a net repulsive force between them.  When the spacing $d$ between the plates is large enough, the bridge can no longer span the two plates and the net force between the plates is strongly reduced.  Oscillations of the force arise from the crystalline ordering of the particles between the walls: due to the finite radius of the particles, certain values of $d$ are commensurate with integer numbers of particle diameters, and at these spacings well-ordered crystals can form between the plates, producing a stronger effective plate-plate repulsion. \citet{ni2015tunable} also considered the dilute limit and found an exponential attractive interaction between the walls, similar to that obtained by \citet{ray2014casimir}.  The magnitude and range of the force increase with increasing self-propulsion.  \citet{ni2015tunable} also observed that the attraction arises due to a reduction of the particle density between the walls.

Similar effects have also been considered for two spherical objects immersed in an active bath \cite{das2014phase,harder2014role}. Here, the concept of swim pressure has turned out to be helpful \cite{takatori2014swim,solon2014pressure,solon2015pressure,smallenburg2015swim,yan2015force}. Understanding the wall curvature dependence of the swim pressure gives access to a general theory of depletion in an active bath, so that the anaytical Asakura--Oosawa model for a thermal non-interacting bath can be appropriately  generalized to the active case \cite{smallenburg2015swim}. For example, \citet{harder2014role}, using numerical simulations, considered the interactions between large objects immersed in a active bath, as shown in Fig.~\ref{F21}a, and found that the shape of the objects plays an important role in determining the polarity of the force: as shown in Fig.~\ref{F21}b, there is an attractive force between two interacting passive disks due to a depletion effect even when the active motor force is null (i.e. $F_{\rm a}=0$), while, when the disks are active (i.e. $F_{\rm a}\neq 0$), the force between them becomes repulsive.  In this case, the repulsion arises due to the accumulation of active particles in the corner regions between two closely spaced disks. The active particles generate a net outward force that pushes the disks apart.  When the disks are inactive, the accumulation of active particles in the corners does not occur.  When the large passive particles are rod-shaped rather than disk-shaped, there is an attractive force between the rods that increases with increasing activity.  The case of parallel rods is then similar to the Casimir geometry in the dilute limit where attractive forces arise. 

We remark that the basic notion of active depletion is that the presence of obstacles in an active bath causes guided motion of the active particles along the surface of the obstacles, leading to concentration gradients of the bath particles; by contrast, for passive depletion, the obstacles cause gradients by impeding the paths of the bath particles, once the obstacles get close to each other. \citet{ray2014casimir} provides a clean demonstration of depletion which is purely active, with no passive component; in fact, the active particle have zero size and, therefore, passive depletion must play no role, because zero-size particles cannot be excluded by any obstacle.

There are also many other types of fluctuation-induced force effects that occur in active matter systems and can depend on the shape or flexibility of both the active and passive particles. For example, \citet{parra-rojas2014casimir} argued that genuine Casimir effects\footnote{When two plates are placed in some form of fluctuating environment, confinement effects can produce an attractive force between them that is known as the Casimir effect \cite{casimir1948attraction,lamoreaux1997demonstration,munday2009measured,intravaia2013strong}.   As a function of the spacing $d$ between the plates, Casimir forces typically obey a power law $F_{\rm C} \propto d^{-\alpha}$, where $\alpha > 3$.} can occur in microswimmer suspensions since, due to the discreteness of the particles in suspension, the fluctuations in the system depend on the local particle density. 

Adding active particles to an ensemble of passive particles could lead to new routes of rapid self-assembly or tunable self-assembly in which the structures could rapidly disassemble when the activity is reduced.  Additionally, many biological systems may be exploiting active fluctuations to bring objects together, such as within or around cells \cite{machta2012critical}.

\subsubsection{Flexible passive particles and polymers}\label{sec:flexible}

The scaling behavior of polymers or chains under thermal fluctuations has a rich history \cite{degennes1979scaling}.  Methods for characterizing polymers include Flory scaling exponents in which the extension $S$ of a polymer goes as $S\propto N_{\rm m}^\nu$, where $N_{\rm m}$ is the number of units along the chain (or the molecular weight) and $\nu$ is the Flory exponent. Since polymers, chains, and elongated structures are common in biological systems, it is interesting to ask how polymers behave in the presence of an active bath rather than a thermal bath.  

Kaiser and L\"owen \cite{kaiser2014unusual} performed two-dimensional simulations of a flexible polymer chain in a bacterial bath in a relatively diluted limit and found that for very long chains the polymer extension follows a two-dimensional Flory scaling.  This occurs since the chains are considerably longer than the bacterial running lengths, so the polymers sample the bacterial bath fluctuations only at long time scales, at which the bacteria effectively undergo regular diffusion.  When the chains are short, i.e. of the order of the bacterial running length or smaller, the activity becomes important and the chains are expanded or compressed by the bacterial bath.

\begin{figure}[t]
\includegraphics[width=\columnwidth]{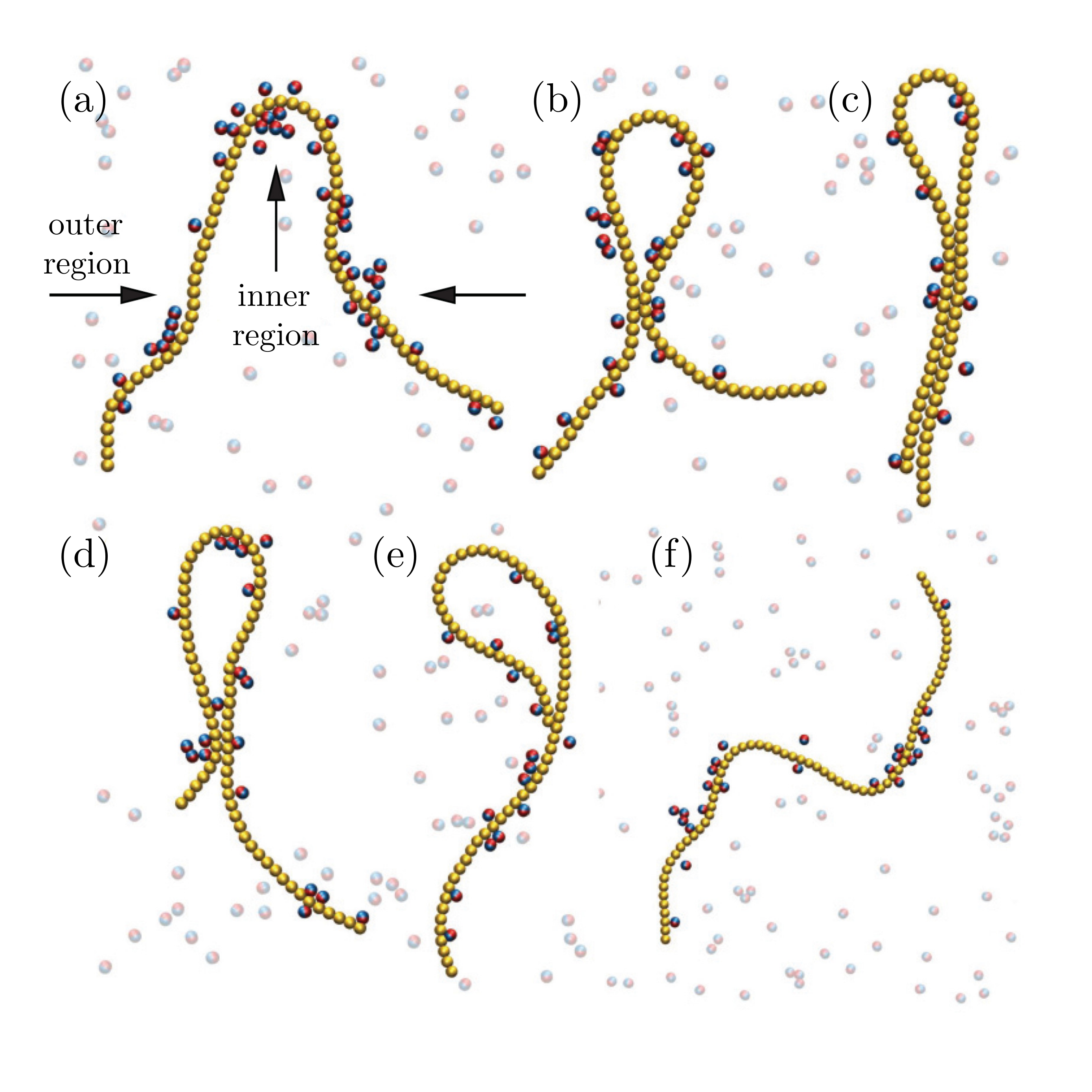}
\caption{(Color online) Flexible chain in an active bath.
(a-f) Sequence of snapshots from simulations depicting a filament at various stages of folding and unfolding. The propelling force acts along the axis connecting the poles of the two hemispheres used to depict the active particles in the blue-to-red (light-to-dark gray) direction. For the sake of clarity, active particles away from the filament have been rendered with a semitransparent filter. In (a), the convention for the definition of the inner and outer regions of a bent filament are explicitly shown.
From \citet{harder2014activity}.
\label{F22}
}
\end{figure}

\citet{harder2014activity} also performed numerical studies of a chain in an active bath and confirmed the results of \citet{degennes1979scaling} in the fully flexible limit.  They then added rigidity to the chains and observed a variety of new types of behaviors, including a crumbling transition to a metastable hairpin structure at intermediate activity levels.  This is illustrated in Fig.~\ref{F22}, which shows fluctuations between a hairpin configuration and an elongated configuration of the polymer.  Protein folding has been a long-standing problem and is generally understood to occur under thermal fluctuation conditions; however, the results of \citet{harder2014activity} indicate that folding configurations may be facilitated by the presence of active fluctuations.  This suggests that active matter dynamics could play an important role in certain biological functions such as the folding of chains or membranes \cite{suzuki2015polar}. Therefore, this is an important direction for future study in active matter systems \cite{mallory2015anomalous,shin2015facilitation,isele2015self}.

\citet{kaiser2015how} numerically and analytically studied the case of a single chain composed of active particles in the absence of any bath, either thermal or active.  They considered a free chain, a chain confined by an external trap, and a chain being dragged by one end, and observed Flory exponents of $\nu=0.5$, $0$, and $1$, respectively.  Here the activity does not change the Flory exponents but it does modify the prefactor of the scaling law.  When self-avoidance is added to the model, the equilibrium Flory exponents still appear, but the chain extension becomes non-monotonic for increasing activity.

\section{Complex Environments}\label{sec:complex}

So far we have only considered individual and collective behaviors of active particles moving in  homogeneous environments without any physical boundaries. Another avenue of research in active matter systems is the study of the interaction of active particles with physical obstacles and boundaries. On the one hand, such situations are relevant for biological microswimmers moving in many natural habitats, from soil and the guts to culture media such as agar \cite{berg2004ecoli}; for example, it is a tantalizing possibility that motile biological systems might have evolved specific rules of motion that facilitate navigation through similar complex environments. On the other hand, understanding how to control the motion of artificial active particles through their interaction with complex environments could prove beneficial also for applications, e.g. inside lab-on-a-chip devices or living organisms \cite{chin2007lab-on-a-chip}.

This section will therefore focus on the role of boundaries on the swimming properties of micro- and nanoswimmers. We will first consider how microswimmers interact with a planar wall and explain the basic particle--obstacle interaction mechanisms  as well as their modeling (Section~\ref{sec:wall}). We will then consider active particles confined within a pore, and show how their behavior differs from that of passive particles in similar confined environments (Section~\ref{sec:confinement}). We will then consider the effects that emerge when microswimmers interact with various kinds of  complex environments (Section~\ref{sec:obstacles}). Finally, we will discuss how microswimmers can be sorted exploiting micropatterned environments (Section~\ref{sec:sorting}).

\subsection{Interaction with a wall}\label{sec:wall}

There are various ways in which a microswimmer can interact with an obstacle. The two most important are arguably hydrodynamic and steric interactions, even though other (e.g. electrostatic, depletion) interactions can also occur. The basic concepts have already been introduced in Section~\ref{sec:hydro} for hydrodynamic effects, and in Section~\ref{sec:interactions} for other kinds of particle--particle interactions.

The simplest situation in which an active particle interacts with an object is arguably when it interacts with a planar (vertical) wall; some examples (where hydrodynamic interactions are important) are reproduced in Fig.~\ref{F10}. In this case, there is a basic asymmetry between approaching and leaving the wall: when approaching the boundary, the particle will remain effectively stuck at the wall until its orientation points aways from it; when leaving, it will just swim away. This asymmetry leads to a tendency for swimmers to accumulate near confining boundaries, which is not observed for particles at thermodynamic equilibrium; in fact, even if the wall is purely repulsive, there is a large accumulation which would require huge attraction strengths for passive particles at equilibrium. This effect has been observed for rods and spheres in a linear channel \cite{wensink2008aggregation,elgeti2009self,volpe2011microswimmers,wysocki2015giant} as well as for bacteria in circular cavities \cite{vladescu2014filling}. Chiral active particles, however, have been shown to behave differently, as they can slide along  a planar wall using their effective driving torque \cite{vanteeffelen2008dynamics,vanteeffelen2009clockwise}.

\begin{figure}[t]
\includegraphics[width=\columnwidth]{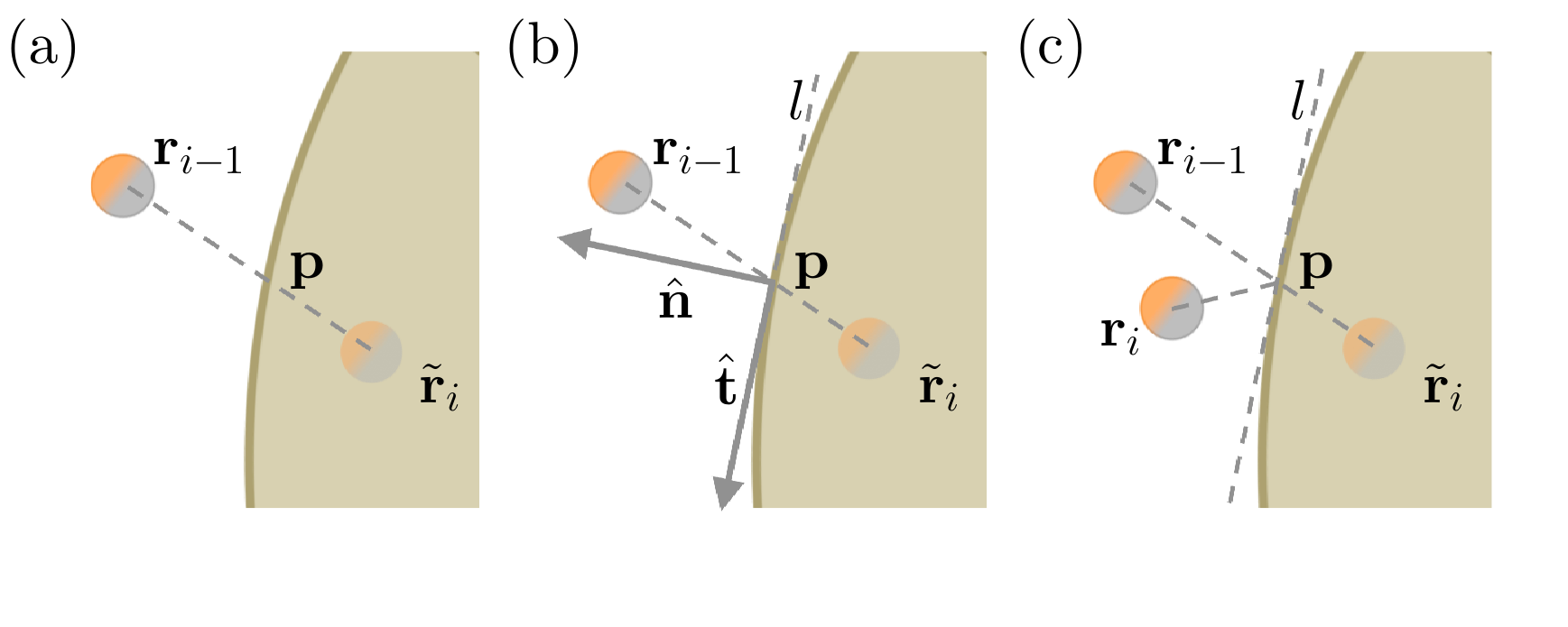}
\caption{(Color online) Numerical implementation of reflective boundary conditions. 
At each time-step, (a) the algorithm checks whether a particles has moved inside an obstacle. If this is the case, (b) the boundary of the obstacle is approximated by its tangent $l$ at the point ${\bf p}$ where the particle entered the obstacle and (c) the particle position is reflected on this line. In (b), $\hat{\bf t}$ and $\hat{\bf n}$ represent the tangential and normal unitary vectors to the surface at point ${\bf p}$. Note that the orientation of the particle is not flipped.
\label{F23}
}
\end{figure}

Both steric and hydrodynamic interactions typically lead to an alignment of the active particle such that its self-propulsion is directed along the wall of the obstacle. 
{\it Steric effects} of linear \cite{wensink2008aggregation} and circle swimmers \cite{vanteeffelen2008dynamics,vanteeffelen2009clockwise,kuemmel2013circular} with a planar wall can be qualitatively understood using the effective force description introduced in \ref{sec:phenom}. When a self-propelled particle hits an obstacle such as a planar wall, the propelling force can be decomposed into two components: one tangential and one normal to the wall. The tangential component leads to sliding along the wall while the normal component is compensated by the steric wall--particle interaction. Numerically, this process can be modeled using reflective boundaries, as shown in Fig.~\ref{F23} \cite{volpe2014simulation}, even though more accurate higher-order algorithms are also available \cite{behringer2011hard,behringer2012brownian}. Importantly, \citet{elgeti2013wall} showed that these steric effects are sufficient to produce the accumulation and alignement of particles with a wall, without the need for hydrodynamic interactions.

{\it Hydrodynamic interactions} between the active particle and the wall are also an important kind of interactions (though frequently neglected). They give rise to stable wall entrapment so that the motion of microswimmers is mainly along the hard boundary of the obstacle even if, in qualitative difference to steric interactions, the object boundary is convex.  This has been seen in experiments and appropriately described by hydrodynamics in \citet{takagi2014hydrodynamic}, \citet{sipos2015hydrodynamic}, and \citet{schaar2014detention}. 
Furthermore, a crucial difference emerges between the cases of pushers and pullers; this is discussed in detail in Section~\ref{sec:boundaries} and in Fig.~\ref{F11}.

\subsection{Active particles in a confined geometry}\label{sec:confinement}

In the previous section, we have seen that active particles tend to preferentially spend some at walls because of the asymmetry between the processes that take them towards and away from boundaries. We will now discuss the consequences of this for active particles confined within a finite space such as a pore. We will first discuss how non-Boltzmann particle distributions emerge in confined pores, which is in fact a signature of the difference between passive and active particles (Section~\ref{sec:nonboltzmann}). We will then introduce the very important (and largely open to future investigation) topic of the derivation of an equation of state for active matter systems (Section~\ref{sec:eos}). Finally, we will deal with the emergence of collective behaviors in confined geometries (Section~\ref{sec:confined}).

\subsubsection{Non-Boltzmann position distributions for active particles}\label{sec:nonboltzmann}

\begin{figure}[t]
\includegraphics[width=\columnwidth]{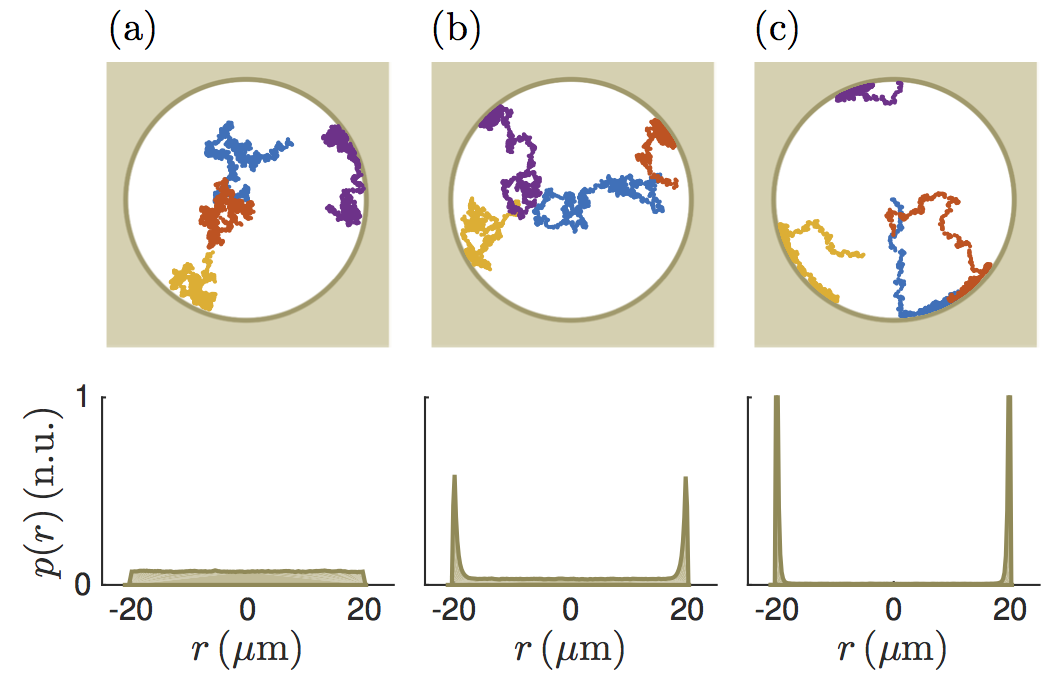}
\caption{(Color online) Non-Boltzmann position distributions for active particles in a pore. (a-c) Simulated trajectories ($10\,{\rm s}$, solid lines) of active Brownian particles (radius $R = 1\,{\rm \mu m}$) moving within a circular pore (radius $20\,{\rm \mu m}$) with reflective boundaries at velocity (a) $v=0\,{\rm \mu m\,s^{-1}}$, (b) $v=5\, {\rm \mu m\,s^{-1}}$, and (c) $v=10\,{\rm \mu m\,s^{-1}}$. The histograms on the bottom show the probability distribution along a diameter of the circular pore: while the probability is uniform across the whole pore in the case of passive Brownian particles, the probability increases towards the walls in the case of active Brownian particles together with the particle velocity and the associated persistence length $L$ (Eq.~(\ref{eq:L})).
\label{F24}
}
\end{figure}

We consider a microswimmer confined within a circular pore, as shown in Fig.~\ref{F24}. Figure~\ref{F24}a shows four 10-s trajectories of passive Brownian particles ($v=0\,{\rm \mu m\,s^{-1}}$): they explore the configuration space within the pore uniformly. Instead, active particles, shown in Figs.~\ref{F24}b ($v=5\,{\rm \mu m\,s^{-1}}$) and \ref{F24}c ($v=10\,{\rm \mu m\,s^{-1}}$), tend to spend more time at the pore boundaries. When a microswimmer encounters a boundary, it keeps on pushing against it and diffusing along the cavity perimeter until the rotational diffusion orients the propulsion of the particle towards the interior of the pore. The chances that the active particle encounters the pore boundary in one of its straight runs increases as its velocity (and thus its persistence length $L$ given by Eq.~(\ref{eq:L})) increases. These observations can be made more quantitative by using the particle probability distribution. The histograms on the bottom of Figs.~\ref{F24}a-c show a section along a diameter of the pore of the probability distribution of finding the particle at the boundaries: it increases together with the particle velocity. This accumulation has been seen in experiments, e.g., by \citet{bricard2015emergent}.

The fact that active particles tend to accumulate at the boundaries of a pore is a consequence of their out-of-equilibrium nature. For a passive Brownian particle at thermodynamic equilibrium with its environment, the probability distribution $p(x,y)$ is connected to the external potential $U(x,y)$ by the Boltzmann relation $p(x,y) \propto \exp\left\{ -\frac{U(x,y)}{k_{\rm B}T} \right\}$. In the case presented in Fig.~\ref{F24}, there are no external forces acting on the particle and, therefore, $U(x,y)$ and the corresponding Boltzmann distribution are homogeneous, as shown in Fig.~\ref{F24}a. However, the fact that the distributions in Figs.~\ref{F24}b and \ref{F24}c are not homogeneous, despite the homogeneous potential, is a clear deviation from the Boltzmann distribution, i.e. from the behavior of matter at thermodynamic equilibrium.

While the Boltzmann distribution holds for every equilibrium system, a generalized form for the stationary probability distribution characterizing all types of active particles does not exist. However, for some specific classes of active particles analytic solutions can be found, and the traditional Boltzmann equation can also be generalized to some active systems \cite{schnitzer1993theory,tailleur2008statistical,tailleur2009sedimentation,thuroff2014numerical,thuroff2013critical}. The first case is that of non-interacting run-and-tumble particles moving over a one-dimensional, conservative force field $f(x)=-\frac{dU(x)}{dx}$. It can be shown that, in a close system with no fluxes, run-and-tumble particles will distribute in space with probability \cite{schnitzer1993theory,tailleur2008statistical}
\begin{equation}
p(x)=p(x_0)\frac{D(x_0)}{D(x)}\exp\left[\int_{x_0}^{x}\frac{\mu\;f(s)}{D(s)}\;ds \right] \; ,
\end{equation}
where $\mu = \gamma^{-1}$ is the particle's mobility and 
\begin{equation}
D(x)=v^2\tau-\mu^2 f(x)^2 \tau
\end{equation}
is an effective space-dependent diffusion coefficient that is smaller when the magnitude of the external force $f(x)$ is larger. Taking the limit of a vanishing tumble rate $\tau$ while keeping the free-space diffusion $D_0=v^2\tau$ finite, a Boltzmann-type distribution is recovered with the effective temperature $T_{\rm eff}=D_0/(\mu k_{\rm B})$. In the simple case of a harmonic potential well with stiffness $k$, i.e. $U(x)=k x^2/2$, run-and-tumble particles will be confined in a region $|x|<v/(\mu k)$ with a probability density \cite{tailleur2009sedimentation}
\begin{equation}
p(x)=p(0)\left[1-\left(\frac{\mu k x}{v}\right)^2\right]^{{\tau \over 2\mu k} - 1} \; ,
\end{equation}
which can be very different from the equilibrium Gaussian shape and, in particular, becomes bimodal for $\tau>1/(2\mu k)$ with particles accumulating at the edges of the allowed region. The above results are exact but only valid for one-dimensional and non-interacting particle systems. For the more generic case of interacting particles in any dimension $d$, a good approximation can be found in the case of colored Gaussian noise. Calling $U(\mathbf x)$  the potential energy function of an ensemble of $N$ active particles described by the $N d$ coordinate vector $\mathbf{x}$, a flux-free stationary probability distribution can be obtained within the unified colored noise approach \cite{maggi2015multidimensional,marconi2015towards} and it reads
\begin{equation}
p(\mathbf x)=\frac{1}{Q}\exp\left[-\frac{U(\mathbf x)}{D_0}-\frac{\tau|\nabla U(\mathbf x)|^2}{2 D_0}\right] ||\mathbb I+\tau\nabla\nabla U(\mathbf x)||
\end{equation}
with $Q$ a normalization factor, $\mathbb I$ the $N d$-dimensional identity matrix, and $||\cdot||$ representing the absolute value of the determinant. Interestingly,  within the same approximation, also an analytical expression for the velocity distribution can be found establishing an explicit link between the mean square velocity and the system's configuration \cite{marconi2015velocity}. Again, in the limit of vanishing $\tau$, the distribution reduces to the Boltzmann form for an effective temperature $T_{\rm eff}=D_0/(\mu k_{\rm B})$. Contrary to run-and-tumble particles, the one-dimensional case of an external harmonic potential  always presents a Gaussian shape although with an effective temperature $T_{\rm eff}=D_0/(\mu k_B) (1+2\mu k\tau)^{-1}$ that depends on the potential curvature. The coexistence of multiple effective temperatures associated with different curvatures in a two-dimensional external potential landscape has been observed in \citet{maggi2014generalized}.

Recent studies have generalized the results about the dynamics, diffusivity, and density distributions of active particles in circular confined geometries to geometries of general shape \cite{sandoval2014effective,fily2015dynamics}. \citet{ezhilan2015distribution} examined the swim pressure in the regions between two planar walls for run-and-tumble swimmers and found that for large widths the pressure obeys the ideal gas law; at lower widths, however, the pressure deviates from the ideal gas law and decreases at the center since the particles  spend more time confined near the walls rather than moving in the bulk.

\subsubsection{Active matter forces and equation of state}\label{sec:eos}

When considering active-depletion forces in Section~\ref{sec:casimir}, it was shown that active particles can generate (attractive or repulsive) forces between passive plates. This raises the questions of whether such forces depend on the microscopic nature of the interactions of the active particles with the plates and, more generally, of whether an equation of state exists for active particles.

At thermodynamic equilibrium, an ideal gas in a container of volume $V$ can be described by the standard ideal gas  equation of state: $P=\rho K_{\rm B} T$, which relates the pressure $P$ to the temperature $T$ and the number density of gas molecules $\rho$. As the temperature increases, the pressure experienced by the container walls increases, independent of the nature or shape of the walls.  To address the question of what happens to such a relation when the thermal gas particles are replaced by self-propelled active particles, \citet{mallory2014anomalous} performed numerical simulations to extract an equation of state for self-propelled repulsively interacting disks in two- and three-dimensional systems, and found a nonmonotonic dependence of the pressure on the temperature. By generalizing to active dynamics the definition of  pressure as the trace of the microscopic stress tensor, \citet{yang2014aggregation} and \citet{takatori2014swim} found that mechanical pressure in active systems can be decomposed as the sum of the usual stress component arising from interparticle interactions and a swim pressure component analogous to kinetic pressure in equilibrium systems. In contrast to equilibrium kinetic pressure, however, where particles speed is solely controlled by temperature, swim pressure depends on the actual average speed of the particles that is indirectly controlled  by density and persistence length \cite{solon2015pressure}. In the special case where  particles are not subject to any external torque, either from other particle or from container walls, an equation of state can be derived \cite{solon2015pressure} connecting pressure to bulk properties that is not sensitive to the details of the particles' interaction with the wall. It can also be shown that such an equation of state does not exist in the general case where particle are not torque-free  \cite{solon2015pressure}.

Experimentally, equations of state in colloidal systems can be extracted from sedimentation profiles. \citet{palacci2010sedimentation} performed experiments with dense sedimenting suspensions of ${\rm H_2O_2}$-activated Janus particles, which show a strong change in the sedimentation profile compared to passive systems: after ${\rm H_2O_2}$ is added to the solution, the solid phase remains but the gas phase spreads to much higher heights. \citet{ginot2015nonequilibrium} performed sedimentation experiments and simulations for active Janus particles and found that activity strongly alters the standard thermodynamic equation of state; however, it was possible to model these alterations using ideas developed through equilibrium concepts.  In the dilute case, there is an activity-dependent effective temperature, while, at higher densities, an increase in the activity can be modeled as an effective increased adhesion between equilibrium particles.

\subsubsection{Collective behaviors in confined geometries}\label{sec:confined}

For microscopic matter that is not at thermodynamic equilibrium, collective behaviors can also emerge in confined geometries due to complex interplay with the surrounding fluid. This can lead to the formation of complex patterns, such as vortices \cite{hernandez2005transport,yang2014aggregation}. These behaviors have been experimentally observed for both swimming bacteria \cite{lushi2014fluid,wioland2014confinement} and colloidal rollers \cite{bricard2015emergent}. Similar behaviors have also been reported for vibrated polar granular disks and rods \cite{deseigne2012vibrated,kudrolli2008swarming}.

\subsection{Interaction with obstacles}\label{sec:obstacles}

The interaction of active particles with obstacles within complex environments lends itself to be exploited in a wealth of potential applications. In general, these may take advantage of the fact that the motion of active particles is non-thermal in order to use the features of the environment to perform complex tasks such as the separation, trapping, or sorting of active particles on the basis of their swimming properties. In this section we will explore these possibilities; we will, in particular, consider how active particles can be gathered by wedges (Section~\ref{sec:concentration}),  directed by ratcheted walls (Section~\ref{sec:ratchet}), funnelled along patterned channels (Section~\ref{sec:channel}), sorted in periodic arrays of obstacles (Section~\ref{sec:arrays}), and trapped within random environments (Section~\ref{sec:subdiffusion}).

\subsubsection{Capture and concentration of active particles}\label{sec:concentration}

\begin{figure*}
\includegraphics[width=\textwidth]{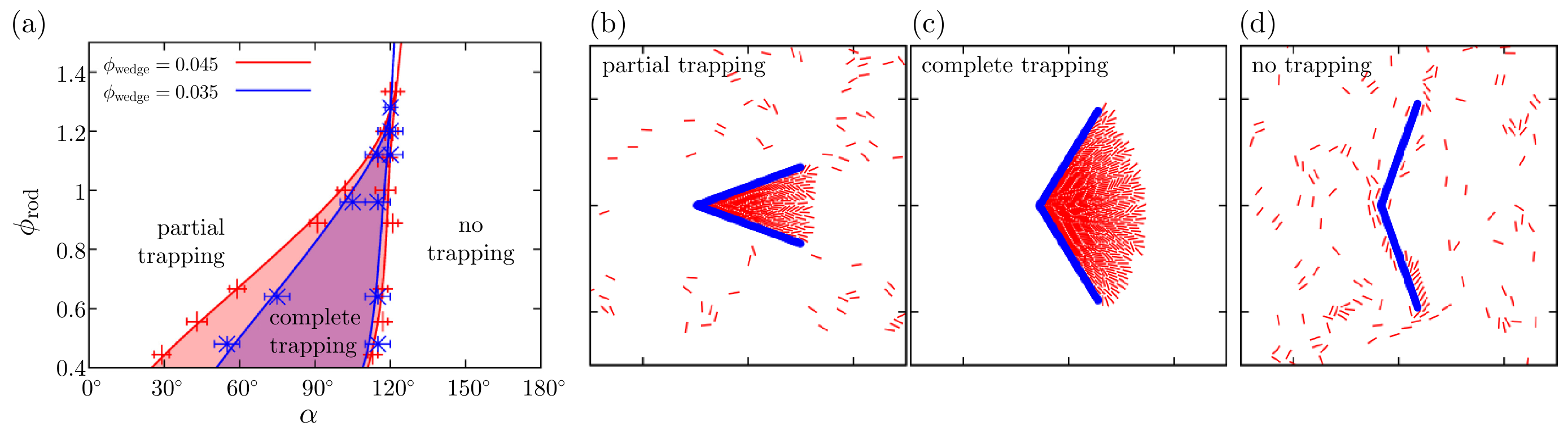}
\caption{(Color online) Capturing active particles with a wedge.
(a) Phase diagram marking three different collective trapping states of self-propelled rods at a wedge upon variation of the reduced rod packing fraction ($\phi_{\rm rod}$): no trapping at large apex angle $\alpha$, complete trapping at medium $\alpha$, and partial trapping at small $\alpha$. Phase boundaries are shown for two different values of the area fraction occupied by the wedge ($\phi_{\rm wedge}$). The region of complete trapping is bounded by a triple point at larger rod concentration beyond which a smooth transition from no trapping to partial trapping occurs. (b-d) Snapshots depicting the examples of the three stationary states.
From \citet{kaiser2012capture}.
\label{F25}
}
\end{figure*}

\begin{figure*}
\includegraphics[width=\textwidth]{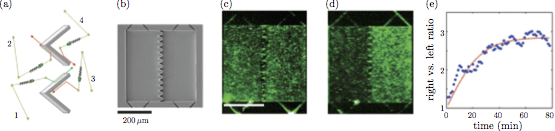}
\caption{(Color online) Concentration of active particles.
(a) Schematic drawing of the interaction of bacteria with a funnel opening: on the left side, bacteria may (trace 1) or may not (trace 2) get through the gap, depending on the angle of attack; on the right, almost all bacteria colliding with the wall are diverted away from the gap (traces 3 and 4). 
(b) Scanning electron micrograph of the device.
(c) Uniform distribution of bacteria in a structure with a funnel wall after injection and
(d) steady-state distribution after $80\,{\rm minutes}$. 
(e) Ratio of densities in the right and left compartments versus time: the circles are experimental data, and the dashed line is a fit to the relative theory.
From \citet{galajda2007wall}.
\label{F26}
}
\end{figure*}

In many applications, it is important to catch groups of autonomously navigating microbes and man-made microswimmers in a controlled way. This can be achieved, for example, by using some obstacles in the shape of wedges. The condition for trapping is a pretty sharp cusp on the length scale of the swimmer extension. For example, \citet{kaiser2012capture} considered active self-propelled rods interacting with stationary wedges as a function of the wedge angle $\alpha$, finding three regimes of behavior, whose phase diagram is shown in Fig.~\ref{F25}a:  for small angles, a partial trapping of the active particles occurs (Fig.~\ref{F25}b for $\alpha=40^\circ$); for intermediate angles, complete trapping of the particles occurs (Fig.~\ref{F25}c for $\alpha=116^\circ$); and, for large angles, there is no longer any trapping of the particles (Fig.~\ref{F25}d for $\alpha=140^\circ$). We remark that the trapping by wedges is distinct from carrier motion discussed in Fig.~\ref{F18} because it is triggered and self-amplified by the strong aligning excluded-volume forces between the rods, while carrier motion already occurs for an ideal gas of swimmers. In a subsequent work, \citet{kaiser2013capturing} found enhanced trapping when the wedge is moved in certain orientations (see Section~\ref{sec:gears}). This trapping behavior can be enhanced employing a system of multi-layered asymmetric barriers \cite{chen2015enhancing}. The self-trapping behavior of active rods was then confirmed in experiments on artificial rod-like swimmers \cite{restrepo-perez2014trapping} and sperm cells \cite{guidobaldi2014geometrical}.

\citet{galajda2007wall} placed swimming bacteria in a confined area containing an array of funnel shapes (Figs.~\ref{F26}a and \ref{F26}b), and observed that the bacteria concentrated in the portion of the container toward which the funnel apertures pointed, indicating that the bacteria were undergoing a ratchet motion (Figs.~\ref{F26}c-e).  \citet{wan2008rectification} performed simulations of a simplified model of the system in which the bacteria are treated as run-and-tumble particles that move along any wall they encounter.  For small run lengths, the system behaves thermally and no ratchet effect occurs; however, for longer run lengths a ratchet effect emerges and, just as in the experimental case, there is a build-up of simulated bacteria on one side of the container.  The simulations also showed a build-up of active particle density in the tips of the funnel barriers.  \citet{tailleur2009sedimentation} studied run-and-tumble particles interacting with walls and observed a build-up of particle density along the walls as well as a ratchet effect in the presence of funnel-shaped walls.  They found that changing the interaction of the particles with the wall, e.g. by introducing a reflection of the particles from the wall, can destroy the ratchet effect.  \citet{galajda2008funnel} constructed a macroscale version of the ratchet system and observed the same behavior as in the bacterial version, indicating that detailed hydrodynamic interactions with the walls are not solely responsible for the ratchet effect.

\subsubsection{Ratchet effects and directed motion}\label{sec:ratchet}

\begin{figure}
\includegraphics[width=\columnwidth]{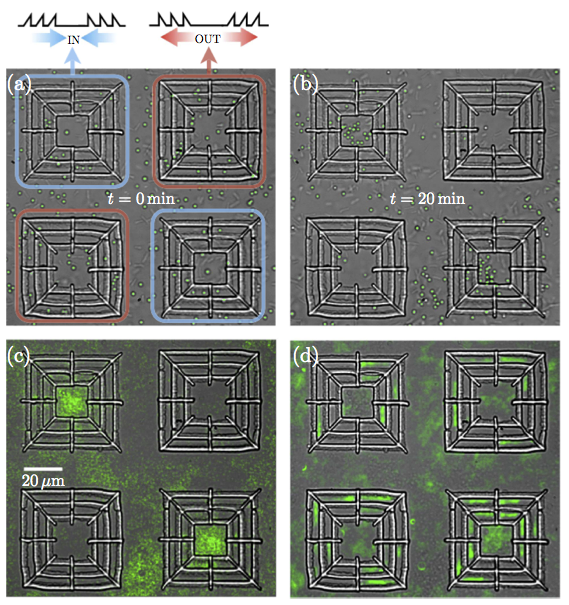}
\caption{(Color online) Observation of particle concentration and depletion by bacteria. (a-b) Snapshots of particles and bacteria (a) at $t=0$ (the initial state), where particles are randomly distributed, and (b) at $t=20\,{\rm min}$, where particle distributions have been strongly affected by bacterial transport over asymmetric barriers. The colloidal particles that are not stuck on the surface are highlighted. (c-d) Particle distributions averaged over a steady state (c) for particles in the bacterial bath between $t_1=15\,{\rm min}$ and $t_2=20\,{\rm min}$ ($\Delta t = 5\,{\rm min}$) and (d) for particles in an experiment without bacteria, undergoing simple Brownian motion for $\Delta t = 10 \,{\rm min}$. In the absence of bacteria, the colloidal particles remain trapped within the structures' compartments.
From \citet{koumakis2013targeted}.
\label{F27}
}
\end{figure}

Thermal Brownian particles moving over an asymmetric substrate do not spontaneously drift in one direction (if the spatial average of the force is zero). If, however, an external periodic drive is added, or if the substrate is flashed on and off to create nonequilibrium conditions, a net continuous drift of particles known as a \emph{ratchet effect} can arise \cite{reimann2002brownian}. 

Numerous studies of ratchet effects have been performed for a variety of active matter systems in the presence of asymmetric barriers, including swimming eukaryotes \cite{kantsler2013ciliary}, swimming sperm cells \cite{guidobaldi2014geometrical}, active particles moving in corrugated channels \cite{pototsky2013rectification,ai2013rectification,yariv2014ratcheting,koumakis2014directed}, active polymers \cite{wan2013directed}, active ellipsoids \cite{ai2014transport}, crawling cells \cite{mahmud2009directing}, and systems containing a variety of asymmetric objects with varied swimming strategies \cite{berdakin2013influence,ghosh2013self,potiguar2014self,ai2014entropic}.   

Variations of the active ratchet effect include active drift ratchets, in which an external continuous drive is applied to the active particles as they pass through an array of asymmetric obstacles \cite{volpe2011microswimmers,reichhardt2013active}.  There are also chiral ratchet effects that arise for active particles that have an intrinsic swimming asymmetry oriented in either the clockwise or counterclockwise direction \cite{reichhardt2013dynamics,mijalkov2013sorting}.  When such particles are placed in an asymmetric substrate, particles with different chiralities move in different directions, permitting the chiral species to be separated. The net step along this research line is to (experimentally) realize substrate geometries to sort chiral active particles on the nanoscale, which could have numerous applications in biological and medical sciences, as we will see in more detail in Section~\ref{sec:sorting}.  

It is also possible to obtain a reversed ratchet effect in which the particles move in the forward ratchet direction for one set of parameters but in the reverse direction for a different set of parameters.  \citet{drocco2012bidirectional} demonstrated a reversible ratchet effect for active flocking particles moving through an array of funnel barriers.  They simulated a variant of the Vicsek flocking model in which the particles have an additional short-range steric repulsion in order to give the flocks a finite size.  In the dilute or high-noise limit, the particles are disordered and ratchet in the forward or normal direction through the funnels; however, for densities or noise levels in the collective or flocking regime, flocks become effectively jammed when they attempt to pass through the funnel due to the incompressibility of the flock, which behaves as a rigid solid.  In contrast, a flock approaching the funnel array from the hard flow direction can split in two and lose a portion of its members to the other side of the array, resulting in a net reversed ratchet motion of the particles.  Experimental studies of crawling cells also showed that it is possible for one species of cells to ratchet in one direction through an asymmetric environment while another species of cells ratchets in the opposite direction \cite{mahmud2009directing}.

\citet{koumakis2013targeted} demonstrated that a ratchet effect can also be induced on passive colloidal particles that jump over asymmetric barriers when pushed by swimming bacteria. As shown in Fig.~\ref{F27}, the escape rate over an asymmetric barrier was found to be higher when the same barrier is approached from the small slope side. By surrounding a target region with asymmetric walls, colloidal particles can accumulate over the target when the high slope sides of the barriers point inside, or clear the target region when higher slopes point outside. This effect can be interpreted in terms of a non-uniform effective temperature \cite{koumakis2014directed}. \citet{ghosh2013self} performed numerical simulations of active Janus particles in asymmetric funnel geometries containing additional passive particles and found that even a small number of active particles can induce a directed motion of the passive particles.  

The works described in this section indicate that active ratchet effects can be employed in techniques aiming at the separation, mixing, or directed flow of passive particles.  

\subsubsection{Motion rectification in a microchannel}\label{sec:channel}

\begin{figure}[t!]
\includegraphics[width=\columnwidth]{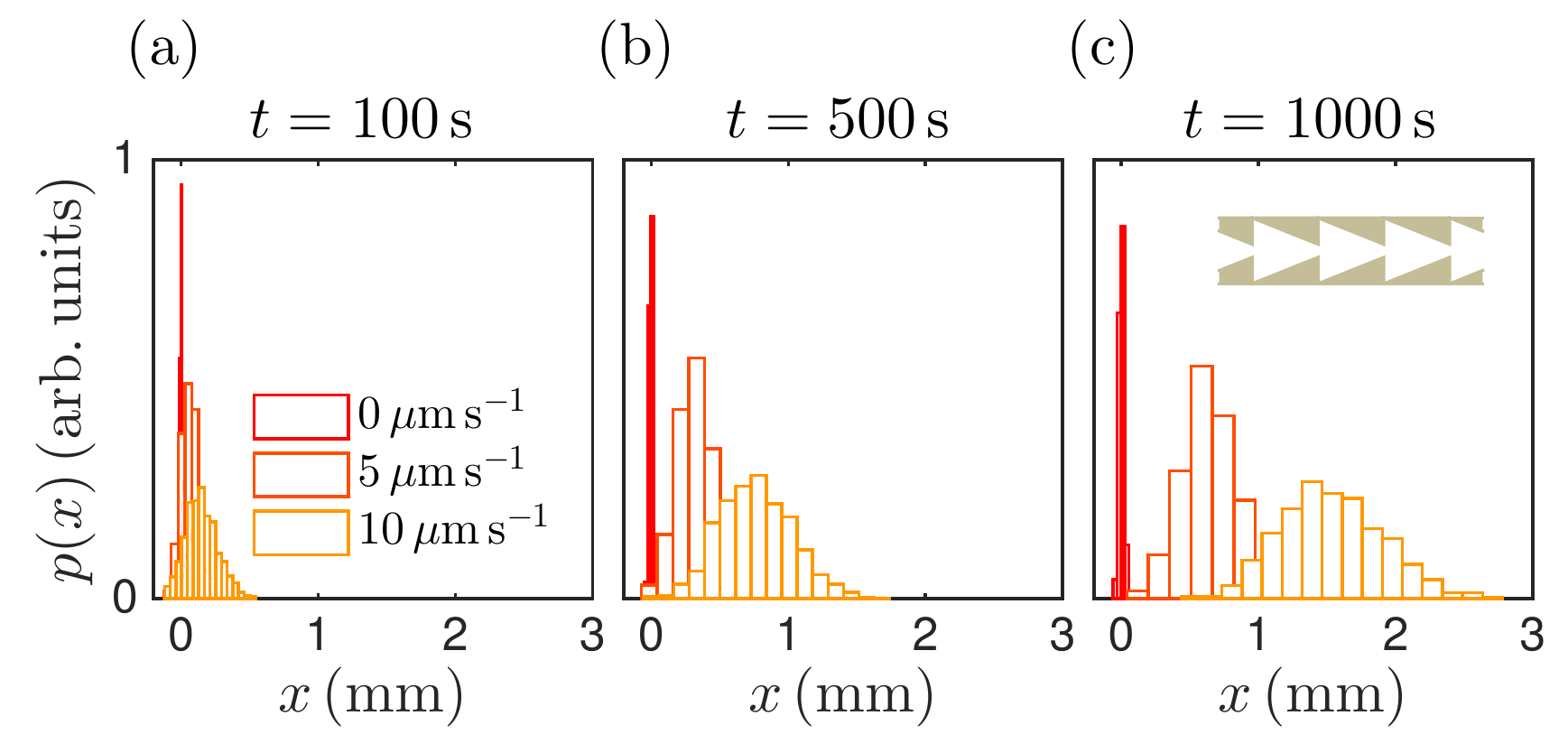}
\caption{(Color online) Rectification of active Brownian motion in an asymmetric ratchet-like microchannel. The distribution of passive (red/dark gray histograms) and active (yellow/light gray histograms) Brownian particles (radius $R=1\,{\rm \mu m}$) released at time $t=0\,{\rm s}$ from position $x=0\,{\rm mm}$ are plotted at times (a) $t=100\,{\rm s}$, (b) $t=500\,{\rm s}$, and (c) $t=1000\,{\rm s}$. The higher the active particle velocity, the farther the active particles travel along the channel. Every histogram is calculated using 1000 particle trajectories. A segment of the channel, whose dent is $10\, {\rm \mu m}$ long, is represented by the gray structure in the inset of (c).
\label{F28}
}
\end{figure}

The motion of active particles can be rectified by a patterned microchannel (a recent topical review is \citet{ao2014active}). The inset of Fig.~\ref{F28}c shows an example of such a microchannel decorated with a series of asymmetric dents on both its walls. Differently from a group of passive Brownian particles released at time $t=0\,{\rm s}$ from position $x = 0\,{\rm \mu m}$ which diffuse symmetrically around the initial position (filled histograms in Figs.~\ref{F28}a-c), a group of active Brownian particles is funneled by the channel in such a way that an average directed motion is imposed on the particles, as can be seen in the open histograms in Figs.~\ref{F28}b-c. The rectification is the more pronounced the higher the velocity of the active particle. This and similar effects have been proposed to sort microswimmers on the basis of their velocity or size \cite{hulme2008using,mijalkov2013sorting}, to trap microswimmers in moving edges \cite{kaiser2013capturing}, to deliver microscopic cargoes to a given location \cite{koumakis2013targeted}, to convey a strategic advantage in trespassing channels against external biases \cite{locatelli2015active}, and even to control pedestrian flows \cite{oliveira2016keep}.

\subsubsection{Extended landscapes of obstacles}\label{sec:arrays}

\begin{figure}
\includegraphics[width=\columnwidth]{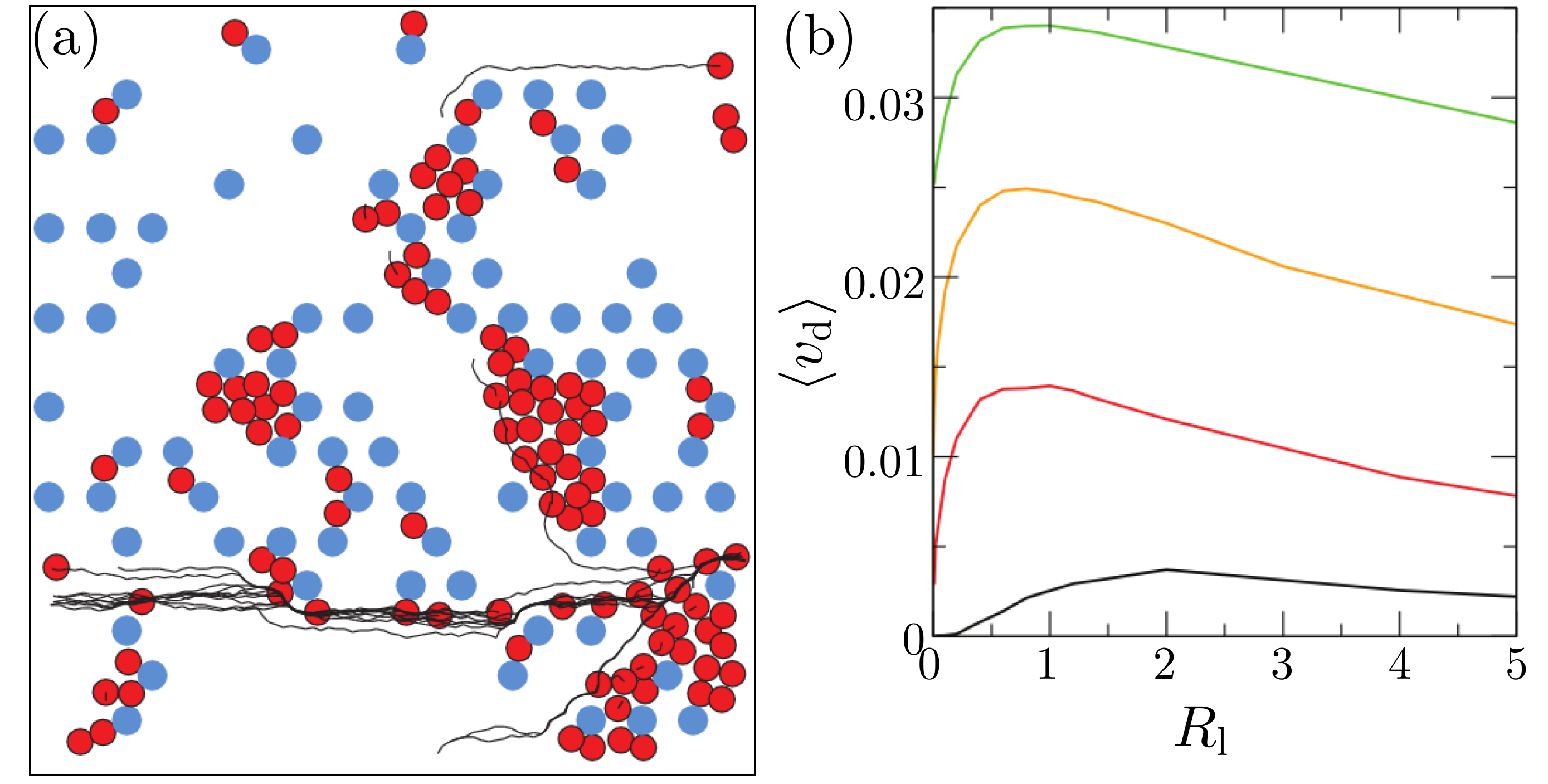}
\caption{(Color online) Effect of a disordered landscape on the driving velocity of active particles.
(a) Positions of pinned disks (blue/light gray circles), active disks (red/dark gray circles), and disk trajectories over a period of time (lines) in a small section of a sample with obstacle area fraction $\phi_{\rm p} = 0.0235$, total area fractiton $\phi = 0.188$, and run length $R_{\rm l} = 0.004$.
(b) The average drift velocity $\langle v_{\rm d}\rangle$ as a function of $R_{\rm l}$ passes through a maximum for all the curves; $\phi_{\rm p} = 0.055$, $0.094$, $0.1413$, and $0.188$, from top to bottom.
From \citet{reichhardt2014active}.
\label{F29}
}
\end{figure}

\begin{figure*}
\includegraphics[width=\textwidth]{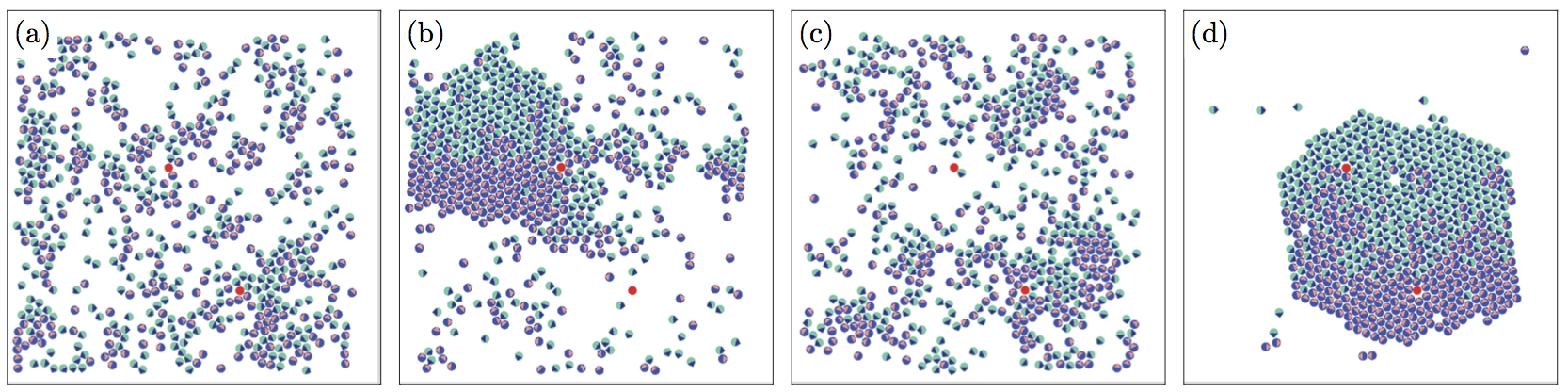}
\caption{(Color online) Cluster formation induced by obstacles. Snapshots of the positions of active disk and obstacles in a system with area fraction $\phi=0.363$ and number of obstacles $N_{\rm p} = 2$. The arrows indicate the direction of the motor force for each disk. Dark particles with light arrows have a net motion in the positive $y$ (vertical) direction; light particles with dark arrows have a net motion in the negative $y$ direction. Red disks are immobile obstacles. (a) The initial fluctuating state contains small transient clusters. (b) After some time, transient clusters are nucleated by the obstacles. (c) At a still later time, the transient cluster shown in (b) has broken apart. (d) Finally, the dynamically frozen steady state appears where a faceted crystal forms around the obstacles.
From \citet{reichhardt2014absorbing}.
\label{F30}
}
\end{figure*}

The motion and properties of active particles can also be influenced by the presence of extended potential landscapes or by the presence of obstacles in the environment.

\citet{reichhardt2014active} performed a numerical study of interacting run-and-tumble disks of radius $R$ moving under a drift force $F_{\rm d}$ through an assembly of immobile disks that serve as a disordered landscape of obstacles. This is similar to a system of ordinary charge-stabilized colloidal particles driven with an electric field over an obstacle array, where the transport can be characterized by the average velocity $\langle v_{\rm d} \rangle$ of the particles in the direction of the applied drift force.  For the passive particles, when no obstacles are present, $\langle v_{\rm d} \rangle$ increases linearly with the magnitude $F_{\rm d}$ of the drift force, i.e. $\langle v_{\rm d} \rangle \propto F_{\rm d}$, indicating Ohmic behavior where the damping arises from the Stokes flow.  When obstacles (or pinning) are present, there can be a finite critical external driving (or depinning threshold) force $F_{\rm c}$ required to set the particles in motion, so that $\left< v_{\rm d}\right>=0$ when $F_{\rm d} < F_{\rm c}$.  Once the particles are moving, $\left< v_{\rm d} \right>$ may increase linearly with $F_{\rm d}$ or it may follow a power law $\left<v_{\rm d}\right> \propto F_{\rm d}^\alpha$.  When thermal fluctuations are also present, the true critical depinning threshold is lost due to particle creep, and the velocity follows the form $\left<V_{\rm d}\right> \propto v \exp(-U_{\rm p}/(k_{\rm B} T))$, where $v$ is the velocity of the particle in the obstacle-free limit and $U_{\rm p}$ is the effective trapping potential created by the disorder.  Here the velocity decreases with increasing $U_{\rm p}$ or decreasing $T$, so that increasing the magnitude of the thermal fluctuations always produces a larger drift velocity.  Similarly, increasing $T$ always leads to a larger diffusion of particles in the presence of random disorder.  If the fluctuations are active rather than thermal, the velocity can behave very differently. Figure~\ref{F29}a shows an image of a system where active particles are moving through an assembly of obstacles. Some plots of $\left<v_{\rm d}\right>$ versus run length $R_{\rm l}$ for the same system are shown in Fig.~\ref{F29}b.\footnote{Because these plots show a crossing due to the non-monotonicity of the drift velocity for varied run lengths, they are more informative than plotting the drift velocity versus force since those curves are generally linear.} At $R_{\rm l}=0$, most of the particles become trapped and $\left<v_{\rm d}\right>$ is low.  As $R_{\rm l}$ increases, the fluctuations assume a thermal character and local clogs in the assembly break apart, allowing particles to flow in the drift direction and increasing $\left<v_{\rm d}\right>$.  When $R_{\rm l}$ is further increased, however, the drift velocity begins to decrease, as shown in Fig.~\ref{F29}b by the fact that $\left<v_{\rm d}\right>$ reaches its maximum value near $R_{\rm l}=1$.  The drop in velocity is correlated with an onset of clustering (or phase-separation effect) that occurs in active disk systems for long run lengths or high activity.  In the regime of low but finite activity, the particles form a liquid state, so an individual particle is at most temporarily trapped behind an obstacle before diffusing around it.  When the activity level increases and the system enters the phase-separation regime, the formation of clusters of particles makes it possible for an individual obstacle to trap an entire cluster. Further work can be envisaged on this topic; in particular, it would be interesting to determine how the depinning transition is rounded by the particle's activity as well as to study the system's behavior at finite but large run times.

It is possible to create a random energy landscape for active particles using speckle patterns which can be generated, e.g., by an optical field \cite{volpe2014brownian,volpe2014speckle,pesce2015step}.\citet{paoluzzi2014run} performed molecular dynamics simulations of swimming bacteria interacting with a speckle pattern; the swimmers are modeled as dumbbell particles and interact through short-range repulsive interactions;  a crossover is observed from non-trapping to trapping of the particles as the speckle intensity is increased. \citet{pince2016disorder} demonstrated that the presence of spatial disorder can alter the long-term dynamics in a colloidal active matter system, making it switch between gathering and dispersal of individuals: at equilibrium, colloidal particles always gather at the bottom of any attractive potential; however, under non-equilibrium driving forces in a bacterial bath, the colloids disperse if disorder is added to the potential; the depth of the local roughness in the environment regulates the transition between gathering and dispersal of individuals in the active matter system, thus inspiring novel routes for controlling emerging behaviors far from equilibrium.

\citet{reichhardt2014absorbing} considered an active matter system composed of run-and-tumble repulsive disks in the limit of an infinite run time.  Using a periodic system, they found that for a sufficiently large density of active particles, most of the particles become trapped in a single large clump and all dynamical fluctuations in the system vanish.  The large clump can still be translating, but all trajectories of all particles are periodic.  The transient time required to reach such a state diverges as a power law at the critical particle density, suggesting that this is an example of a nonequilibrium phase transition from a strongly fluctuating state to a dynamically frozen state.  Such transitions are known as absorbing phase transitions, and they have been studied experimentally and numerically for actin motion, where the system can transition from a fluctuating state to an ordered spiral state containing almost no fluctuations \cite{schaller2011frozen,reichhardt2011dynamical}.  In \citet{reichhardt2014absorbing}, obstacles are added to the system of infinitely running particles as illustrated in Fig.~\ref{F30}.  These obstacles act as nucleation sites for clusters, as shown in Fig.~\ref{F30}a.  Inside a cluster, particles are swimming against each other and are held in the cluster configuration by steric repulsion.  In Fig.~\ref{F30}d the particles are colored according to the average $y$ component of the active force, making it clear that the upper half of the cluster is moving downward in opposition to the lower half of the cluster which is moving upward, resulting in a completely pinned cluster.  When obstacles are present, clusters can form even for very low active particle densities.  Since the active disks are monodisperse in size, triangular ordering appears inside the cluster, which can lead to faceted crystals, as shown in Fig.~\ref{F30}d.  When the obstacle density is higher, the system reaches the frozen state much more rapidly and the cluster becomes increasingly disordered, as shown in Figs.~\ref{F30}b and \ref{F30}c.  In the absence of obstacles, the dynamically frozen state forms for active disk densities $\phi >0.5$; however, as the obstacle density $\phi_{\rm p}$ increases, the density at which the dynamically frozen state appears decreases. 
\citet{kumar2011symmetry} considered the large-deviation function of a single polar active particle in a crowded structured environment and \citet{kumar2014flocking} considered the flocking of active polar objects in a crowded non-motile background, where the crowdedness is crucial to producing the ordering.

\begin{figure*}
\includegraphics[width=\textwidth]{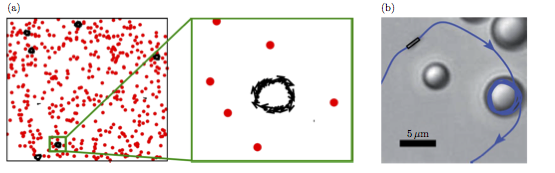}
\caption{(Color online) Trapping of active particles. 
(a) In a system where the particle orientation is affected by the presence of obstacles, spontaneous particle trapping occurs for large values of the turning speed and of the obstacle density. Obstacles are indicated by gray/red dots while black arrows correspond to active particles.
From \citet{chepizhko2013diffusion}.
(b) Sample trajectory of a self-propelled rod orbiting around a passive sphere. The sphere has diameter $6\,{\rm \mu m}$, the rod length is $2\,{\rm \mu m}$, and its speed is on the order of $20\,{\rm \mu m\,s^{-1}}$.
From \citet{takagi2014hydrodynamic}.
\label{F31}
}
\end{figure*}

In addition to swimming particles, there have been studies of flocking particles interacting with obstacle arrays.  \citet{chepizhko2013optimal} considered a variant of the Vicsek model for self-propelled particles interacting with a heterogeneous environment of obstacles. When a particle interacts with an obstacle, it turns its velocity vector away from the obstacle once it comes within a certain distance from it. There is a stochastic noise term added to the particle alignment direction which affects not only its flocking behavior but also its interaction with the obstacles.  In the absence of obstacles, the system reduces to the Vicsek model, in which increasing the magnitude of the noise term results in a transition from a coherent state in which all particles move in the same direction to a disordered state in which the direction of motion is randomized. When obstacles are present, \citet{chepizhko2013optimal} find that the system is disordered when the noise term is small; however, as the noise term increases in magnitude, the system transitions into a state with quasi-long-range order, while for the highest values of noise the system is disordered again.  This result indicates that the addition of some noise can induce ordering or coherent motion in an active matter system, and can be regarded as an example of producing order through disorder.  This effect arises since the noise tends to wash out the effect of the quenched disorder array and to make the particle density more homogeneous; however, when the noise is large enough, the system becomes disordered again.

Another system that has been explored is chiral active particles in random landscapes. \citet{nourhani2015guiding} considered chiral self-propelling particles moving on a periodic potential where they found that particles of different chirality can be separated; they also observed a variety of distinct sub-classes of dynamic states in which the particles form orbits that either  have a translational propagation or are  localized. Another feature of this system is that it is possible to steer particles to arbitrary locations by changing the strength of the substrate. \citet{schirmacher2015anomalous} considered a model of noninteracting particles undergoing circular motion in a random landscape to model electrons in a magnetic field moving through random disorder; however, the same model can be used to represent circularly swimming particles; when the swimming radius decreases, there is a transition from a delocalized to a localized state.

\subsubsection{Subdiffusion and trapping of microswimmers}\label{sec:subdiffusion}

Various emergent phenomena have been observed when active particles interact with a disordered environment such as subdiffusion and trapping.

\citet{chepizhko2013diffusion} considered the diffusion and trapping of active particles in the presence of obstacle arrays using the same flocking model described in \citet{chepizhko2013optimal} (see previous section). As shown in Fig.~\ref{F31}a, they studied the system as a function of the turning rate of the particles. They found that the motion is diffusive when the turning rate is small; however, the diffusion constant depends non-monotonically on the particle density.  For high obstacle densities and large turning rates, the obstacles induce particle trapping, producing subdiffusive motion.  By exploiting this effect, it could be possible to perform a filtering of active particles using an obstacle array, where a portion of the particles with specific properties would become trapped in the array while other types of particles could diffuse freely across it.

\citet{quint2013swarming} studied a swarming model in the presence of static disorder.  They observed that even a small amount of disorder can suppress the ordering of the swarm when only aligning interactions are included; however, when additional repulsive forces between particles are added, they found a transition from a collectively moving state to a gas-like state at a finite amount of disorder.  This transition resembles a percolation transition; however, the transition occurs at densities well below those at which standard percolation would be expected to appear. \citet{chepizhko2015active} also numerically considered swarming particles in heterogeneous media.

Studies of active particles interacting with obstacles are often performed in the limit where hydrodynamic effects can be neglected.  \citet{takagi2014hydrodynamic} experimentally studied chemically propelled microrods interacting with large spheres, and found that the rods become trapped in circles around the spheres as illustrated in Fig.~\ref{F31}b.  The behavior of the trapped rods can be explained using a model that includes the hydrodynamic interactions between the rod and the surface of the sphere.  \citet{spagnolie2015geometric} performed a numerical and theoretical analysis of the capture of microswimmers interacting with obstacles, showing that the time during which the swimmers are trapped by the obstacles before escaping has a distribution with a long tail. Similar hydrodynamic trapping effects were observed in bacteria swimming near large convex obstacles \cite{sipos2015hydrodynamic}.

\subsection{Sorting of microswimmers}\label{sec:sorting}

Sorting microswimmers based on their swimming properties (e.g. velocity, angular velocity, and chirality) is of utmost importance for various branches of science and engineering. Genetically engineered bacteria can be sorted based on phenotypic variations of their motion \cite{berg2004ecoli}. Velocity-based spermatozoa selection can be employed to enhance the success probability in artificial fertilization techniques \cite{guzik2001sperm}. Considering the intrinsic variability of microfabrication techniques, the efficiency of artificial microswimmers for a specific task, e.g. drug-delivery or bioremediation, can be increased by selecting only those with the most appropriate swimming properties. In this section we will explore how complex environments can be used to sort microswimmers based on their swimming style; we will, in particular, consider static patterns (Section~\ref{sec:staticsorting})
and chiral sorting (Section~\ref{sec:chiralsorting}).

\subsubsection{Static patterns}\label{sec:staticsorting}

\begin{figure*}
\includegraphics[width=\textwidth]{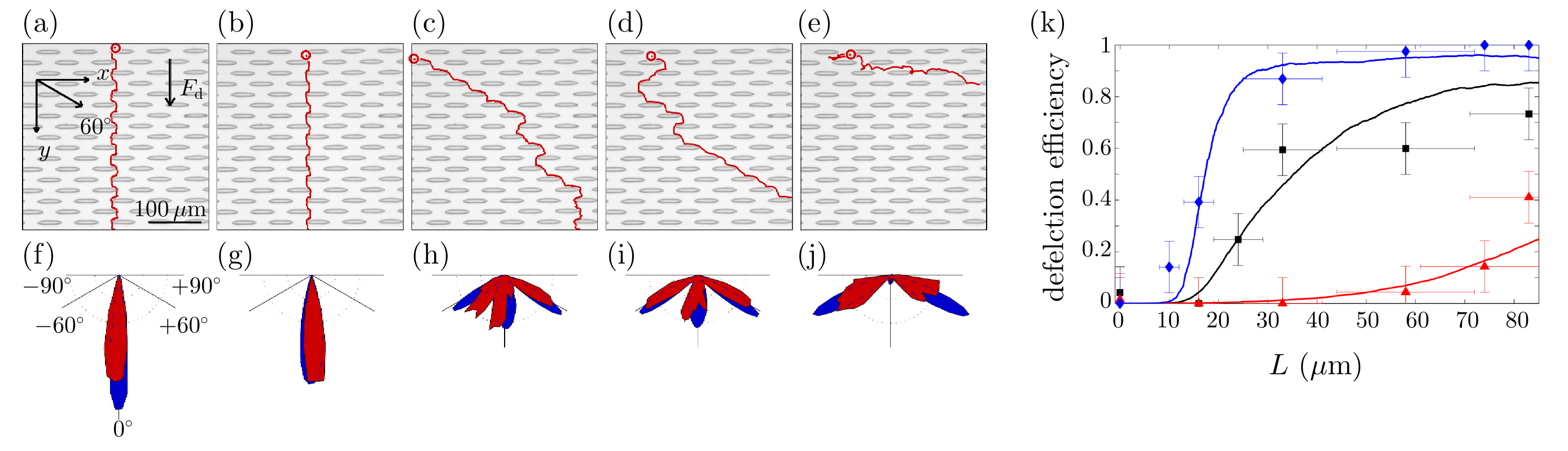}
\caption{(Color online) Sorting of microswimmers in a periodic environment.
(a-e) Typical trajectories of self-propelled particles moving through a triangular lattice (lattice constant $L_{\rm c} = 35\,\mathrm{\mu m}$) of elliptical obstacles when a drift force $F_{\rm d}=0.12\, \mathrm{pN}$ is applied along the $y$-direction. The characteristic swimming length is (a) $L = 0$ (Brownian particle, no propulsion), (b) $16$, (c) $24$, (d) $33$, and (e) $83\, \mathrm{\mu m}$. (f-j) Corresponding histograms of the experimentally measured (red/light gray) and simulated (blue/dark gray) directions of the particle trajectories as defined by two points in the trajectory separated by $100\, \mathrm{\mu m}$. The experimental histograms were obtained considering more than 100 trajectories in each case. 
(k) Measured probability that particles are deflected by more than $30^{\circ}$ after a travelling length of $100\, \mathrm{\mu m}$ as a function of $L$ for various imposed drift forces $F_{\rm d}=0.06\pm0.02$ (diamonds), $0.12\pm 0.05$ (squares), and $0.28\pm 0.12\, \mathrm{pN}$ (triangles). The solid lines are the results of numerical calculations.
From \citet{volpe2011microswimmers}.
\label{F32}
}
\end{figure*}

Two-dimensional periodic patterns can be used to select active particles based on their swimming style. \citet{volpe2011microswimmers} investigated the behavior of self-propelled particles in the presence of a two-dimensional periodic pattern where straight unlimited swims are only possible along certain directions. The structure was made of a series of ellipsoidal pillars arranged in a triangular lattice (lattice constant $L_{\rm c}=35\, \mathrm{\mu m}$, Fig.~\ref{F32}a). Within such structure, long swimming runs are only possible along two main directions: at $\pm 60^{\circ}$ and $\pm 90^{\circ}$ with respect to the $y$-axis. Otherwise the motion is strongly hindered due to collisions with the obstacles. The addition of a drift force $F_{\rm d}$ along the $y$-direction leads to strong differences in the particle trajectories depending on their swimming length.

The typical trajectory of a Brownian particle is shown in Fig.~\ref{F32}a. When compared to the effect of the drift force, the effect of the diffusion is rather weak so that the particle meanders almost deterministically through the structure in the direction of $F_{\rm d}$. For increasing swimming lengths $L$, however, significant changes in the trajectories are observed. These become particularly pronounced for $L>L_{\rm c}$, where the particles perform swimming runs of increasing length along the diagonal channels (Figs.~\ref{F32}c,d). For $L=83\, \mathrm{\mu m}$ the propulsion becomes so strong that the particles partially move perpendicular to the drift force (Fig.~\ref{F32}e); occasionally even motion against the drift force can be observed.

The direction of the particle motion through the structure is characterized by the direction (with respect to the $y$-axis) of the line connecting points of the trajectory separated by a distance of $100\, \mathrm{\mu m}$. The probability distributions of these angles are shown by the red/light gray polar histograms in Figs.~\ref{F32}f-j. One clearly observes that with increasing $L$ the propagation of particles along the direction of the applied drift becomes less likely, while trajectories along $\pm 60^{\circ}$, i.e. along the directions that permit long swimming events, become more frequent. Differently from the deflection of Brownian particles in a periodic potential \cite{macdonald2003microfluidic,reichhardt2004dynamic,huang2004displacement}, this mechanism relies on the dynamical properties of the microswimmers. These results were also compared with numerical simulations (blue/dark gray polar histograms in Figs.~\ref{F32}f-j), which show good agreement with the experimental data.

With the additional possibility of varying the drift force, these observations can be exploited to spatially separate self-propelled particles with small differences in their individual swimming behavior (Fig.~\ref{F32}k).  The sorting mechanism discussed here can be directly applied to other self-propelled objects; in these cases, drift forces can be created, e.g., by electric fields or by a solvent flow through the device. 

\subsubsection{Chiral particle separation}\label{sec:chiralsorting}

\begin{figure*}
\includegraphics[width=\textwidth]{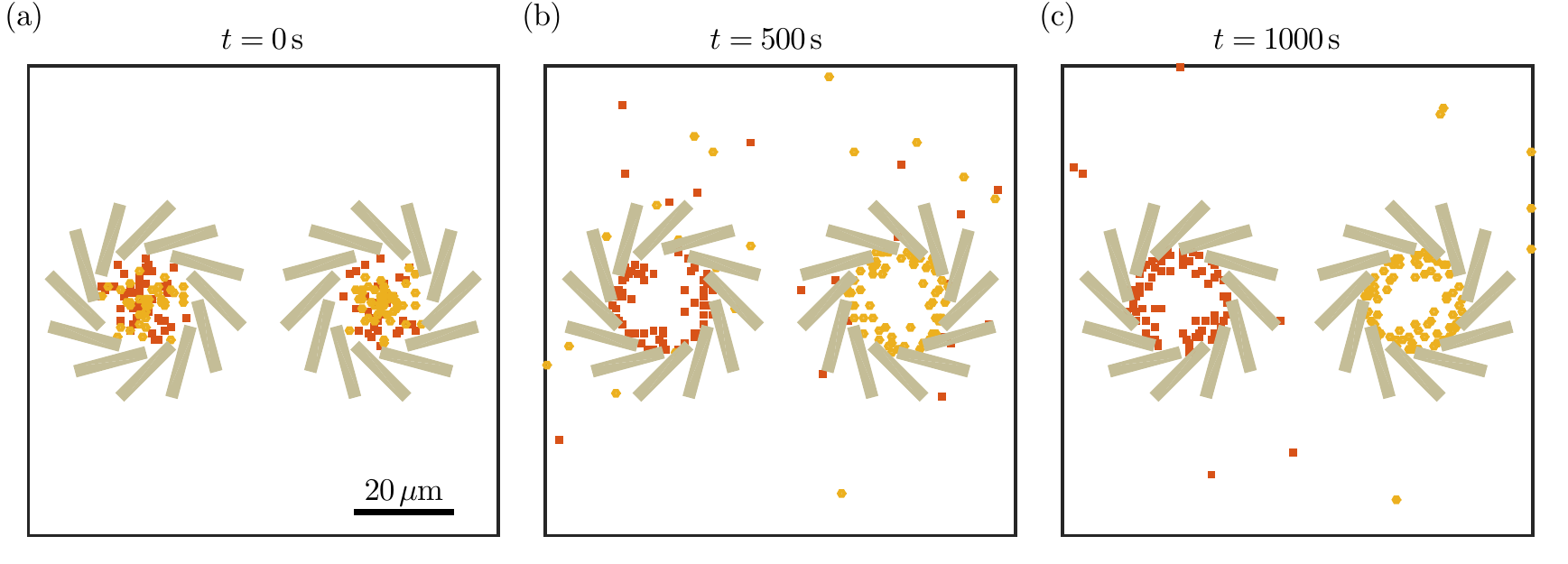}
\caption{(Color online) Sorting of chiral microswimmers ($R = 1\,{\rm \mu m}$, $v=31\,{\rm \mu m\,s^{-1}}$, and $\omega=\pm 3.14\,{\rm rad\,s^{-1}}$) with chiral flowers (rectangles). (a) At $t=0\,{\rm s}$ a racemic mixture of active particles is released inside two chiral flowers with opposite chiralities. (b-c) As time progresses, the levogyre (red/dark gray squares) active particles are trapped in the left chiral flower, while the dextrogyre ones (yellow/light gray circles) are trapped in the right chiral flower.
\label{F33}
}
\end{figure*}
 
Active particles can also be separated on the basis of their chirality \cite{mijalkov2013sorting,reichhardt2013dynamics,ai2015chirality,wu2015longitudinal,chen2015sorting}. This is a particularly interesting option because it may provide a better technique to separate levogyre and dextrogyre chiral molecules by chemically coupling them to chiral propellers, sorting the resulting chiral microswimmers, and finally detaching the propellers, as theoretically suggested by \citet{mijalkov2013sorting}; this is interesting because often only one specific chirality is needed by the biochemical and pharmaceutical industries \cite{ahuja2011chiral}, and separation can hardly be achieved by mechanical means due to the extremely small Reynolds numbers \cite{marcos2009separation}. Figure~\ref{F33} shows a possible approach to sorting active particles based on the sign of their motion chirality in the presence of some chiral patterns in the environment \cite{mijalkov2013sorting}, such as arrangements of tilted rectangles along circles forming \emph{chiral flowers}. Two chiral flowers with opposite chiralities are enclosed in a $100\,{\rm \mu m}$-side box where the particles can move freely. At time $t=0\,{\rm s}$ a racemic mixture is placed inside each flower (Fig.~\ref{F33}a). With time passing by, most of the levogyre (dextrogyre) microswimmers escape the right (left) chiral flower, while the ones with the opposite chirality remain trapped. At $t = 1000\,{\rm s}$ most of the microswimmers are stably trapped, as can be seen in Fig.~\ref{F33}c. Other approaches have also been proposed to separate chiral microswimmers, e.g. periodically patterned channels \cite{mijalkov2013sorting,li2014manipulating,ao2015diffusion,ai2015chirality} and rotary obstacles \cite{chen2015sorting}.

\section{Towards the Nanoscale}\label{sec:nano}

Scaling down microswimmers to the nanoscale and, therefore, implementing effective nanoswimmers is an open technological and scientific challenge \cite{peplow2015march}. Being successful in this task would be extremely beneficial: it would in fact allow the scaling down of several technologies both increasing their efficiency and decreasing their footprint. The main problem on the way towards the nanoscale is that, as the size of an active particle decreases, its motion becomes more random and less directed mainly because of an increase in the particle's rotational diffusion. In fact, by comparing Eq.~(\ref{eq:Dt}) and Eq.~(\ref{eq:Dr}), we can see that, while $D_{\rm T}$ scales according to $R^{-1}$, $D_{\rm R}$ scales according to  $R^{-3}$ so that, when we approach nanometric sizes, $\tau_{\rm R}$ becomes very short. For instance, for a nanoparticle with hydrodynamic radius $R = 50\,{\rm nm}$, $\tau_{\rm R} = 760\,{\rm \mu s}$. Thus, self-propulsion does not have enough time to alter the way in which a nanoswimmer interacts with its environment; its main effect is, instead, to enhance the particle's diffusion, generating what has been refereed to as \emph{hot Brownian motion} \cite{rings2010hot}.

Nevertheless, not everything is lost. In fact, different solutions can be brought into play to outbalance the increased rotational diffusion at nanometric swimmers' sizes, e.g. by exploiting:
\begin{description}
\item[Taxis] Taxis is the movement of a system in response to an external stimulus such as light or the presence of a chemical plume. For example, nanoswimmers can be designed to respond to the gradient in an external (e.g. chemical, optical, acoustic) field that powers their propulsion mechanism so that a long-term drift appears in their motion as a response to the gradient. The gradient can even counteract the rotational diffusion by inducing alignment of the swimmers. Due to the increased rotational motion at the nanoscale, however, there exists a size limit of $\approx 600\,{\rm nm}$ below which tactic strategies (e.g. swim-and-tumble motion) are no longer efficient \cite{dusenbery1997minimum}.
\item[Environmental properties] Nanoswimmers can be steered through non-Newtonian media exploiting their interaction with the fine structure of the surrounding media. For example, some screw-propellers that have a filament diameter of about 70 nm have been shown to be capable of moving through high-viscosity solutions with velocities comparable to those of larger micropropellers, even though they are so small that Brownian forces suppress their actuation in pure water: when actuated in viscoelastic hyaluronan gels, these nanopropellers appear to have a significant advantage because they are of the same size range as the gel's mesh size \cite{schamel2014nanopropellers}. 
\item [Swarming effects] Finally, collective effects in the interaction of several nanoswimmers can also lead to controllable directed motion due to short-range coupling of the motion of the individual entities \cite{saha2014clusters}. 
\end{description}

\section{Outlook and Future Directions}\label{sec:conclusions}

Active particles have the potential to have far-reaching impact on many fields of science and technology. From the fundamental side, their study can shed light on the far-from-equilibrium physics underlying the adaptive and collective behavior of biological entities such as chemotactic bacteria and eukaryotic cells. From the more applied side, they provide options to perform tasks not easily achievable with other available techniques, such as the targeted localization, pick-up, and delivery of microscopic and nanoscopic cargoes, e.g. in healthcare, environmental sustainability, and security.

However, there are still several open challenges that need to be tackled in order to achieve the full scientific and technological potential of microswimmers in real-life settings. In our opinion, the main scientific and technological challenges are: 
\begin{enumerate}
\item to understand more in depth and exploit their behavior in complex and crowded environments; 
\item to learn how to engineer emergent behaviors;
\item to identify biocompatible propulsion mechanisms and energy supplies capable of lasting for the whole particle life-cycle; and
\item to scale down their dimensions towards the nanoscale.
\end{enumerate}
These challenges will necessarily take years or decades to be met and will need to involve concerted effort from academy and industry: in fact, although significant research progress has been accomplished over the past decade, the applicability of microswimmers to real-life problems is still in its infancy since current proof-of-concept research efforts on micro- and nanoswimmers need to be transformed into larger-scale pilot studies, and eventually into field applications. In all the future envisioned applications, in particular, understanding the physical interaction of micro- and nanoswimmers with their complex environment (e.g. tissues in living organisms, microchannels in lab-on-a-chip devices, obstacles and physical barriers in soils) represents an enabling step for the design of autonomous micro- and nanomachines that can navigate efficiently in realistic conditions of operation.

Concretely, a series of so far unexplored (and potentially extremely interesting) research directions can be identified --- going from the most straightforward to the most imaginative:
\begin{description}

\item[Addressing fundamental questions]
Some fundamental problems are still open and need to be addressed in order for the field to be able to progress. These include, in particular, to understand how self-propelled Janus particles swim and interact; to explore under what conditions can thermodynamic equilibrium concepts be applied to detailed-balance-violating systems such as active matter; and to make synthetic microswimmers that will behave well at high densities avoiding problems with, e.g., the formation of gas bubbles and fuel starvation.

\item[Time-varying environments] 
Studying the behavior of microswimmers in time-varying potentials and complex environments is an interesting future direction of both fundamental and applied research.

\item[Colloidal molecules] 
Formation of colloidal molecules between active particles (or passive particles in an active bath) with matching shapes is an interesting possibility. This has already been explored for the case of passive colloidal particles, e.g. the depletion-driven lock-and-key mechanism presented in \citet{sacanna2010lock} and the self-assembly of nanostructures tunable using critical Casimir forces presented in \citet{faber2013controlling}. The concept of passive colloidal molecules \cite{manoharan2003dense,van2003chemistry,kraft2009colloidal} was indeed recently generalized to active colloidal molecules. The latter comprise clusters of either diffusiophoretic particles \cite{soto2014self,soto2015self}, or tightly bound \cite{babel2015dynamics} or magnetically attractive microswimmers \cite{kaiser2015active}. In this context some new directions emerge from the possibility of using critical Casimir forces to fine-tune the particle assembly \cite{hertlein2008direct,paladugu2016nonadditivity}.

\item[Hybrid systems]
An interesting possibility is to use a combination of microfabricated and biological elements to generate motion on the nanoscale. For example, \citet{gao2015using} trapped smooth swimming bacteria in micron-sized structures and used their rotating flagella to generate a flow capable of transporting materials along designed trajectories.

\item[Emergent collective organization] 
Albeit recently some studies have found interesting novel forms of aggregation in crowded suspensions of microswimmers \cite{palacci2013living,buttinoni2013dynamical}, they still consider only one kind of microswimmer at a time. Nevertheless, we know from recent experiments in robotics \cite{halloy2007social,rubenstein2014programmable,mijalkov2015engineering,volpe2016effective} and biology \cite{gravish2015glass,tennenbaum2016mechanics} that complex and robust collective behaviors can arise also in the presence of very simple constituent agents. It is therefore appealing the possibility of translating these results from robotic swarms to microswimmers and of exploring how several different kinds of microswimmers can synergistically interact to perform a task.

\item[Stochastic thermodynamics in active baths]
Stochastic thermodynamics is an emergent field of research \cite{jarzynski2011equalities,seifert2012stochastic}. Until now mostly systems coupled to a thermal bath and, therefore, satisfying Boltzmann statistics have been considered. However, systems satisfying these conditions are limited and, in particular, do not include many systems that are intrinsically far from equilibrium such as living matter, for which fluctuations are expected to be non-Gaussian. For example, biomolecules within the cell are coupled with an active bath due to the presence of molecular motors within the cytoplasm, which leads to striking and largely not yet understood phenomena such as the emergence of anomalous diffusion \cite{barkai2012strange}. Also, protein folding might be facilitated by the presence of active fluctuations \cite{harder2014activity} and active matter dynamics could play a central role in several biological functions \cite{suzuki2015polar,mallory2015anomalous,shin2015facilitation}. It is therefore an open and compelling question to assess whether and to what degree stochastic thermodynamics can be applied to systems coupled to active baths \cite{argun2016experimental}.

\item[Nanoswimmers]
Active matter systems have so far  been confined to microscopic (or larger) scales; however, with advances in creating artificial swimmers it could be possible to realize active matter systems  on much smaller scales.  In these regimes, thermal effects will be much more pronounced at room temperature and could mask the effects of the activity, as we have seen in Section~\ref{sec:nano}; however, experiments could be performed at lower temperatures and a new field of low-temperature active matter systems could be realized. At these smaller scales new effects may arise such as van der Waals forces, charging, or other fluctuation forces, which can lead to new types of active matter emergent behaviors. In terms of applications, activity on the nanoscale can be useful for creating new methods for self-assembly or pattern formation, changing the nature of fluids on the nanoscale, or serving as the first step to nanoscale probes and machines for medical applications.

\item[Active polymers] The interaction of passive or active polymers with an active bath will be a very interesting field.  Open questions include how reptation dynamics might change when a chain is active rather than thermal, whether active polymer chains could produce unusual rheological properties, whether entanglement would increase or decrease under different activity levels, what would happen if the polymers were breakable, and whether glassy or jamming behavior can arise in dense active polymer mixtures.  Extensions of such studies could be applied to understanding membranes, sheets, or flexible tubes in the presence of active matter baths.

\item[Underdamped active particles]
In most active matter systems the swimmers are considered to be moving in an overdamped medium. An alternative area of exploration would be to consider cases of active systems where the damping is significantly reduced so that inertial effects  can  play an important role. One example might be to make active Janus particles that move through a dusty plasma \cite{morfill2009complex}, air, or even a  vacuum. In this context, an important class of systems are dusty plasma bilayers governed by effective nonreciprocal wake forces \cite{ivlev2015statistical}, which under appropriate conditions form self-propelled pairs \cite{bartnick2015emerging}. Another example are self-propelling microdiodes \cite{chang2007remotely,sharma2015remote}. Examples of questions that could be asked are whether activity-induced clustering transitions remain robust when  inertial effects are present, or whether clustering can happen in a vacuum where dissipation only occurs at the particle--particle interaction level. When inertia is present, many nonlinear systems can exhibit a range of collective dynamical behaviors  such as solitons, nonlinear waves, and shock phenomena \cite{infeld2000nonlinear}. If the medium is active, then new kinds of active soliton behaviors could occur, as well as new classes of self-propagating waves or shock-like phenomena.

\item[Swimmers in non-viscous fluids]
Another question to ask is how artificial swimmers would behave in different kinds of fluids. Such effects could include shear-thickening or shear-thinning fluids.  It is also possible to explore anisotropic swimmers moving in a superfluid \cite{leggett1999superfluidity}, where new  effects could arise due to the frictionless nature of the flow, as well as the generation of quantized vortex-anti-vortex couples when the velocity of the active particles is above the superfluid velocity \cite{bewley2006superfluid}. Possible experiments could be performed using swimmers in  liquid helium \cite{bewley2006superfluid} or in superfluids created in Bose-Einstein condensates \cite{madison2000vortex}.

\item[Non-local interactions to mimic quantum effects]
There have been a variety of efforts to create collective classical systems that mimic quantum phenomena \cite{couder2006single,bush2015new}. Another future direction for active matter systems would be to explore the feasibility of making artificial swimmers that have effective non-local interactions generated via coupling or feedback responses, such as by using light fields or arrays of swimming robots with transmitters. For the correct choice of coupling rules, the active matter assembly could mimic the behavior of quantum systems.  In this way swarms of swimmers could coordinate to form an effective wavefunction-type object. On a more theoretical side it could be interesting to consider how activity would affect tunnelling, interference, and entanglement.

\item[Topological active matter systems on lattices]
Recently there has been growing interest in  topological effects in both quantum \cite{hasan2010colloquium,ran2009one} and classical systems \cite{kane2014topological,paulose2015topological}, where the behavior at  the edge of the system is different from that in the bulk. Active particles moving on various kinds of finite lattices offer a system in which such topological effects could be investigated.  For example, the activity could be enhanced at or occur only on the edges of the system.  Additionally, in dense active matter systems higher order topological objects such as dislocations could arise that could have dynamical properties that are significantly different from those of similar objects in purely thermal systems. For active matter on substrates, certain constraints could be imposed by the geometry of the substrate lattice that would allow only certain topological modes to propagate.

\item[Active matter on active substrates]
So far, active matter systems have been studied in the absence of a substrate or interacting with rigid substrates; however, in many real-world systems, particularly biological systems, the substrate is flexible rather than rigid, and in some cases the substrate itself is active. Thus,  another future direction would be understanding how active matter systems couple to flexible environments where the activity could in principle change the environment or cause some form of self-organization to occur. If the substrate itself is active, then some type of resonance could arise when the time scales of the substrate and the active matter match, and it may be possible that biological active matter systems could harness the activity of a coupled system to enhance certain functions.

\end{description}

\section{Acknowledgments}

We have benefited immensely from discussions with many colleagues and friends in the last few years. We would like to thank especially Aykut Argun, Ivo Buttinoni, Agnese Callegari, Frank Cichos, Peer Fisher, Gerhard Gompper, Andreas Kaiser, Felix K\"ummel, Mite Mijalkov, Roland Netz, Cynthia J. Olson-Reichhardt, Fernando Peruani, Roberto Piazza, Maurizio Righini, Holger Stark, Borge ten Hagen, Sabareeh K. P. Velu, Jan Wehr, Roland Winkler, and Raphael Wittkowski.

Clemens Bechinger and Hartmut L\"owen acknowledge funding from the SPP 1726 of the DFG.
Roberto Di Leonardo acknowledges funding from the European Research Council Grant Agreement No. 307940.
Giorgio Volpe acknowledges support from COST Action MP1305.
Giovanni Volpe acknowledges funding from Marie Curie Career Integration Grant (MC-CIG) PCIG11GA-2012-321726,  a Distinguished Young Scientist award of the Turkish Academy of Sciences (T\"UBA), and COST Actions MP1205, IC1208 and MP1305.


%

\end{document}